\documentclass[prx,tightenlines,10pt,notitlepage,nofootinbib,twocolumn,superscriptaddress, floatfix]{revtex4-2}
\usepackage{amsmath,epsfig,amssymb}
\usepackage{bbm}
\usepackage{bm}
\usepackage{times}
\usepackage{color}
\usepackage{amsthm}
\usepackage{color}
\usepackage[caption=false]{subfig}
\usepackage{multirow}
\usepackage[normalem]{ulem}
\usepackage{qcircuit}
\usepackage{braket}
\usepackage{amsfonts,amssymb,dsfont}
\usepackage{graphicx}
\usepackage{hyperref}
\usepackage{inputenc}
\usepackage{dsfont}
\usepackage{float}
\usepackage{array}
\usepackage{tabularx}
\usepackage[capitalise]{cleveref}

\usepackage[acronym,shortcuts]{glossaries}

\hypersetup{
    colorlinks=true,       
    linkcolor=red,          
  citecolor=magenta,        
    filecolor=magenta,      
    urlcolor=cyan,           
    runcolor=cyan
}

\def\sx{\sigma^x}
\def\sy{\sigma^y}
\def\sz{\sigma^z}

\def\Tr{\text{Tr}}

\def\Tr{\mathrm{Tr}}

\def\>{\rangle}
\def\<{\langle}
\newcommand{\ketbra}[1]{|{#1}\>\!\<#1|}

\newcommand{\ketb}[2]{|{#1}\>\!\<#2|}

\newcolumntype{C}[1]{>{\centering\arraybackslash}p{#1}}

\definecolor{amethyst}{rgb}{0.6, 0.4, 0.8}

\newcommand{\beq} {\begin{equation}}
\newcommand{\eeq} {\end{equation}}
\newcommand{\bes} {\begin{subequations}}
\newcommand{\ees} {\end{subequations}}

\begin{document}
\title{Dynamically Generated Decoherence-Free Subspaces and Subsystems on Superconducting Qubits}

\author{Gregory Quiroz}
\affiliation{Johns Hopkins University Applied Physics Laboratory, Laurel, Maryland 20723, USA}
\affiliation{William H. Miller III Department of Physics \& Astronomy,
Johns Hopkins University, Baltimore, Maryland 21218, USA}

\author{Bibek Pokharel}
\affiliation{Department of Electrical Engineering, University of Southern California, Los Angeles, CA 90089}
\affiliation{Center for Quantum Information Science \& Technology,
University of Southern California, Los Angeles, CA 90089, USA}

\author{Joseph Boen}
\affiliation{Johns Hopkins University Applied Physics Laboratory, Laurel, Maryland 20723, USA}

\author{Lina Tewala}
\affiliation{Johns Hopkins University Applied Physics Laboratory, Laurel, Maryland 20723, USA}
\affiliation{Thomas C. Jenkins Department of Biophysics,
Johns Hopkins University
Baltimore, MD, 21218, USA
}

\author{Vinay Tripathi}
\affiliation{Center for Quantum Information Science \& Technology,
University of Southern California, Los Angeles, CA 90089, USA}
\affiliation{Department of Physics and Astronomy, University of Southern California, Los Angeles, CA 90089}

\author{Devon Williams}
\affiliation{Johns Hopkins University Applied Physics Laboratory, Laurel, Maryland 20723, USA}
\affiliation{William H. Miller III Department of Physics \& Astronomy,
Johns Hopkins University, Baltimore, Maryland 21218, USA}

 \author{Lian-Ao Wu}
 \affiliation{Department of Theoretical Physics and History of Science,
 The Basque Country University (EHU/UPV), 48008, Spain}
 \affiliation{IKERBASQUE, Basque Foundation for Science, 48011 Bilbao, Spain}
 \affiliation{EHU Quantum Center, University of the Basque Country UPV/EHU, Leioa, Biscay 48940, Spain}

\author{Paraj Titum}
\affiliation{Johns Hopkins University Applied Physics Laboratory, Laurel, Maryland 20723, USA}

\author{Kevin Schultz}
\affiliation{Johns Hopkins University Applied Physics Laboratory, Laurel, Maryland 20723, USA}

\author{Daniel Lidar}
\affiliation{Department of Electrical Engineering, University of Southern California, Los Angeles, CA 90089}
\affiliation{Center for Quantum Information Science \& Technology,
University of Southern California, Los Angeles, CA 90089, USA}
\affiliation{Department of Physics and Astronomy, University of Southern California, Los Angeles, CA 90089}
\affiliation{Department of Chemistry, University of Southern California, Los Angeles, CA 90089}

\begin{abstract}
Decoherence-free subspaces and subsystems (DFS) preserve quantum information by encoding it into symmetry-protected states unaffected by decoherence. An inherent DFS of a given experimental system may not exist; however, through the use of dynamical decoupling (DD), one can induce symmetries that support DFSs. Here, we provide the first experimental demonstration of DD-generated DFS logical qubits. Utilizing IBM Quantum superconducting processors, we investigate two and three-qubit DFS codes comprising up to six and seven noninteracting logical qubits, respectively. Through a combination of DD and error detection, we show that DFS logical qubits can achieve up to a 23\% improvement in state preservation fidelity over physical qubits subject to DD alone. This constitutes a beyond-breakeven fidelity improvement for DFS-encoded qubits. Our results showcase the potential utility of DFS codes as a pathway toward enhanced computational accuracy via logical encoding on quantum processors.
\end{abstract}
\maketitle

%
%
\section{Introduction}
Scalable quantum computation relies on the ability to perform high-fidelity quantum logic operations. The path toward such operations is challenging due to inherent system-environment interactions and systematic errors. Ultimately, both induce noise processes that degrade qubit coherence and gate accuracy. Therefore, addressing noise in quantum systems is paramount to attaining viable and reliable quantum computation.

Broadly, approaches designed to manage noise in quantum systems seek to suppress, correct, or avoid errors~\cite{lidar_brun_2013}. Error suppression approaches (e.g., dynamical decoupling (DD)~\cite{Viola:98,viola1999dynamical,zanardiSymmetrizingEvolutions1999}) rely on the application of appropriately modulated control fields~\cite{Gordon:2008:010403} such as fast and strong pulses to effectively average out noise~\cite{Suter:2016aa}. In contrast, quantum error correction (QEC) leverages logical encodings of a collection of physical qubits to actively detect and correct errors~\cite{shor1995scheme,Steane:96a,Gottesman:1996fk,Gaitan:book}. As a passive alternative, decoherence-free subspaces (DFSs) and noiseless subsystems (NS) form a special class of quantum codes that provide error avoidance by exploiting symmetries in the system-environment interaction~\cite{Alicki:88,Palma:96,zanardiNoiselessQuantumCodes1997,Duan:1997aa,lidar1998decoherence,Knill:2000dq}. The three approaches can be unified under a single, symmetry-based framework~\cite{Zanardi:99d}. Error mitigation, the newest category of quantum error management, utilizes information from an ensemble of quantum experiments to reduce noise-biasing in expectation values~\cite{cai2022quantum}. While in principle, each class of protocols can be employed on its own, it has long been appreciated that practical quantum error management schemes are likely to necessitate multiple approaches working in concert~\cite{lidar1999qec-dfs,Lidar:PRA00Exchange,violaDynamicalGenerationNoiseless2000,Alber:01,Alber:02a,KhodjastehLidar:02,KhodjastehLidar:03,ngCombiningDynamicalDecoupling2011,Paz-Silva:2013tt} to achieve utility-scale quantum computation and, eventually, fault tolerance~\cite{Campbell:2017aa}.

Despite the elusiveness of fault tolerance, utility-scale quantum computing may be on the horizon in part due to advancements in error management. Demonstrations on currently available noisy quantum processors have showcased the potential for classes of protocols to be executed independently and simultaneously. For example, confirmation of quantum error mitigation's effectiveness has been shown for quantum algorithms, such as the variational quantum eigensolver~\cite{Dumitrescu2018vqe, kandala2019error} and quantum dynamics simulations~\cite{kim2023scalable}. Relatedly, 
a quantum algorithmic scaling advantage enabled by error suppression via DD has recently been demonstrated in superconducting systems
\cite{pokharelDemonstrationAlgorithmicQuantum2022}
building on longstanding experimental evidence of its utility~\cite{carr1954effects, meiboom1958modified, haeberlen1968coherent, biercuk2009experimental,Biercuk:09,Sagi:2010aa,naydenov2011dynamical,Sar:2012km,souza2012robust,pokharelDemonstrationFidelityImprovement2018, jurcevicDemonstrationQuantumVolume2020,Aharony:2021aa, tripathiSuppressionCrosstalkSuperconducting2022, ezzellDynamicalDecouplingSuperconducting2022,zhouQuantumCrosstalkRobust2022}. Combining the two approaches has also been shown to be fruitful for enhancing quantum algorithm performance~\cite{kim2023scalable}.

\begin{figure*}[t]
    \centering
    \includegraphics[width=\textwidth]{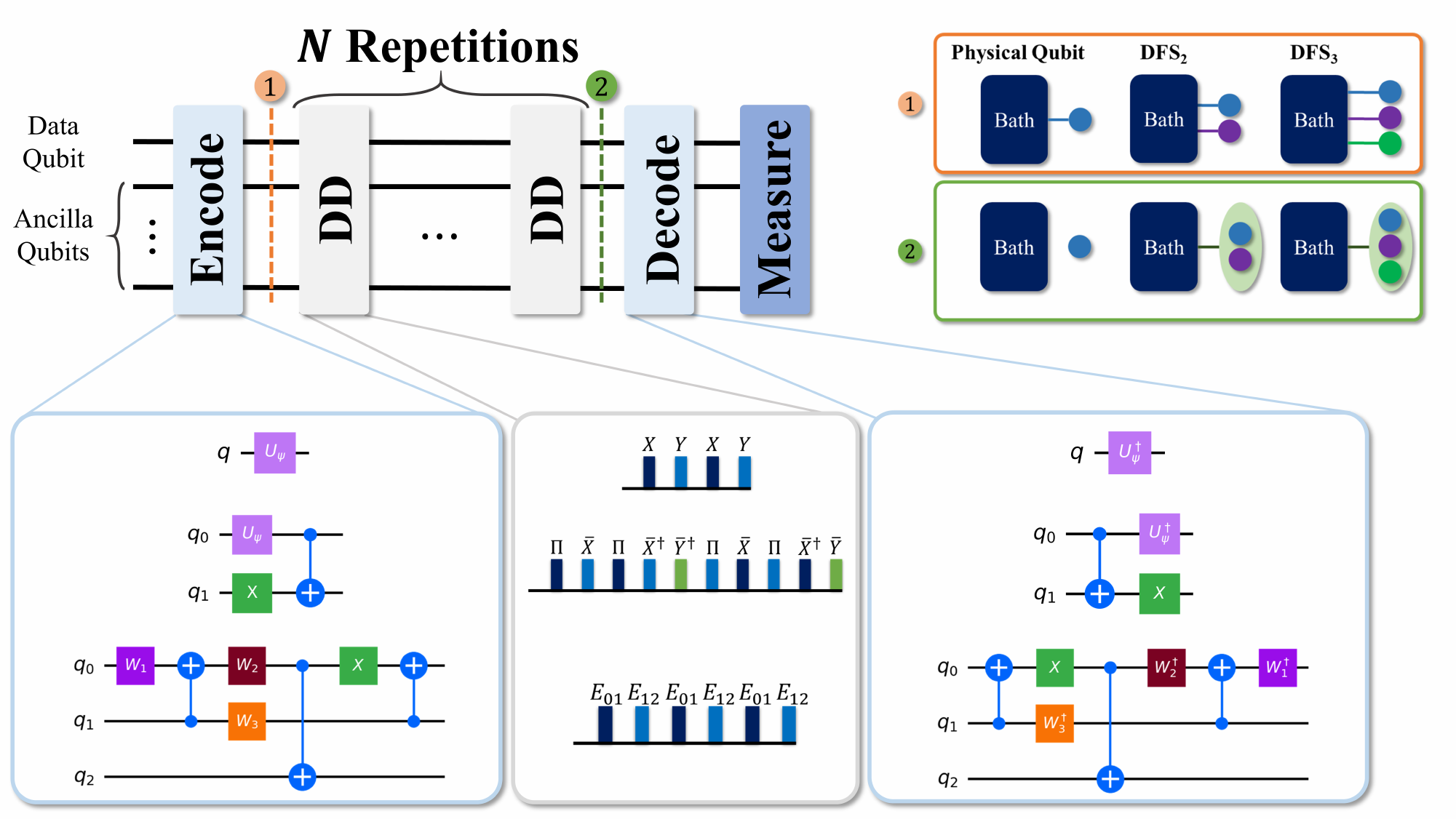}
        \caption{Schematic of the quantum state preservation experiment for all encoding schemes considered: unencoded (physical), 2-qubit DFS (DFS$_2$), and 3-qubit DFS (DFS$_3$). The qubits are initially prepared according to the encoding scheme and then subjected to $N$ rounds of DD or the equivalent free evolution duration, where the system is allowed to evolve according to its internal dynamics. The inverse of the encoding procedure is then used to decode the qubit states prior to measuring in the computational basis. The encoding step for the DFS$_2$ code utilizes the unitary $U_{\psi}$ to prepare the desired single-qubit state $\ket{\psi}$ for the data qubit. In the 3-qubit encoding, $W_j$, $j=1,2,3$ are unitaries dependent upon the specifications of $\ket{\psi}$. Each encoding scheme is associated with a specific DD protocol. Unencoded evolution utilizes the XY4 universal decoupling sequence, while the 2-qubit DFS symmetrization sequence is composed of logical rotations about the $x$ and $y$-axes of the logical qubit Bloch sphere. Noise symmetrization is achieved in the 3-qubit case using a sequence of SWAP operators $E_{ij}$. Further information regarding the encoding circuits and DD sequences can be found in the appendix.} 
    \label{fig:setup}
  \end{figure*}

Noisy quantum devices have further led to proof-of-principle demonstrations of QEC~\cite{PhysRevLett.94.130501,harper2019ftgates, andersen2020repeated, chen2021qec, krinner2022realizing, sivak2023real, miao2022overcoming,ai2023suppressing}. This has included instances of error detection utilized to protect variational quantum algorithms~\cite{urbanek2020vqe}. Furthermore, verification of the added benefits of DD has been observed. For example, it has been incorporated into error correcting codes to protect idle qubits during long syndrome measurement acquisition and reset periods~\cite{chen2021qec, krinner2022realizing,ai2023suppressing,postler2023demonstration,bluvstein2023logical}.
Error mitigation, DD, and quantum error detection have all been combined in a recent demonstration of better-than-classical execution of Grover's algorithm~\cite{Pokharel:better-than-classical-Grover}.
    
Despite early experimental instantiations~\cite{kwiat2000experimental, viola2001experimental, PhysRevLett.91.217904,mohseni2003dfs-exp, fortunato2003ns-exp, altepeter2004dfs-exp, pushin2011dfs-exp}, DFSs have yet to be examined as a scalable approach to error management in the current quantum computing era~\cite{Preskill2018}. Their ability to circumvent measurement-based feedback gives them a potential advantage over their error-correcting counterparts. Furthermore, DFS codes can be readily integrated with error suppression to dynamically engineer the required symmetries of the code~\cite{violaDynamicalGenerationNoiseless2000,wuCreatingDecoherenceFreeSubspaces2002,wuEfficientUniversalLeakage2002,Byrd:2002:047901,LidarWu:02,lidarQuantumComputersDecoherence2003,ByrdLidarWuZanardi:05,fongUniversalQuantumComputation2011,west2012exchange}. Given their relative economy of qubit- and control-resource requirements, the question arises: what role can DFSs play in practical error management schemes for near-term and future quantum processors?

We address this question specifically for error-protected quantum memory and idle gates. DFSs are employed in conjunction with DD and error detection procedures to preserve logical qubit states on the IBM Quantum Platform (IBMQP) superconducting qubit processors. Specially designed DD sequences are used to engineer an effective noise environment and enforce symmetry conditions conducive to DFSs and NSs composed of two and three physical qubits, respectively. We present evidence for the existence of dynamically-generated DFSs and demonstrate their ability to surpass the fidelity of physical qubits subject to DD alone. The scalability of DFSs is showcased through the simultaneous generation and preservation of multiple logical qubits, where logical qubit fidelity up to 23\% better than those achieved by the physical qubit constituents is observed. Together, these results highlight the potential utility of DFSs when used in conjunction with error suppression and detection procedures to enhance logical error management on quantum processors.

An overview of our methodology is presented in \cref{fig:setup}. Physical and logical protocols are compared by first encoding the qubits and then subjecting them to an error management protocol. As the objective is to evaluate each protocol's ability to preserve qubit states, we apply repetitions of said protocols for equivalent time durations. State fidelity is then determined by applying a decoding operation and measuring in the computational basis. Physical qubits are protected by error suppression, while logical DFS qubits undergo error suppression and detection. We focus on the 2 and 3-qubit DFS code, with encoding operations and dynamical symmetry-generating sequences for each shown in the bottom-left and bottom-middle of \cref{fig:setup}, respectively. Error detection is applied after measurement via post-selection based on the state of the ancilla qubits.

This paper is organized as follows. In \cref{sec:dfs-dd}, background on DFS codes and DD is presented. \cref{sec:evidence-col-sym,sec:preserve-qubits} showcase the results of the study. Evidence of collective interactions generated by logical DD is exhibited in \cref{sec:evidence-col-sym}, while time-dependent preservation is the focus of \cref{sec:preserve-qubits}. In the latter, we assess the performance of logical encodings against physical error suppression and highlight instances where DFS codes prevail. \cref{sec:concl} summarizes the results and concludes.

%
%
\section{Decoherence-Free Subspaces and Dynamical Decoupling}
\label{sec:dfs-dd}
\subsection{Decoherence-Free Subspaces and Noiseless Subsystems}
\label{subsec:dfs-ns}
As passive error-correcting codes, DFSs leverage intrinsic symmetries in the system-environment interaction. Their construction is based on identifying subspaces of the system Hilbert space unaffected by noise. More concretely, consider an open quantum system described by the Hamiltonian
\begin{equation}
    H = H_S + H_B + H_{SB},
    \label{eq:H}
\end{equation}
where $H_S$ is the pure system Hamiltonian and $H_B$ is the pure bath Hamiltonian, generating dynamics only within the system and bath Hilbert spaces $\mathcal{H}_S$ and $\mathcal{H}_B$, respectively. The interaction between the system and its environment is captured by $H_{SB}$,
which can be generically expressed as $H_{SB}=\sum_\alpha A_\alpha \otimes B_\alpha$, where $\{A_{\alpha}\}$ and $\{B_{\alpha}\}$ act exclusively on the system and bath, respectively. 

DFS encoding relies on finding a ``good" subspace $\mathcal{H}_G \subset \mathcal{H}_S$ that is unaffected by $H_{SB}$. As long as the system is initialized in $\mathcal{H}_G$ and no operations are performed by $H_S$ to take the system out of $\mathcal{H}_G$, the time evolution will be unaffected by the decoherence resulting from $H_{SB}$~\cite{lidar1998decoherence}. Group theoretic arguments show that under reasonable mathematical assumptions about the system operators $A_\alpha$, $\mathcal{H}_{G}$ always exists and it is possible to perform scalable, universal quantum computation in the DFS~\cite{kempe2001theory}. More precisely, the system Hilbert space can be decomposed as a direct sum over the irreducible representations (irreps) $J$ of the associative algebra $\mathcal{A}$ generated by the set $\{ A_\alpha \}$: \(\mathcal{H}_{S}=\bigoplus_{J} \mathbb{C}^{n_{J}} \otimes \mathbb{C}^{d_{J}}\), such that each noiseless subsystem $\mathbb{C}^{n_{J}}$, where $n_J$ is the degeneracy of irrep $J$, is invariant under the effects of $\mathcal{A}$~\cite{Knill:2000dq}. That is,
\begin{equation}
    A_\alpha \ket{a}\otimes \ket{b} = \ket{a} \otimes M_\alpha \ket{b}\quad \forall \alpha,
    \label{eq:dfs-cond}
\end{equation}
where the states $\ket{a}\in \mathbb{C}^{n_{J}}$ remain invariant under the error algebra $\mathcal{A}$, while $\ket{b}\in \mathbb{C}^{d_{J}}$ can be altered by the arbitrary operator $M_\alpha$ without consequence to the computation. Quantum information is stored in $\mathbb{C}^{n_J}$. When the irrep dimension $d_J=1$, $\mathbb{C}^{n_{J}} \otimes \mathbb{C}^{d_{J}}$ reduces to a DFS, i.e., a ``good" subspace $\mathcal{H}_G = \mathbb{C}^{n_{J}}$. 

In this study, we employ DFS protection against collective interactions, which arises when the system-bath coupling is invariant under qubit permutation. The $N$-qubit system-environment interaction under such permutation symmetry is fully described in terms of the total spin operator $A_{\alpha} = \sum_{i=1}^{N} \sigma^{\alpha}_{i}$, where $\alpha \in \{ +,-,z \}$; thus, $H^{\rm col}_{SB} = \sum_{\alpha \in\{+,-, z\}} A_{\alpha} \otimes B_{\alpha}$. Below, DFS codes consisting of $N$ qubits will be denoted as DFS$_N$ for brevity.

\subsubsection{Collective Dephasing DFS}
\label{subsubsec:col-dfs}
The simplest type of collective system-bath interaction for which a DFS can be identified is collective dephasing, i,e., $H^{\rm col}_{SB}$ with $B_\pm\equiv 0$. In turn, one can identify the smallest logical qubit encoded within two physical qubits. The subspace invariant under collective dephasing, which we refer to as DFS$_2$, is spanned by the logical states $\ket{0_L}=\ket{01}$ and $\ket{1_L}=\ket{10}$. A circuit describing the generation of an arbitrary state $\ket{\psi}$ within the logical subspace is shown in \cref{fig:setup}. Logical manipulations that preserve the DFS consist of all Hermitian operators that belong to the commutant of the error algebra, $\mathcal{A}_z\equiv\{O:[O,S_z]=0\}$, i.e., all operators $O$ that commute with the noise. One such set of encoding operators is given by
\bes
    \label{eq:dfs-logi-ops}
\begin{align}
    \bar{\sigma}^x&=\frac{1}{2}(\sx_1\sx_2 + \sy_1\sy_2)\\
    \bar{\sigma}^y&=\frac{1}{2}(\sy_1\sx_2 - \sx_1\sy_2),
\end{align}
\ees
where $\bar{\sigma}^z=i[\bar{\sigma}^x,\bar{\sigma}^y]/2$~\cite{kempe2001theory,fortunato2002implementation}. Logical rotations within the DFS are therefore defined by $\bar{R}_{\hat{n}}(\theta) = \exp(-i\theta \hat{n} \cdot\vec{\bar{\sigma}})$, where $\vec{\bar{\sigma}}=(\bar{\sigma}^x, \bar{\sigma}^y, \bar{\sigma}^z)$.

\subsubsection{Collective Decoherence DFS}
\label{subsubsec:col-ns}
Decoherence-free subsystems, or NSs, build upon the notion of noise-invariant subspaces to more generally define subsystems corresponding to preserved degrees of freedom. Such a subsystem can be constructed when a quantum system is subject to collective decoherence, i.e., $B_z\neq0$ and $B_\pm\neq 0$, using a minimum of three physical qubits~\cite{Knill:2000dq,Yang:01}. The logical space is constructed from four orthonormal states: 
\beq
\label{eq:4NS-states}
\begin{aligned}
\ket{\bar{1}}&=\ket{S_0}\ket{0},\quad  \ket{\bar{2}}=\ket{S_0}\ket{1}\\
\ket{\bar{3}}&=\left(\sqrt{2}\ket{T_+}\ket{1}-\ket{T_0}\ket{0}\right)/\sqrt{3}\\ 
\ket{\bar{4}}&=\left(\ket{T_0}\ket{1}-\sqrt{2}\ket{T_-}\ket{0}\right)/\sqrt{3}.
\end{aligned}
\eeq 
These states are composed of the singlet state $\ket{S_0}=\left(\ket{01}-\ket{10}\right)/\sqrt{2}$ and triplet states $\ket{T_+}=\ket{00}$, $\ket{T_0}=(\ket{01}+\ket{10})/\sqrt{2}$, and $\ket{T_-}=\ket{11}$. The 3-qubit code (DFS$_3$) is defined by the logical states 
\beq
\label{eq:DFS3-states}
\begin{aligned}
\ket{0_L}&=\gamma\ket{\bar{1}}+\delta \ket{\bar{2}}\\
\ket{1_L}&=\gamma\ket{ \bar{3}}+\delta \ket{\bar{4}} ,
\end{aligned}
\eeq 
where $\gamma$ and $\delta$ specify the gauge. Note that here the gauge degrees of freedom are that of a single qubit and thus, $d_J=2$. \cref{fig:setup} displays the encoding circuit for the DFS$_3$ code with $\gamma=1$ and $\delta=0$; see \cref{app:subsec:ns-state-prep} for an extension to arbitrary $\gamma$ and $\delta$. Computations on the three-qubit code are generated by the logical operators~\cite{kempe2001theory}:
\begin{align}
\bar{\sigma}^x &= \frac{1}{\sqrt{3}}(E_{23}-E_{13})\\
\bar{\sigma}^z &= \frac{1}{3}(E_{13}+E_{23}-2 E_{12}),
\end{align}
where $E_{ij}$ denotes a SWAP operation between the $i$th and $j$th physical qubits. We note as an aside, that this forms the basis for universal quantum computation using just the Heisenberg interaction~\cite{kempe2001theory,Bacon:Sydney,Kempe:01}, specifically in quantum dot systems~\cite{DiVincenzo:2000kx,west2012exchange,weinstein2022universal}.

\subsection{Error Detection in DFS Codes}
\label{subsubsec:error-detect}
Passive quantum codes share many commonalities with their active correcting counterparts. Specifically, DFSs can be described as a highly degenerate quantum error correcting code with infinite distance when \emph{all} operations are restricted to the code space. Of course, in practice, logical operations are not ideal and leakage outside of the code space can occur. It is in this domain that the stabilizer properties of the DFS codes can be employed for an additional layer of protection.

Under the stabilizer formalism, continuous (non-Abelian) stabilizers can be defined based on the DFS condition, \cref{eq:dfs-cond}~\cite{kempe2001theory}.  Collective dephasing yields $Z_{\rm col}=\bigotimes^{N}_{i=1}Z_i$ as one of the stabilizer elements, while collective decoherence includes additional collective Pauli operations as stabilizer elements: $X_{\rm col}=\bigotimes^{N}_{i=1} X_i$ and $Y_{\rm col}=\bigotimes^{N}_{i=1} Y_i$. As a result, the DFS can detect any odd number $<N$ of single-qubit bit-flips under collective dephasing or of arbitrary single-qubit Pauli errors under collective decoherence. We utilize the error detection properties of the code here by performing post-selection (PS) after decoding. Measurement outcomes of data qubits are post-selected based on the state of specific ancilla qubits. Below, we show that the PS procedure boosts the code's ability to maintain logical states.

\subsection{Dynamically Generated DFSs}
\label{subsec:dg-dfs-ns}

\subsubsection{Dynamical Decoupling}
\label{subsubsec:dd}
Quantum processors rarely possess intrinsic noise environments with the ideal permutation symmetry of collective interactions. However, through the use of DD, such symmetries can be effectively engineered. DD sequences generally comprise control pulses applied at predetermined time intervals to modify the system-environment interaction $H_{SB}$. Given a unitary evolution subject to the total Hamiltonian $H$ [\cref{eq:H}],
\begin{equation}
f_\tau\equiv e^{-iH\tau},
\end{equation}
DD sequences with delta-function-like pulses result in the evolution 
\begin{equation}
U_{DD}(T) = P_{K} f_{\tau}  P_{K-1} f_{\tau}  \dots   P_{1} f_{\tau}. \nonumber
\end{equation}
The total evolution time $T=K\tau$ and $\{P_j\}$ are the control pulses. Conventionally, DD sequences are designed to effectively cancel the system-bath interaction (i.e., $H_{SB} =0 $) up to a certain order in $T$. More precisely, an $\ell$th order decoupling sequence yields an effective time evolution given by $U_{DD}(T)=e^{-i (H_S+H_B)T} + \mathcal{O}[(\lambda T)^{\ell+1}]$, where $\lambda$ depends on both $\|H_{SB}\|$ and $\|H_B\|$~\cite{ngCombiningDynamicalDecoupling2011,Xia:2011uq}. A notable example, which will be relevant in this study, is the universal decoupling sequence~\cite{viola1999dynamical}
\begin{equation}
    XY_4 = Yf_\tau X f_\tau Y f_\tau X f_\tau.
\end{equation}
Utilizing $X$ and $Y$ pulses, representing $\pi$-rotations about the $x$ and $y$ axes of the single qubit Bloch sphere, respectively, $XY_4$ offers first order ($\ell=1$) decoupling for general single qubit noise.

Beyond suppression of system-environment interactions, DD can be used to selectively average out components of $H_{SB}$ to create the necessary conditions for a DFS. The group-theoretic foundations for such ``symmetrizing" sequences are given in Refs.~\cite{zanardiSymmetrizingEvolutions1999,violaDynamicalGenerationNoiseless2000}. Specific sequences for generating collective interactions are derived in Ref.~\cite{wuCreatingDecoherenceFreeSubspaces2002} and are elaborated upon below. In principle, achieving symmetrization conditions conducive to a DFS can require fewer pulses than complete suppression of general multi-qubit system-environment interactions~\cite{lidarReviewDecoherenceFree2014}; this is a potential advantage of combining DFSs with DD.

\subsubsection{Dynamically Generated Collective Dephasing}
\label{subsubsec:dg-dfs}
Two-qubit collective dephasing is generated by DD sequences consisting of logical operations. In the most general two-qubit setting, where single and two-qubit couplings to the environment are allowed, two-qubit collective dephasing is created by a concatenation of three sequences consisting of rotations on the logical Bloch sphere (see \cref{sec:DD-Protocol}):
\begin{align}
&\bar{Y} \circ \bar{X} \circ \Pi \circ f_\tau = \notag \\
&\quad \bar{Y} \bar{X}^{\dagger} f_{\tau} \Pi f_{\tau} \bar{X} f_{\tau} \Pi f_{\tau} \bar{Y}^\dagger \bar{X}^{\dagger} f_{\tau} \Pi f_{\tau} \bar{X}  f_{\tau} \Pi f_{\tau}.
\label{eq:dfs-dd-seq}
\end{align}
The inner-most sequence composed of $\Pi=\bar{R}_x(2\pi)$ operations is used to suppress leakage out of the DFS. In contrast, concatenating logical operators $\bar{Y}=\bar{R}_y(\pi)$ and $\bar{X}=\bar{R}_x(\pi)$ enables cancellation of all logical single-qubit errors. Note that the suppression properties of the sequence are independent of concatenation ordering. All variations lead to an effective Hamiltonian $H^{\rm col}_{SB}$ with $B_{\pm}\equiv0$.

\subsubsection{Dynamically Generated Collective Decoherence}
\label{subsubsec:dg-ns}
The three-qubit collective decoherence condition can be generated using the sequence
\begin{equation}
 E_{12} f_{\tau} E_{01} f_{\tau} E_{12} f_{\tau} E_{01} f_{\tau} E_{12} f_{\tau} E_{01} f_{\tau}.
 \label{eq:ns-dd-seq}
\end{equation}
Intuitively, the resulting evolution is akin to rapidly swapping the states of the qubits such that the environment cannot distinguish between them~\cite{west2012exchange}. The sequence assumes the underlying noise model is given by $H_{SB}=\sum^{N}_{j=1}\vec{\sigma}_j \cdot \vec{B}_j$, with $\vec{\sigma}_j=(\sigma^+_j, \sigma^-_j, \sigma^z_j)$ and $\vec{B}_j=(B^+_j, B^-_j, B^z_j)$. The effective Hamiltonian resulting from the DD sequence is $H^{\rm col}_{SB}$.

Practical implementation of the above sequences on the IBMQP requires composite pulses consisting of multiple noisy two-qubit gates. For example, $E_{jk}$ includes three CNOTs, each of which demands two faulty cross-resonance gates~\cite{mckayQiskitBackendSpecifications2018}; see \cref{app:subsec:dfs-logi-pulses} for further details. As such, realizations of these sequences are quite far from the noiseless, delta-function-pulse idealization from which they were derived. Nonetheless, as we show below, dynamically generating DFSs with these composite operations is achievable despite the imperfections inherent in the two-qubit gates.

%
%
\section{Evidence of Collective Symmetry}
\label{sec:evidence-col-sym}

\subsection{Logical State Invariance}
\label{subsec:log-state-inv}
\begin{figure}[t]
    \centering
    \includegraphics[width=\columnwidth]{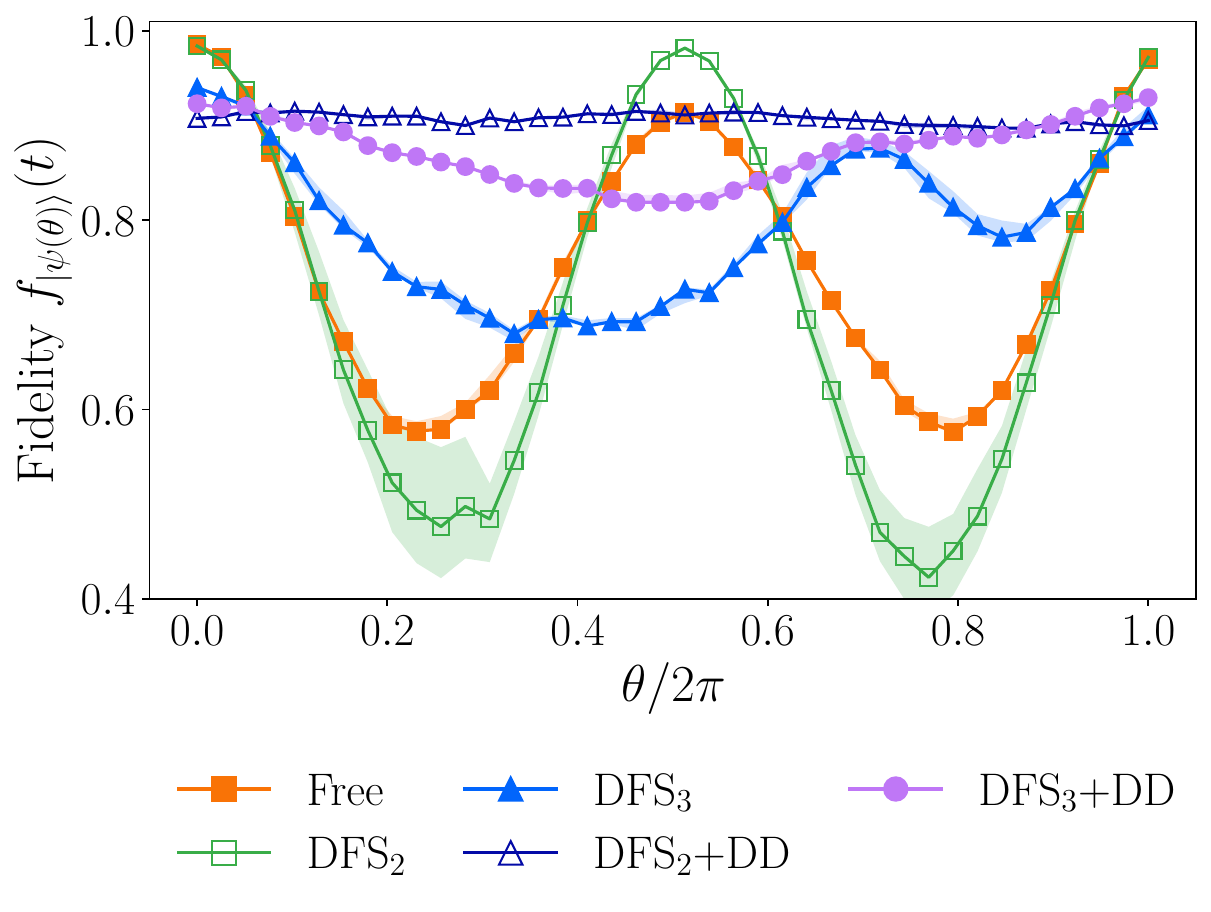}
    \caption{Fidelity as a function of elevation angle $\theta$ for unprotected and protected states. Initial states $\ket{\psi(\theta)}$ are chosen to be a subset of states that lie in the $(x,z)$-plane of the Bloch sphere. Each data point corresponds to the evolution of an unprotected or protected state for $T=14.89\mu$s, or the equivalent of one repetition of the DFS$_2$+DD sequence with state encoding and decoding. DFS protocols (with post-selection) correspond to free evolution bookended by encoding and decoding circuits. State invariance is not observed, indicating that the native noise does not satisfy the DFS conditions. DFS$_2$+DD and DFS$_3$+DD utilize the collective symmetrizing DD and post-selection to achieve the expected state invariance (independence on $\theta$). Data points represent estimated means and shaded regions denote 95\% CIs, all of which are determined from bootstrapping over five realizations of the experiment. All data were collected from the five-qubit Manila device.}
    \label{fig:theta_scan}
  \end{figure}
  
We investigate the presence of native and dynamically generated collective decoherence on the IBMQP. A single logical qubit is compared against its physical qubit constituents using a state-dependent fidelity analysis. The system is prepared in a quantum state lying in the $(x,z)$-plane of the Bloch sphere, such that $\ket{\psi(\theta)} = \cos \left( \frac{\theta}{2} \right) \ket{0} + \sin \left( \frac{\theta}{2} \right) \ket{1}$ for physical qubits and equivalently $\ket{\psi_L(\theta)}=\cos \left( \frac{\theta}{2} \right)\ket{0_L}+ \sin \left( \frac{\theta}{2} \right)\ket{1_L}$ is defined for DFS codes. For the DFS$_3$ code [\cref{eq:DFS3-states}], we first focus on the particular gauge $\gamma=1$ and $\delta=0$ for this comparison. An investigation of gauge dependence is presented below.
We do not scan over the azimuthal angle $\phi$ as previous work has shown that the free evolution fidelity depends almost entirely on the elevation angle $\theta$~\cite{pokharelDemonstrationFidelityImprovement2018, ezzellDynamicalDecouplingSuperconducting2022} (see the discussion following Eq.~(15) in Ref.~\cite{tripathi2023modeling} for an explanation of this effect).

Following physical or encoded state preparation, the system is allowed to freely evolve or is subjected to DD; the resulting state is denoted by $\rho_{\rm out}(t)$. An inverse state preparation completes the evolution prior to measurement in the computational basis. This sequence of operations is used to evaluate a physical or logical protocol's ability to preserve an initial state via the state fidelity
\begin{equation}
    f_{\ket{\psi(\theta)}}(t) =  \braket{\psi(\theta) | \rho_{\rm out}(t) |  \psi(\theta)},
    \label{eq:fid}
\end{equation}
where ideally $\rho_{\rm out}(t)\equiv \ketbra{\psi(\theta)}$. We investigate this fidelity as a function of $\theta$ for both unprotected states undergoing free evolution and protected states. The term ``protected'' refers to both DD and logical DFS encodings, or their combination.

Unprotected states are obtained from idle (free) evolution, where the system is allowed to evolve according to its internal dynamics after state preparation. In order to equalize qubit resources between unprotected and encoded states, we report the best fidelity of three adjacent physical qubits. Unprotected states are compared against DFS encodings without DD (DFS$_2$, DFS$_3$) and with DD (DFS$_2$+DD, DFS$_3$+DD). All results, unless otherwise specified, include PS. We find that PS leads to an overall increase in fidelity for both logical encodings whether or not DD is employed. This is discussed further in \cref{subsec:one-qubit-pres}, with additional details presented in \cref{app:sec:post-select-analysis}.

In \cref{fig:theta_scan}, we show the state fidelity $f_{\ket{\psi(\theta)}}(t)$ as a function of $\theta$ for a total evolution time of $t\approx 14.89\,\mu$s, or one repetition of the DFS$_2$+DD sequence, including logical state encoding/decoding. Experiments are performed on the five-qubit Manila device using qubits (2, 3, 4). While this subset of qubits yields the highest fidelity for logical encodings, we observe qualitatively similar behavior for alternative configurations; see \cref{app:subsec:alt-q-subsets}. 
Estimates of fidelity (solid lines) and 95\% confidence intervals (CIs; shaded regions) are determined by bootstrapping over five realizations of the demonstration using $8000$ measurement shots.

\begin{figure}[t]
    \centering
    \includegraphics[width=\columnwidth]{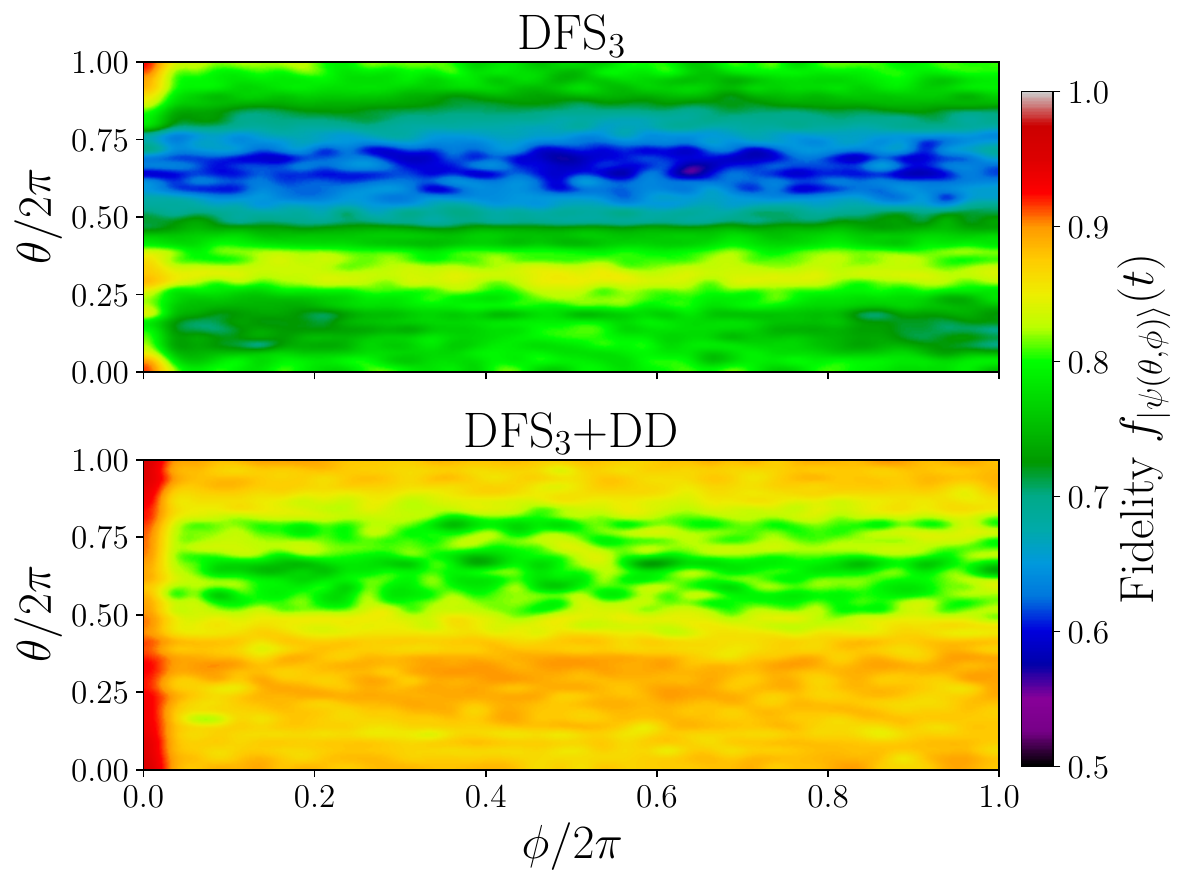}
    \caption{Investigation of gauge invariance for the DFS$_3$ code. State fidelity as a function of initial state $\ket{\psi(\theta,\phi)}$ is plotted for the DFS$_3$ (top) and DFS$_3$+DD (bottom) protocols. In the former, the system is encoded in the DFS$_3$ code, subject to free evolution equivalent to one repetition of the DFS$_3$ DD sequence, and returned to the ground state prior to measurement. The latter follows a similar procedure, except the free evolution is replaced by the symmetrizing DD sequence. Both protocols include PS. Plots show mean fidelities estimated from bootstrapping data collected over a five-day period on Manila; $8000$ measurement shots are used. Results indicate an enhanced gauge invariance when using the DFS$_3$ encoding in conjunction with PS and DD.}
    \label{fig:gauge_scan}
\end{figure}

Logical encodings alone are not sufficient to observe collective dephasing/decoherence. The DFS$_2$ code exhibits a state-dependent fidelity that is more consistent with free evolution. The DFS$_3$ displays improved state fidelity, yet still a clear dependence upon the initial logical state. Thus, unsurprisingly, the IBMQP does not exhibit evidence of intrinsic collective interactions. 

The inclusion of DD leads to substantially different behavior. The DFS$_2$+DD protocol produces a near-state-invariant behavior consistent with collective dephasing. Similar results are observed for the DFS$_3$+DD protocol, though some residual $\theta$-dependence remains.
Overall, DD substantially improves logical state fidelity and state invariance, consistent with dynamically generated collective interactions.

\subsection{Gauge Invariance}
\label{subsec:gauge-inv}
Noiseless subsystems possess a gauge invariance that enables the logical computational basis states to be defined as subsystems, rather than subspaces. As an additional verification of collective decoherence generation, we examine the existence of this gauge invariance for the DFS$_3$ code. We perform a state-dependent analysis similar to the previous subsection but also permit rotations within the gauge subspace. More concretely, we consider states of the form
\begin{equation}
    \ket{\psi(\theta,\phi)}=\cos\left(\frac{\theta}{2}\right)\ket{0_L(\phi)} + \sin\left(\frac{\theta}{2}\right)\ket{1_L(\phi)},
\end{equation}
where 
\beq
\begin{aligned}
\ket{0_L(\phi)}&=\cos(\phi/2)\ket{\bar{1}} + \sin(\phi/2)\ket{\bar{2}}\\
\ket{1_L(\phi)}&=\cos(\phi/2)\ket{\bar{3}} + \sin(\phi/2)\ket{\bar{4}},
\end{aligned}
\eeq 
with the four constituent states defined in \cref{eq:4NS-states}.
The DFS$_3$ code is prepared via the above state and subject to a single repetition of the DFS$_3$+DD sequence or allowed to freely evolve for an equivalent duration. The state decoding procedure is performed before measurement in the computational basis and subsequent PS.

A comparison between a freely evolving DFS$_3$ and the DFS$_3$+DD protocol is shown in \cref{fig:gauge_scan}. Experiments are performed on the five-qubit Manila device using $8000$ measurement shots. Estimates of mean fidelity [\cref{eq:fid}] are determined by bootstrapping over five realizations of the demonstration.

Evidence of gauge invariance is observed for both the DFS$_3$ encoding alone and with the inclusion of the collective-symmetry-generating DD protocol. In the top panel of \cref{fig:gauge_scan}, the DFS$_3$ fidelity determined via \cref{eq:fid}, is shown as a function of $\theta$ and $\phi$. While there is a clear dependence of fidelity on the logical state, signatures of invariance to the gauge are more prominent. We quantify this invariance by examining the standard deviation of the fidelity over the gauge states and averaged over initial logical states. For the DFS$_3$ encoding, the average standard deviation in fidelity is 0.023. In contrast, the DFS$_3$+DD protocol exhibits an average standard deviation of 0.018. 
The absence of a significant difference in gauge invariance between the DFS$_3$ and DFS$_3$+DD protocols indicates that the gauge degree of freedom is robust under this device's intrinsic decoherence mechanisms.

Consistent with the analysis in the previous subsection, DD improves the fidelity for all NS logical states. One of the most prominent features is the significant boost in fidelity for the gauge state with $\phi=0$; see the bottom panel of \cref{fig:gauge_scan}. This feature is easily explained by the state preparation circuit which requires an additional CNOT gate between the ancilla qubits for all $\phi\neq 0$. Ultimately, due to the topology of the hardware, this requires an additional SWAP operation as well; hence, the notable degradation in fidelity. Further details regarding the $\phi\neq0$ state preparation are discussed in \cref{app:subsec:ns-state-prep}.

%
%
\section{Preservation of Logical Qubits}
\label{sec:preserve-qubits}

Ultimately, the goal of our protection protocols is 
to extend the preservation of arbitrary quantum states.
In this section, we evaluate each protocol's performance by
allowing the system to evolve under free or controlled evolution, i.e., the experiment depicted schematically in \cref{fig:setup}. Under unprotected  (free) evolution, we prepare the state $U\ket{0}$ (we define $U$ below), let the system evolve according to its internal dynamics for fixed periods of time, apply $U^\dagger$, and measure in the computational basis. In the protected case,
after encoding followed by the application of $\bar{U}$ (logical-$U$),
the system is subject to $M$ repetitions of a DD sequence for a total evolution time of $T_{DD}(M)=M t_{DD}$, where $t_{DD}$ is the time for a single DD cycle. As in the case of free evolution, the experiment is completed by applying $\bar{U}^\dagger$, decoding, and a measurement of all qubits in the computational basis. Below, we examine arbitrary state preservation for logical qubit encodings, starting with a single logical qubit and then scaling the protocols up to seven logical qubits. Unprotected states are compared against protected states using equivalent physical resources.

\begin{figure}[t]
    \centering
    \includegraphics[width=\columnwidth]{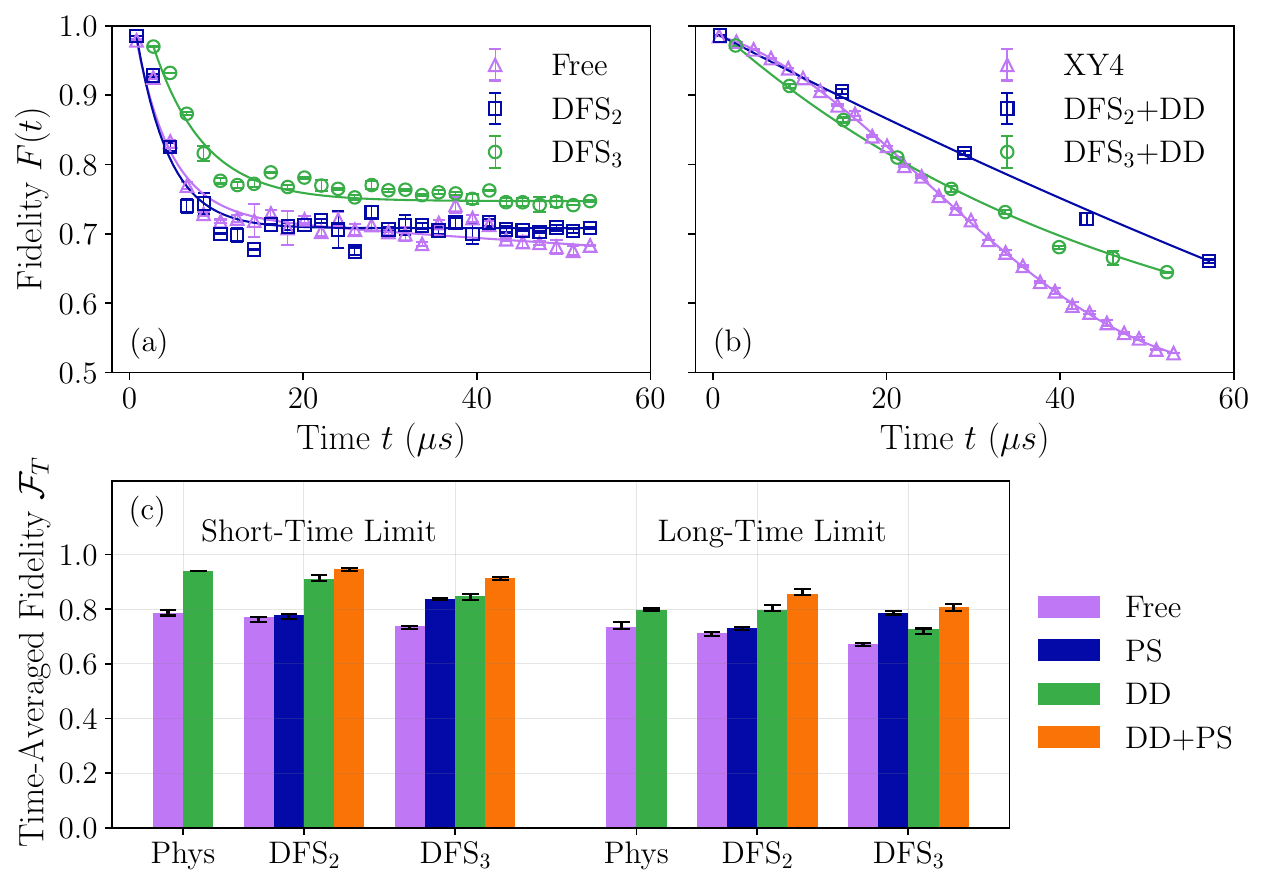}
    \caption{Time-dependent arbitrary state preservation of a single logical qubit. DFS encodings are compared against unencoded qubits using equivalent physical resources. Panel (a) shows the decay of logical states encoded within the DFS codes and subject to PS. The DFS$_3$ code appears to benefit the most from PS, achieving an improvement in performance over physical qubits subject to free evolution alone. In (b), logical encodings are used in conjunction with DD and PS and compared against XY4 on a physical qubit. Logical protocols outperform physical protocols particularly for $M>2$ repetitions of logical DD. The behavior of each protocol is further captured by panel (c) in the short and long-time limit via the time-averaged fidelity [\cref{eq:t-avg-fid}]. All experimental data is collected from the 5-qubit Manila device using $8000$ measurement shots. Estimates of means and CIs are determined by bootstrapping five realizations of the data collected over five days.}
    \label{fig:dfs_compare}
  \end{figure}

\subsection{Preservation of One Logical Qubit}
\label{subsec:one-qubit-pres}
First, we focus on the time-dependent state preservation of a single logical DFS qubit. Ideally, the preservation of an arbitrary state would be determined by sampling over the Haar distribution and calculating the average fidelity $\mathbb{E}_{\rm Haar}[f_{\ket{\psi}}(t)]=\int d\psi f_{\ket{\psi}}(t)$~\cite{Nielsen2011quantumcomputation}. We estimate the Haar fidelity via
\begin{equation}
F(t) = \frac{1}{L} \sum^L_{i=1} f_{\ket{\psi_i}}(t),
\label{eq:avg-fid}
\end{equation}
using an ensemble of $L=20$ states consisting of 14 Haar random states and the six eigenstates of the Pauli matrices, which we refer to as the poles of the Bloch sphere. We find this set to for estimating $\mathbb{E}_{\rm Haar}[f_{\ket{\psi}}(t)]$.

The average fidelity as a function of time is shown in \cref{fig:dfs_compare} for experiments performed on the five-qubit Manila device. Mean fidelities and error bars are determined by bootstrapping over five realizations of the experiment and $8000$ measurement shots. As in \cref{subsec:log-state-inv}, results are shown for the configuration of qubits with the highest average fidelity for logical encodings, i.e., qubits (2, 3, 4). Results for physical qubits consider the highest average fidelity among the three physical qubits used for the DFS$_3$ code. While the DFS$_2$ code only requires two physical qubits, we do not find a significant difference in performance when selecting the best performing qubit among two or three physical qubits. This holds for both the Free and XY4 cases. In the latter, the best performing qubit is typically the one subject to DD; hence, an additional neighbor qubit evolving freely does not alter the best performance.

Solid lines designate fits to the data using
\begin{align}
    \begin{split}
        F(t) &= C_1 f(t) + C_2, \\ 
        f(t) &= e^{-t/\tau_1}\cos(\omega t) + e^{-t/\tau_2}.
    \end{split}
    \label{eq:fit}
\end{align}
$C_1, C_2$ denote dimensionless weight parameters, such that 
\begin{equation}
    C_1 = \frac{F(T_{\max})-F(T_0)}{f(T_{\max})-f(T_0)}, \quad
    C_2 = F(T_0) - C_1 f(T_0).
\end{equation}
The time required to evolve the system for $M_{\max}$ repetitions of a DD sequence (or the free evolution equivalent) is given by $T_{\max}=T_0 + T_{DD}(M_{\max})$, where $T_0$ is the time for state encode, decode, and measurement. Short and long-term coherence times are determined by $\tau_1$ and $\tau_2$, respectively, while $\omega$ is the oscillation frequency. Additional information regarding the fitting procedure can be found in \cref{app:data}.

Figure~\ref{fig:dfs_compare}(a) displays the time-dependent state-averaged fidelity for unprotected physical (Free) and logical (DFS$_2$, DFS$_3$) qubits subject to PS alone. The fidelities of the Free and DFS$_2$ cases are statistically indistinguishable. However, the  DFS$_3$ case has a notably longer short-term decay time, while its long-time decay is essentially infinite; see \cref{tbl:fitting-params}. The longer decay time is accompanied by a higher fidelity. We attribute PS to the enhancement in fidelity, and substantiate this claim 
via an examination of the time-averaged fidelity:
\begin{equation}
    \mathcal{F}_T=\frac{1}{T-T_0}\int^{T}_{T_0} F(t) dt.
    \label{eq:t-avg-fid}
\end{equation}
In Figure~\ref{fig:dfs_compare}(c), the time-averaged fidelity is shown for all protocols in two regimes designated by the DD repetition time of the sequence used for the DFS$_2$ code. The short-time limit is set by the single repetition time $T=T_0+t_{DD}\approx 14.89\mu s$, where $T_0\approx 0.81\mu s$, while the long-time limit corresponds to $M=3$, i.e, $T=T_0+3\,t_{DD}\approx 43.05\mu s$. All estimates of $\mathcal{F}_T$ are calculated by cubic spline integration, with subsequent means and CIs determined by bootstrapping.

The DFS$_3$ code benefits substantially from PS in both the short and long-time limit. PS results in approximately 14\% and 17\% improvement in the DFS$_3$ time-averaged fidelity over the bare encoding, respectively. The DFS$_2$ code obtains a mere 1.3\% and 3\% in comparison. This result is consistent with the stabilizer properties of the code in the sense that the DFS$_2$ code cannot detect all single-qubit errors and is limited to bit-flips. The DFS$_3$ possesses a larger set of detectable errors and therefore, obtains a more robust code space.

\begin{table*}[t]
\begin{tabular}{p{0.18\linewidth}p{0.18\linewidth}p{0.18\linewidth}p{0.18\linewidth}p{0.18\linewidth}}
\hline\hline
Protocol & $\tau_1$ ($\mu$s) & $\omega$ (Hz)& $\tau_2$ ($\mu$s)  & $F(T_0)$ \\\hline
Free & $3.95\pm 0.55$ & 0 & $319.98\pm88.35$  & $0.98\pm 0.02$\\
XY4 & $100.49\pm 3.26$ & $265.03$ & $\infty$ & $0.99$\\
DFS$_2$ w/PS & $3.79\pm 0.36$ & 0 & $\infty$ & $0.99\pm0.01$\\
DFS$_2$ w/DD+PS & $229.91\pm 29.41$ & 0 & $\infty$ & $0.99\pm0.01$\\
DFS$_3$ w/PS & $5.86\pm0.42$ & 0 & $\infty$ & $0.97\pm0.01$\\
DFS$_3$ w/DD+PS & $47.35\pm 3.94$ & 0 & $\infty$ & 0.97\\
\hline\hline
\end{tabular}
\caption{Fit parameters and standard error estimates for the fidelity decay of various logical and physical protocols. Corresponding fits are shown in \cref{fig:dfs_compare}, while the fit is given by \cref{eq:fit}.}
\label{tbl:fitting-params}
\end{table*}

In examining the enhancement afforded by error detection as a function of the initial state, we find that the DFS$_3$ particularly benefits from PS for states approaching the poles of the logical Bloch sphere; see \cref{app:sec:post-select-analysis}. States near the logical ($x,y$)-plane suffer from logical errors that typically render lower fidelities with PS. In contrast, DFS$_2$ logical states only gain from PS near the logical $\ket{+}$ state. Ultimately, this behavior leads to state-averaged fidelities that favor DFS$_3$ when error detection is employed.

Despite the utility of PS, DD typically results in greater improvement in fidelity. In the case of DFS$_2$, the short-time decay rate is greatly enhanced by the inclusion of DD. For example, we observed an approximately 60\% increase in $\tau_1$ when DD is used in conjunction with PS as compared to PS alone. Qualitatively similar behavior appears for the time-averaged fidelity as well. DD results in a short-time limit increase of 17\% and long-time limit increase of 3\% relative to PS. Further incorporating PS with DD boosts $\mathcal{F}_T$ by 3\% and 7\%, respectively, indicating that (1) DD is most impactful in the short-time limit (consistent with DD's propensity to suppress non-Markovian but not Markovian errors) and (2) both protocols perform best when used together.

The DFS$_3$ code exhibits similar characteristics, however, PS plays an important role in the short-time limit as well. On their own, PS and DD enable similar short-time fidelities; DD yields a 15\% increase over the DFS alone, with PS further improving the DFS$_3$+DD fidelity by 7\%. This is accompanied by a nearly 8$\times$ improvement in $\tau_1$ from $5.86\ \mu$s for DFS$_3$ with PS to $47.35\ \mu$s when DD is incorporated. In contrast, DD has a less appreciable effect on $\mathcal{F}_T$ in the long-time limit. DD enhances the time-averaged fidelity of the DFS$_3$ code (without PS) by approximately 8\%. DD and PS again prove to be more beneficial when used together, achieving a 21\% increase over the DFS$_3$ encoding alone.

\begin{figure*}[t]
    \centering
    \includegraphics[width=\textwidth]{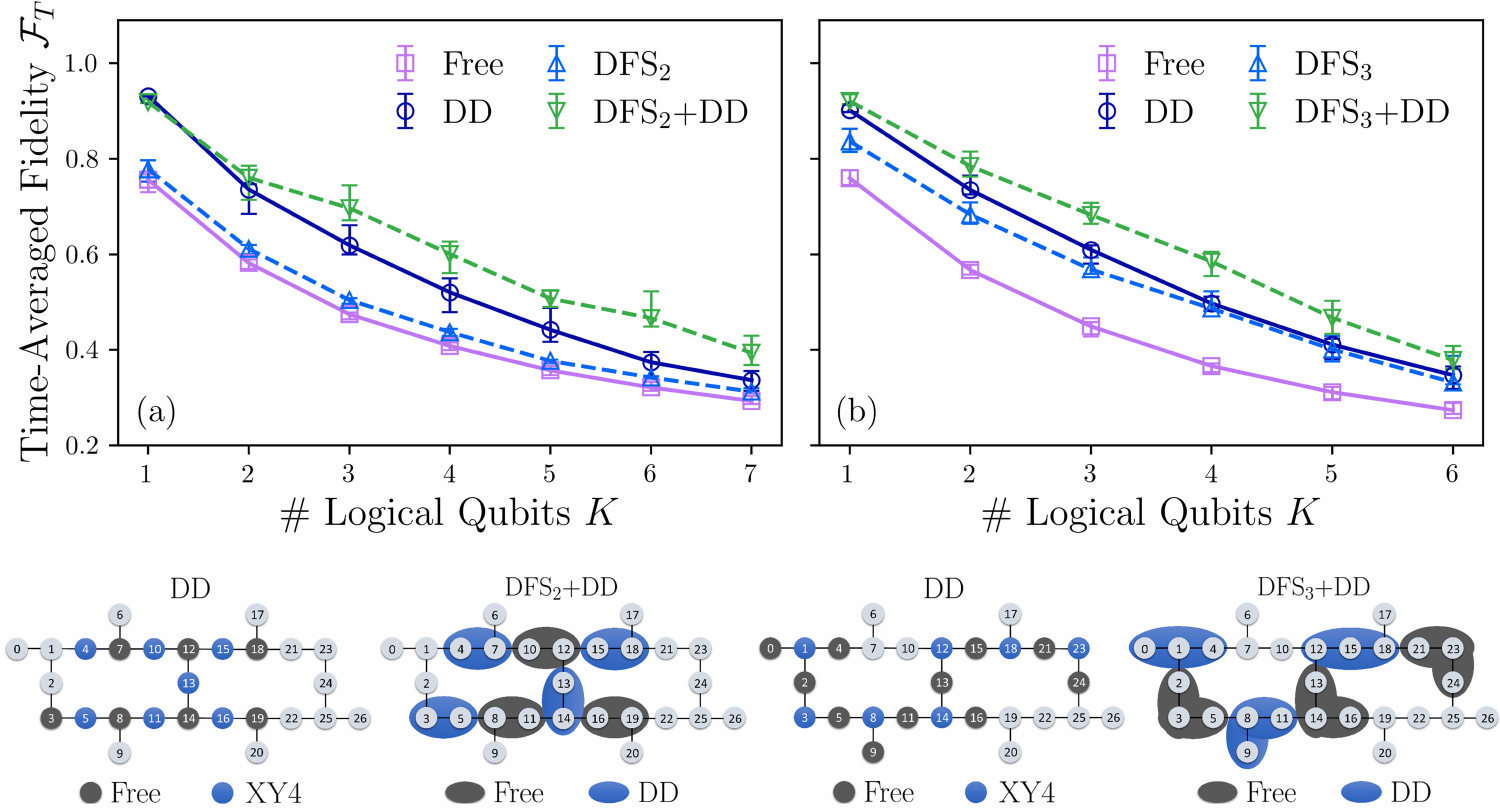}
    \caption{Time-averaged fidelity $\mathcal{F}_T$ as a function of the number of logical qubits. Panels (a) and (b) show results comparing physical and logical protocols for the DFS$_2$ and DFS$_3$ encodings, respectively. The DD and DFS$_2$+DD protocol is shown below panel (a), while a similar illustration for the DFS$_3$ comparison is shown below panel (b). Each panel includes free-evolution (purple squares; solid line), physical qubit DD (dark-blue circles; solid line), the DFS$_N$ (light blue up-triangles; dashed line), and DFS$_N$+DD (green down-triangles; dashed line) protocols. For the DFS$_2$ code, $N=14$ physical qubits are simultaneously prepared into $K'=7$ physically adjacent logical qubits. The time-averaged fidelity is examined for the best $K\leq K'$ logical qubits using a total time of $T=26.65\,\mu$s, or one repetition of the longest DFS$_2$+DD sequence. An identical analysis is performed for the DFS$_3$ using $N=18$ physical qubits to prepare $K'=6$ logical qubits. Physical qubit fidelities are determined from the best $K$ adjacent physical qubits among the available $N$. Experiments are performed on the 27-qubit Montreal device using $8000$ measurement shots. Estimates of $\mathcal{F}_T$ means and CIs are determined from bootstrapping five realizations of the data captured over five days. The results indicate that both the DFS$_2$+DD and DFS$_3$+DD protocols with PS outperform the physical DD protocol, particularly for $K>2$.}
    \label{fig:scalability}
  \end{figure*}
  
In \cref{fig:dfs_compare}(b), the logical encoding schemes with PS and DD are compared against DD on physical qubits. Qubit 3 is subject to XY4, while qubits 2 and 4 are allowed to evolve freely. After each total evolution time, the best qubit fidelity is taken, where the DD cycle time is $t_{DD}=4t_g \approx 142.2$ ns for a gate time of $t_g\approx35.6$ ns. Alternative physical DD protocols were considered as well, e.g., applying XY4 simultaneously on all qubits, however, they typically resulted in worse fidelity. We attribute the degradation in fidelity to quantum crosstalk, which is exacerbated by simultaneous operations on nearest-neighbor qubits. As such, the protocol chosen here is one such approach that suppresses parasitic interactions and combats local environmental noise.

Utilizing XY4 as a benchmark for noise protection in physical qubits, we evaluate each logical protocol's ability to outperform its physical qubit constituents. We find that DFS$_2$ (with DD and PS) and XY4 possess comparable short-time fidelity decay; see \cref{tbl:fitting-params}. Similar results are observed for the time-averaged fidelity in the short-time limit. However, in the long-time limit, the distinction between the protocols becomes more apparent. The DFS$_2$ code with DD and PS achieves a 7.25\% improvement over XY4, while the DFS$_3$ code attains a 1.25\% enhancement. Overall, these results indicate that it is possible for DFS encodings to perform on-par with physical error suppression protocols in the short-time limit, while providing greater long-term protection of quantum memory.

\subsection{Preservation of Multiple Logical Qubits}
\label{subsec:multi-preserve}
In order to justify the resource overhead of logical encoding, one must demonstrate that $K$ logical qubits can yield performance advantages over $N=n K$ physical qubits, where $n$ is the encoding overhead, i.e., $n=2$ and $3$ for the DFS$_2$ and DFS$_3$ cases, respectively. The analysis in the previous subsection sheds light on this comparison for $K=1$, where a single logical qubit performs similarly, if not better than physical qubits alone. We now expand this analysis to determine whether performance advantages persist with increasing $K$. We focus on the preservation of the best $K$ physically adjacent logical (physical) qubits selected from a set of $K^\prime$ simultaneously generated logical (physical) qubits. The time-average fidelity serves as the comparison metric, where $T$ is the total time for encoding/decoding and one repetition of the DFS$_2$+DD sequence, i.e., the so-called ``short-time limit" discussed in \cref{subsec:one-qubit-pres}. 

In \cref{fig:scalability}, we summarize two comparisons in which $N=14$ ($N=18$) physical qubits are encoded into $K'=7$ DFS$_2$ ($K'=6$ DFS$_3$) logical qubits on the 27-qubit Montreal device. The state-average fidelity is estimated for the $L=20$ states discussed in \cref{subsec:one-qubit-pres}, using $8000$ measurement shots. $\mathcal{F}_T$ is determined by cubic spline integration of $F(t)$ up to $T\approx26.65\,\mu$s. This is approximately equivalent to two repetitions of the DFS$_3$ DD sequence, including state encode/decode procedures. Estimates of the mean and CI for the time-averaged fidelity are determined by bootstrapping.

Logical qubits are compared against physical qubits using a combination of DD and free evolution. We consider physical qubit protocols corresponding to free evolution on all qubits and a DD-protection scheme in which every other qubit within the array is subject to XY4. Those qubits not protected by DD are allowed to evolve freely. As in the case of the single logical qubit, this protocol outperforms simultaneously XY4 on all qubits due to its suppression of parasitic crosstalk~\cite{tripathiSuppressionCrosstalkSuperconducting2022,zhouQuantumCrosstalkRobust2022}. A schematic illustration of the DD protocol is shown below panels (a) and (b) in \cref{fig:scalability}. We consider similar logical encoding schemes: (1) the qubits are prepared in the logical subspace and allowed to evolve freely and (2) upon preparing the logical subspaces, every other \emph{logical} qubit is subject to the symmetrizing DD sequence. As in the case of the physical qubit DD-protection protocol, we find that simultaneous logical operations on neighboring qubits typically result in lower fidelity. As before, we suspect this behavior is due to crosstalk between neighboring physical qubits that propagates into logical errors. The logical qubit DD-protection protocols for the DFS$_2$ and DFS$_3$ encodings are summarized at the bottom of \cref{fig:scalability}(a) and (b), respectively.

Above the protocol schematics in \cref{fig:scalability} are comparisons of the time-averaged fidelity as a function of the number of logical qubits for each protocol. In panel (a), results are shown for the DFS$_2$ code. Among the available $27$ qubits, we select the best set of $N=14$ physically adjacent physical qubits based on CNOT gate error rates and decoherence times to perform our demonstrations.
Physical qubit protocols are applied to the $N$-qubit set, while the DFS$_2$ protocols involve encoding the $N$ qubits into $K'=7$ logical qubits. We then ask the question: requiring physical adjacency, how do the best $K$ logical qubits compare against the best $K$ physical qubits? Panel (a) summarizes our findings as a function of $K$, where the DFS$_2$ encoding with PS typically performs similarly to physical qubit free evolution. In contrast, upon incorporating DD, the DFS$_2$ code outperforms the physical qubit DD protocol particularly for $K>2$. Within error bars, the performance advantage of the DFS remains consistent up to $K=7$, where time-averaged fidelities range from $12.5\%$ to $24.7\%$ higher than DD-protected physical qubits.

The performance advantage of DFS$_3$ over physical qubits is even more substantial than DFS$_2$. In \cref{fig:scalability}(b), $N=18$ qubits are again selected based on CNOT gate error rates and decoherence times. Physical qubit protocols on $N$ qubits are compared against the DFS$_3$ protocols on $K'=6$ logical qubits. Contrary to the DFS$_3$ code, the DFS$_3$ code with PS alone yields sizable and consistent advantages over physical qubits subjected to free evolution; note the consistency with the discussion in \cref{subsec:one-qubit-pres}. We find that this performance advantage persists as the number of logical qubits increases. Surprisingly, the benefits of the DFS$_3$ with PS are considerable enough to result in near-equivalent time-averaged fidelity with that of DD on physical qubits, particularly for $K>3$. As in the case of the DFS$_2$ encoding, the inclusion of DD results in improved fidelity for the DFS$_3$ code. The relative improvement in time-averaged fidelity varies from approximately $9.7\%$ to $17.7\%$ over the physical qubit DD protocol beyond $K=1$. If instead the time-averaged fidelity is considered up to one cycle of the DFS$_3$ ($T\approx 12.33\mu$s) then a maximum improvement of $23.6\%$ over physical DD is achievable; a result comparable to the DFS$_2$ case. In both cases, we find that DFS encoding, when combined with error detection and suppression outperforms physical error suppression on its own.

%
%
\section{Conclusions}
\label{sec:concl}
In this work, we investigated the viability of DFS codes on currently available quantum devices. Using IBMQP superconducting qubit devices, we showed that DFS codes can achieve advantages over physical error suppression in the task of quantum state preservation. This was accomplished by devising logical qubit protection protocols that incorporate two key aspects: (1) noise-symmetrizing DD and (2) error detection provided by the stabilizer properties of the code. We established the advantage of these protocols for up to seven logical qubits, hence demonstrating their potential applicability and scalability on current and near-term quantum processors.

In evaluating the efficacy of the codes, we showed that the collective system-bath interactions required by DFS codes do not natively exist on the devices. However, such symmetries can be enforced through the application of specially designed pulse sequences. Despite their considerable gate depth, these sequences enable logical subspace invariance indicative of a DFS code with fidelity gains relative to unencoded qubits.

We showed that state invariance leads to improved logical qubit fidelity over error suppression protocols that are applied directly to physical qubits. We observed an enhancement in fidelity up to seven DFS$_2$ and six DFS$_3$ qubits encoded into $14$ and $18$ physical qubits, respectively. Error detection employed via post-selection resulted in logical qubit fidelities that surpass the XY4 DD sequence, particularly in the case of the DFS$_3$ code. Moreover, when noise-symmetrizing DD was used in conjunction with post-selection, even greater performance gains were attainable. The DFS codes obtained up to a $24\%$ and $17\%$ improvement over XY4 in terms of the time-averaged fidelity, respectively. This constitutes a beyond-breakeven fidelity improvement for DFS-encoded qubits. 

Additional studies are required to investigate the potential practical implementation of encoded quantum computation based on DFS codes. In particular, a demonstration of entangled logical qubits based on the methods we explored here is a natural next step. Nevertheless, we have already found encouraging results for the preservation of quantum states encoded within noninteracting copies of multiple logical qubits. Our protocol integrates error detection, avoidance, and suppression, and thus, highlights the significant potential of constructing error management schemes consisting of numerous techniques working in concert to enhance logical qubit fidelity.

\section{Acknowledgements}
GQ thanks Colin Trout for insightful discussions. BP is grateful to Dr. Namit Anand and Haimeng Zhang for useful comments. This work was supported in part by the U.S. Department of Energy (DOE), Office
of Science, Office of Advanced Scientific Computing Research (ASCR) Quantum Computing Application Teams program under fieldwork proposal number ERKJ34 and the Accelerated Research in Quantum Computing program under Award Number DE-SC0020316.
 This research was supported by the ARO MURI grant W911NF-22-S-0007 and is based in part upon work supported by the National Science Foundation the Quantum Leap Big Idea under Grant No. OMA-1936388. This material is also based upon work supported by the Defense Advanced Research Projects Agency (DARPA) under Contract No. HR001122C0063. This research used resources of the Oak Ridge Leadership Computing Facility, which is a DOE Office of Science User Facility supported under Contract DE-AC05-00OR22725. 

\appendix

\section{DFS$_3$ State Preparation}
\label{app:subsec:ns-state-prep}
\begin{figure}[t]
    \centering
    \includegraphics[width=\columnwidth]{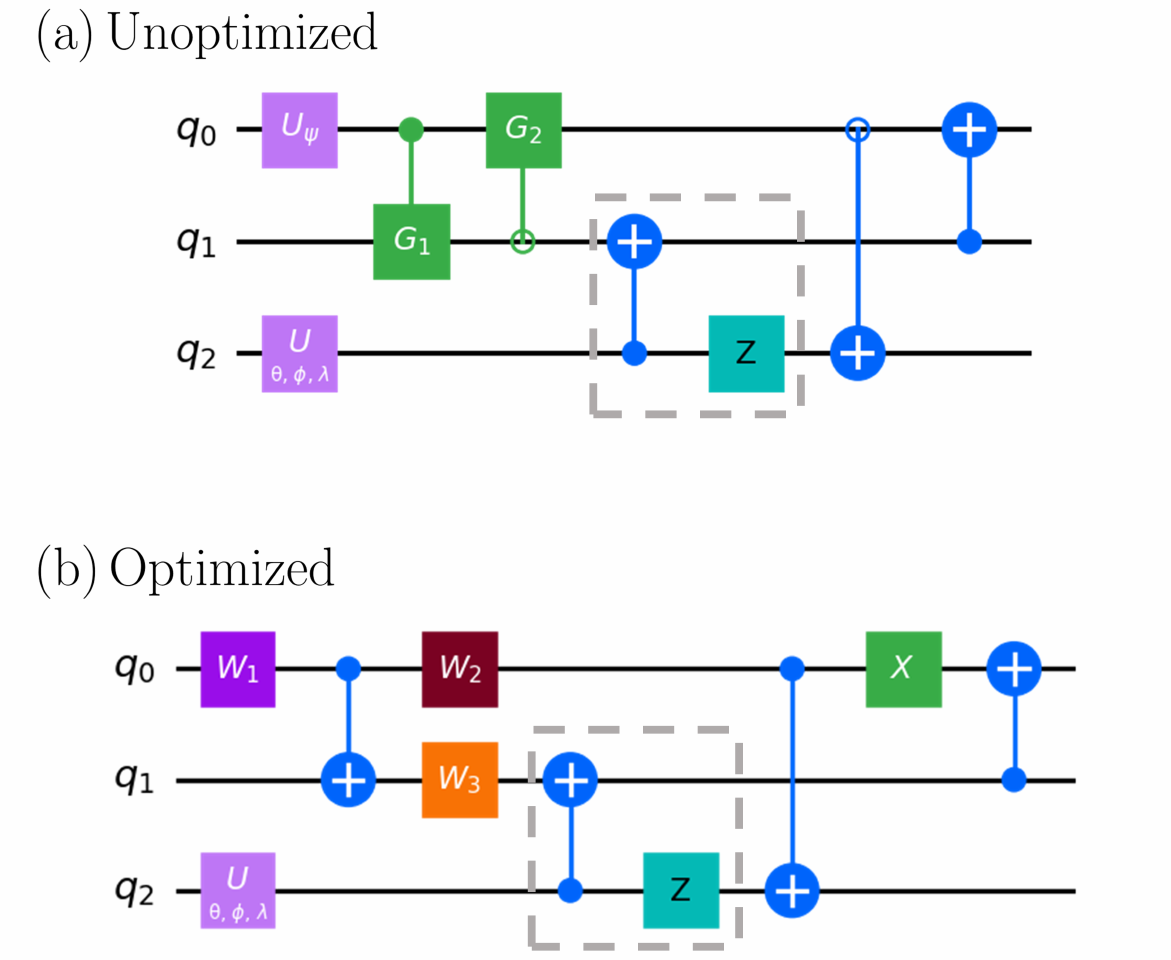}
    \caption{State preparation circuits for NS encoding. Panel (a) shows the unoptimized circuit expressed in terms of controlled gates $CG_1$ and $CG_2$. The optimized circuit resulting from a Schmidt decomposition is displayed in panel (b). The gates encapsulated by the gray, dashed box are only implemented if the gauge qubit (q2) is not prepared in $\ket{0}$.}
    \label{fig:ns-enc-circ}
\end{figure}

The DFS$_3$ encoded states are prepared using the circuit defined in Ref.~\cite{liRecursiveEncodingDecoding2011a} and shown in \cref{fig:ns-enc-circ}. q0 denotes the data qubit, while q1 and q2 designate an additional ancilla qubit and gauge qubit, respectively. The controlled unitaries shown in the unoptimized circuit of \cref{fig:ns-enc-circ}(a) are defined via
\begin{align}
    G_1&=\frac{1}{\sqrt{3}}\left(
    \begin{array}{cc}
        1 & \sqrt{2} \\
        -\sqrt{2} & 1
    \end{array}
    \right)\\
    G_2&=\frac{1}{\sqrt{2}}\left(
    \begin{array}{cc}
        1 & 1 \\
        -1 & 1
    \end{array}
    \right).
\end{align}
The circuit defined in the optimized circuit shown in \cref{fig:ns-enc-circ}(b) within the gray, dashed box is only executed if the gauge qubit is prepared in any state other than $\ket{0}$. Note that by definition of the DFS$_3$ code, the gauge qubit can be initialized in any state while still satisfying the DFS preservation condition.

Compiling the state preparation circuit to the IBMQP gateset requires decomposition of the controlled gates $CG_1$ and $CG_2$. The combined product of these gates can be replaced by a circuit that only requires one CNOT gate. This decomposition is accomplished via the Schmidt decomposition, where $\ket{\Psi}=CG_2CG_1\ket{\psi}\otimes \ket{0}$ is rewritten as
\begin{equation}
    \ket{\Psi}=\alpha \ket{\psi_1}\otimes \ket{\phi_1} + \beta \ket{\psi_2}\otimes \ket{\phi_2},
\end{equation}
where $\braket{\psi_1|\psi_2}=\braket{\phi_1|\phi_2}=0$. The Schmidt decomposition is computed by taking the partial trace of each of the bipartite systems and computing the corresponding eigenvalues. For $\rho_{AB}=\ketbra{\Psi}$, we have $\rho_A=\Tr_B[\rho_{AB}]$ and $\rho_B=\Tr_A[\rho_{AB}]$. We denote by $\alpha,\beta$ the eigenvalues of $\rho_A$. The eigenstates of $\rho_A$ and $\rho_B$ are $\ket{\psi_i}$ and $\ket{\phi_i}$, respectively. The resulting unitaries from the decomposition are:
\begin{equation}
    \tilde{W}_1=\left(
    \begin{array}{cc}
         \alpha & \beta^* \\
         \beta & -\alpha^*
    \end{array}
    \right),
\end{equation}
which can be derived from the Schmidt eigenvalues, and
\begin{align}
    W_2 = \ketb{\psi_1}{0} + \ketb{\psi_2}{1},\nonumber\\
    W_3 = \ketb{\phi_1}{0} + \ketb{\phi_2}{1}\nonumber.
\end{align}
The final optimized circuit is shown in \cref{fig:ns-enc-circ}(b).

\section{DFS$_{2}$ DD Protocol}
\label{sec:DD-Protocol}

The DD sequence in \cref{eq:dfs-dd-seq} generates collective dephasing up to first-order in time-dependent perturbation theory. The sequence can achieve this suppression condition for general two-qubit system-environment interactions of the form
\begin{equation}
  H_{S B}=\sum_{\alpha_{1}, \alpha_{2}}\left(\sigma_{1}^{\alpha_{1}} \sigma_{2}^{\alpha_{2}}\right) \otimes B^{\alpha_{1} \alpha_{2}}, \ \alpha_i \in \{ 0,x,y,z \},
\end{equation}
where $B^{\alpha_{1} \alpha_{2}}$ denotes a bounded bath operator coupled to the system. The interaction Hamiltonian can be rewritten as
\begin{equation}
H_{SB} = H_{\rm Leak} + H_{\rm Logi} + H_{\rm DFS},
\end{equation}
where $H_{\rm Leak}={\rm span}(\mathcal{P}_{\rm Leak})$ denotes leakage errors, i.e., terms that cause transitions between states inside and outside of the DFS. The Hamiltonian $H_{\rm Logi}={\rm span}(\mathcal{P}_{\rm Logi})$ includes operators that form undesirable logic gates on the DFS that couple to the bath and cause decoherence. Lastly, $H_{\rm DFS}={\rm span}(\mathcal{P}_{\rm DFS})$ designates operators that either vanish or are proportional to identity on the DFS. The spanning subspaces for each Hamiltonian, in terms of the two-qubit Pauli basis are
\begin{align}
    \mathcal{P}_{\rm Leak} &=  \left\{\sx_1, \sx_2, \sy_1, \sy_2, \sx_1\sz_2, \sz_1\sx_2, \sy_1\sz_2, \sz_1\sy_2\right\}\nonumber\\
    \mathcal{P}_{\rm Logi} &=  \left\{ \bar{\sigma}^x, \bar{\sigma}^y, \bar{\sigma}^z\right\}\nonumber\\
    \mathcal{P}_{\rm DFS} &=  \Big\{\frac{\sz_1+\sz_2}{2}, \frac{\sx_1\sy_2+\sy_1\sx_2}{2}, \frac{\sx_1\sx_2+\sy_1\sy_2}{2}, \notag \\
    &\qquad\qquad  I_4, \sz_1\sz_2\Big\},
\end{align}
where $I_4$ is the two-qubit identity operator and the logical operators in $\mathcal{P}_{\rm Logi}$ are defined in \cref{eq:dfs-logi-ops}.

Following Ref.~\cite{Byrd:2002:047901}, it can be shown that logical operations can be used to design DD sequences that suppress all terms acting non-trivially on the DFS. Specifically, the sequence shown in \cref{eq:dfs-dd-seq} utilizes a concatenation of three sub-sequences that together suppress $H_{\rm Leak}$ and $H_{\rm Logi}$ in the ideal, instantaneous pulse limit. Each sub-sequence contains two pulses separated by free evolution periods of duration $\tau$. As such, the generic construction of the sub-sequences can be defined by $U \circ f_\tau \equiv U f_\tau U^\dagger f_\tau$, where $f_\tau=e^{-i H_{SB}\tau}$. Note that 
$\Pi=\bar{R}_x(2\pi)$ anticommutes with $\mathcal{P}_{\rm Leak}$, so it suppresses leakage out of the DFS. The logical operators $\bar{Y}=\bar{R}_y(\pi)$ and $\bar{X}=\bar{R}_x(\pi)$ can be used to suppress $\mathcal{P}_{\rm Logi}$ via a standard XY4 sequence, i.e., a concatenation of $\bar{X}$ and $\bar{Y}$. The concatenation of the leakage suppression and logical error suppression sequences yields the following sequence:
\begin{equation}
\begin{aligned}
    &\bar{Y} \circ \bar{X} \circ \Pi \circ f_\tau\\
    &= \bar{Y} \circ \bar{X} \circ (\Pi f_\tau\Pi^\dagger f_\tau)\\
    &= \bar{Y} \circ [\bar{X}(\Pi f_\tau\Pi f_\tau)\bar{X}^\dagger(\Pi f_\tau\Pi f_\tau)]\\
    &= \bar{Y}\bar{X}^\dagger f_\tau\Pi f_\tau\bar{X} f_\tau\Pi f_\tau\bar{Y}^\dagger\bar{X}^\dagger f_\tau\Pi f_\tau\bar{X} f_\tau\Pi f_\tau ,
\end{aligned}
\end{equation}
where in the last line we used $\Pi=\Pi^\dagger$ along with $\bar{X}\Pi=\bar{X}^\dagger$ and $\bar{X}^\dagger\Pi=\bar{X}$.

\section{Implementation of DFS$_{2}$ DD Pulses}
\label{app:subsec:dfs-logi-pulses}

The DD sequence used to generate the collective dephasing DFS the requires logical operations $\bar{R}_x(\theta)$ and $\bar{R}_y(\theta)$. These logical unitaries are generated by logical operators $\bar{\sigma}^x$ and $\bar{\sigma}^y$, respectively, which can take many forms. Belonging to the commutant of the error algebra only requires that the Hermitian operators be expressed as a linear combinations of the identity, $\sigma^z_i$, $\sz_1\sz_2$, and $\vec{\sigma}_1\cdot \vec{\sigma}_2 = \sx_1\sx_2 + \sy_1\sy_2 + \sz_1\sz_2$. The operators given in \cref{eq:dfs-logi-ops} define symmetric logical operators that satisfy the so-called independence property~\cite{kempe2001theory} (i.e., they act entirely within the specified
DFS) and preserve \emph{all} DFSs in the two-qubit Hilbert space~\cite{fortunato2002implementation}. Hence, operators on the logical subspace defined by $\{\ket{01}, \ket{10}\}$ do not enable mixing between the one-dimensional subspaces defined by $\ket{00}$ and $\ket{11}$. While alternative, non-preserving logical operators, such as 
\begin{equation}
\begin{aligned}
    \bar{\sigma}^x &= \sx_1\sx_2\\
    \bar{\sigma}^y &= \frac{1}{2}\left(\sx_1\sy_2 + \sy_1\sx_2\right) \\
    \bar{\sigma}^z &= -\sz_2
\end{aligned}
\end{equation}
can be defined, empirical evidence shows that the symmetric operators yield higher fidelities. 

\begin{figure}[t]
    \centering
    \includegraphics[width=\columnwidth]{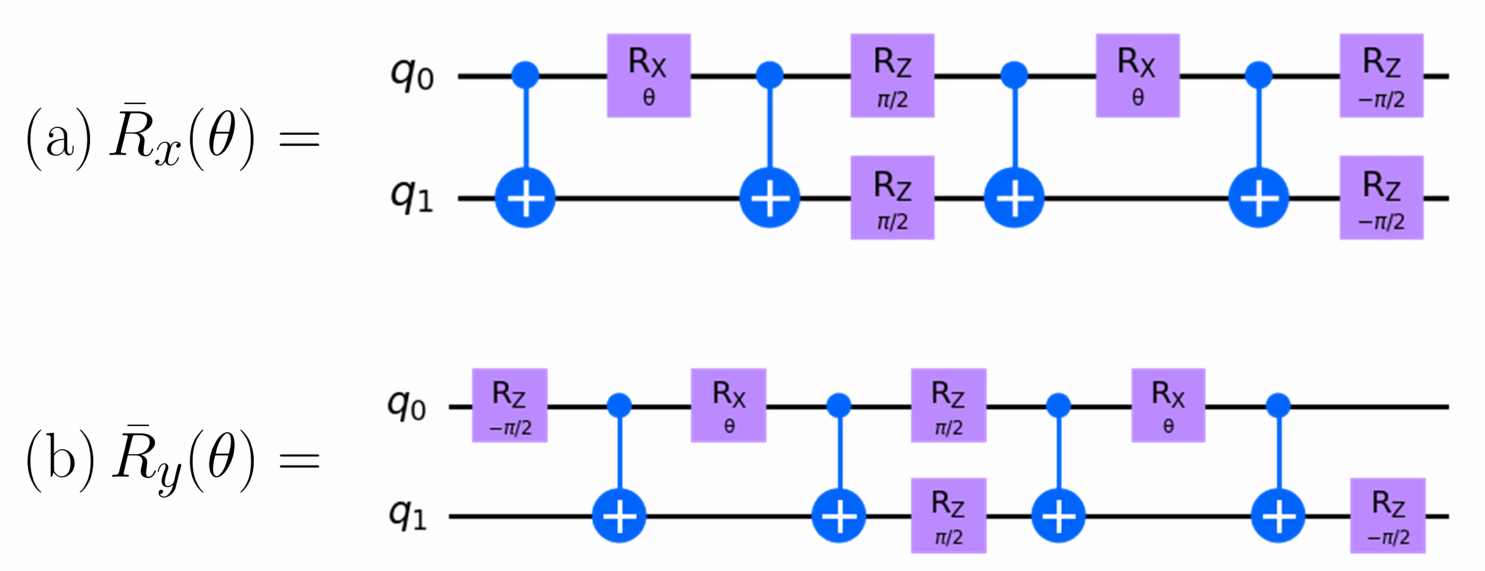}
    \caption{Decompositions for the symmetric logical rotation gates in terms of the IBMQE gateset. (a) shows the logical $x$-rotation gate, while the logical $y$-rotation gate is given in (b).}
    \label{fig:dfs-log-circuits}
\end{figure}

The preference towards the symmetric operators is quite surprising considering that their gate decomposition requires twice as many CNOT operators. This can be seen by utilizing commutativity to decompose the logical operators into products of two-local rotation operators. For example,
\begin{equation}
    \bar{R}_x(\theta)=e^{-i\theta/2 \sx_1\sx_2} e^{-i\theta/2\sy_1\sy_2},
\end{equation}
where each two-qubit interaction unitary can be decomposed into single qubit rotations bookended by CNOT gates; see \cref{fig:dfs-log-circuits}(a). Following a similar decomposition, one obtains the circuit for $\bar{R}_y(\theta)$ in terms of the IBMQP gateset $\mathcal{G}=\{X_{\pi/2}, X_\pi, R_z(\theta), {\rm CNOT}\}$ as shown in \cref{fig:dfs-log-circuits}(b). The average logical gate time across the five data collection periods was $1.47\pm0.30\mu$s on the 5-qubit Manila device. The average logical gate time for the 27-qubit Montreal device was $1.67\pm 0.33\mu$s.

Note that the definitions of non-symmetric logical operators yield unitaries similar to the first two-local interaction in the symmetric decomposition. For example, the non-symmetric logical $x$-rotation operator is $\bar{R}_x(\theta)=e^{-i\theta \sx_1\sx_2}$, which only requires two CNOT gates when decomposed via $\mathcal{G}$. Despite the reduction in gate time and depth, the non-symmetric operators do not outperform their symmetric counterparts.

Moreover, we do not observe improved performance from optimizing the gate decompositions of the symmetric logical operators. The decompositions given in \cref{fig:dfs-log-circuits} utilize four CNOTs, one more than what should be required for the decomposition of any two-qubit gate~\cite{Nielsen2011quantumcomputation}. However, we find that optimizing the number of CNOT gates does not lead to better performance. We suspect that the four CNOT gate decomposition permits some form of inherent noise robustness not afforded by the optimized variation. Further investigation of the inherent hardware noise model is needed to fully understand this effect.

\begin{figure}[t]
    \centering
    \includegraphics[width=0.9\columnwidth]{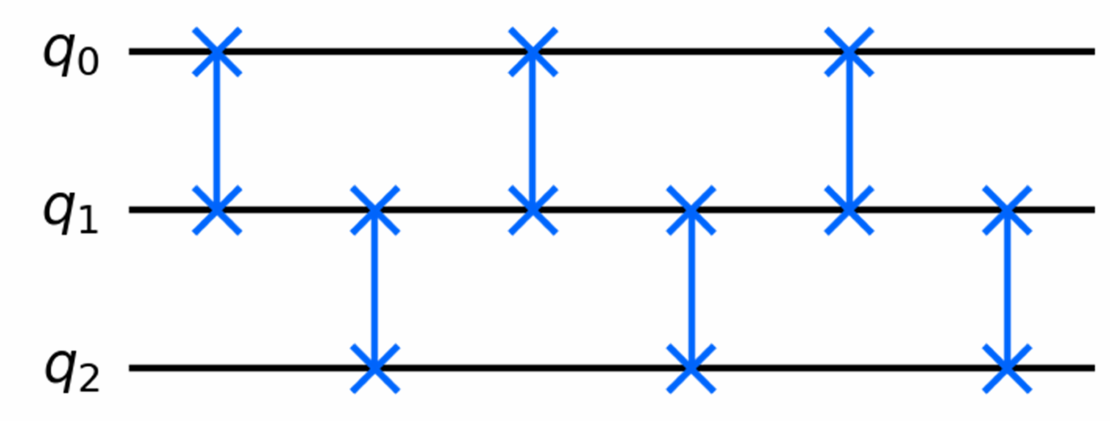}
    \caption{NS symmetrization sequence consisting of six SWAP operations. Each SWAP gate is decomposed into three CNOT gates when implemented on IBMQE processors.}
    \label{fig:ns-dd}
\end{figure}

\section{DFS$_3$ DD Protocol}
The collective decoherence condition for the DFS$_3$ code can be generated by the DD sequence shown in \cref{eq:ns-dd-seq}. This sequence is sufficient to symmetrize any linear system-environment interaction of the form
\begin{equation}
    H_{SB}=\sum^3_{i=1} \vec{\sigma}_i \cdot \vec{B}_i,
    \label{eq:hsb-ns-seq}
\end{equation}
where $\vec{B}_i=(B^x_i, B^y_i, B^z_i)$ and $B^\mu_i$ are bounded bath operators. In the ideal, instantaneous pulse limit, the dynamics after the sequence are governed by 
an effective Hamiltonian of the form $H_{\rm eff} = \left(\sum_i \vec{\sigma}_i\right)\cdot \vec{B}' + \mathcal{O}(\tau^2)$, where $\tau$ is the duration of free evolution between pulses and $\vec{B}'=\vec{B}_1 + 2\vec{B}_2$. This result can be shown by writing the sequence as
\begin{align}
    U(T=6\tau) = \prod^{6}_{j=1} P^\dagger_j f_\tau P_j,
    \label{eq:ns-pulse-conj}
\end{align}
and computing the first-order term in the Magnus expansion with
\begin{equation}
    P_j \in \{I, E_{12}, E_{12}E_{23}, E_{13}, E_{13}E_{12}, E_{23}\}.
\end{equation}
Note that \cref{eq:ns-pulse-conj} can be reconciled with \cref{eq:ns-dd-seq} by using $E_{23}=E_{12}E_{13}E_{12}=E_{13}E_{12}E_{13}$ and $E_{12}=E_{23}E_{12}E_{13}$. The sequence's circuit diagram is shown in \cref{fig:ns-dd}.

It is important to note that the sequence used in this study is not the only choice. An alternative originally proposed in Ref.~\cite{wuCreatingDecoherenceFreeSubspaces2002} also enables collective decoherence at the cost of $14$ SWAP gates. \cref{eq:ns-dd-seq} utilizes only $6$ pulses while still achieving the same first-order effect.


\section{Hardware Specifications}
\begin{figure}[t]
    \centering
    \includegraphics[width=\columnwidth]{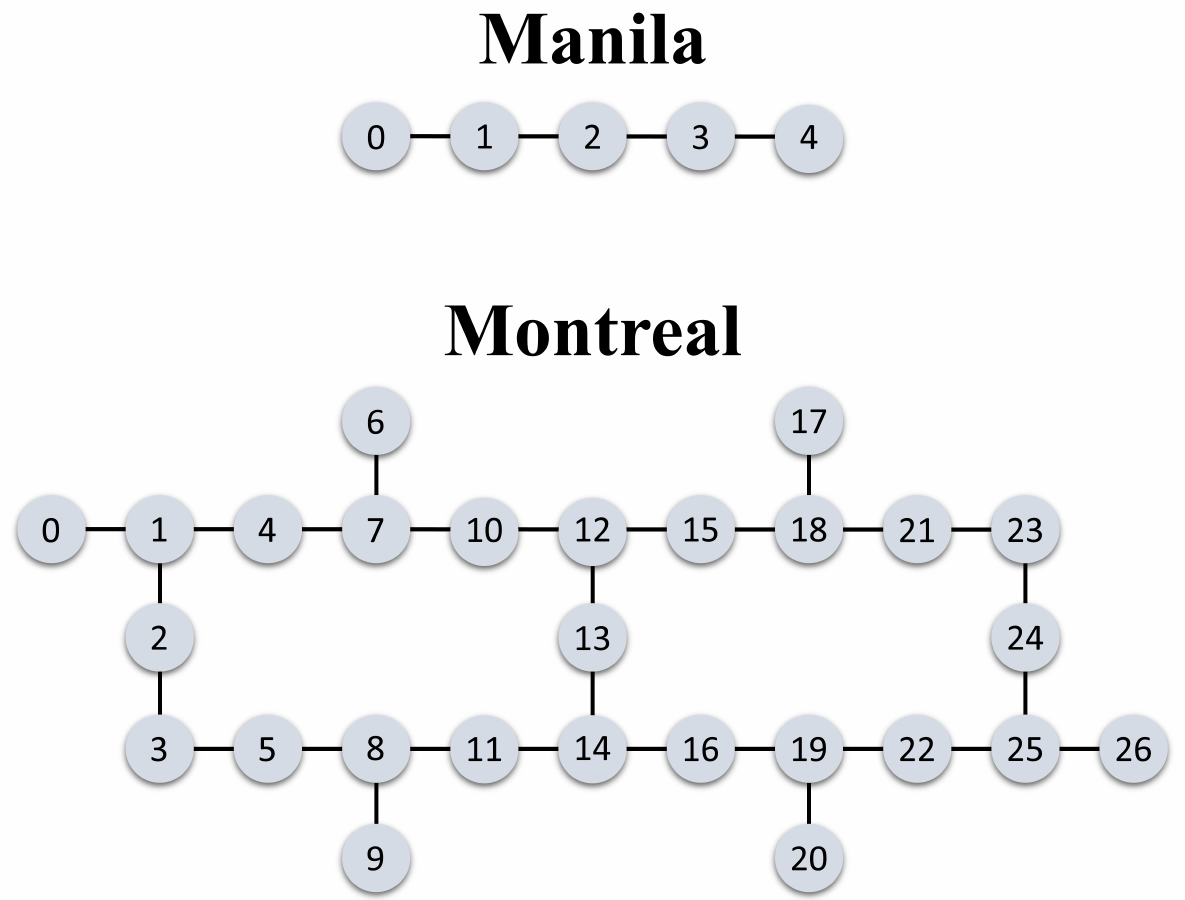}
    \caption{Connectivity graphs for the 5-qubit IBMQE Manila (top) and 27-qubit Montreal (bottom) devices.}
    \label{fig:hw-tops}
\end{figure}

All demonstrations were performed on the IBMQP, a cloud-based quantum computing resource that offers access to superconducting transmon quantum processors. All circuits were written in Python using the Qiskit API created by IBM. We used the 5-qubit Manila and 27-qubit Montreal devices in our experiments. Each processor's connectivity graph is shown in \cref{fig:hw-tops}. The processor type, quantum volume (QV), and circuit layer operations per second (CLOPS) are detailed in \cref{tab:hardware_specs}.

\begin{table}[t]
    \centering
    \begin{tabular}{ccc}
    \hline\hline
    Device & Manila & Montreal \\
    \hline
         Processor & Falcon r5.11L & Falcon r4 \\
         QV & 32 & 128 \\
         CLOPS & 2.8K & 2.7K \\
    \hline\hline
    \end{tabular}
    \caption{Device specification for Manila and Montreal. QV and CLOPS denote quantum volume and circuit layer operations per second, respectively.}
    \label{tab:hardware_specs}
\end{table}

Qubit characteristics, such as $T_1$, $T_2$ times, gate error rates, readout error rates, and gate durations for each device are presented in Tables~\ref{tab:manila_calibration_data}-\ref{tab:manila_cnot_data}. In all cases, qubit characteristics are collected over the duration of data collection, i.e., five realizations of each demonstration collected over multiple days. Values shown in the tables denote averages with error bars indicating one standard deviation. 

Qubit characteristics for Manila are shown in \cref{tab:manila_calibration_data,tab:manila_cnot_data} for experiments performed during September 20-25, 2022. $T_1$ typically ranges from 120.8 $\mu$s to 200 $\mu$s, whereas $T_2$ varies from 43 $\mu$s to 71.3 $\mu$s. Single qubit gate error rates are on the order of $10^{-4}$ and readout error rates on the order of $10^{-2}$. In \cref{tab:manila_cnot_data}, specifications for CNOT gate error rates and durations are shown to range from $5.61\times 10^{-3}$ to $13.78\times 10^{-3}$ and approximately 277 to 469 ns. Data is shown for control-target qubit pairs where the gate duration is shorter. Reverse ordering will incur an additional single qubit gate that increases both the error rate in accordance with \cref{tab:manila_calibration_data} and gate duration by approximately 35.55 ns. 

A similar set of data for Montreal is shown in Tables~\ref{tab:montreal_calibration_data} and \ref{tab:montreal_cnot_data}. We show data specifically for the qubits used in the demonstrations. Averages and standard deviations are determined from calibration data collected during October 11-20, 2022. Qubit relaxation and dephasing times vary across the device as $55.6-105.2$ $\mu$s and $22.7-140.0$ $\mu$s, respectively. Similar to Manila, single qubit gate error rates are on the order of $10^{-3}$ and CNOT error rates are on the order of $10^{-3}$. Readout error rates are on the order of $10^{-2}$ for a majority of the qubits used in the demonstrations. Variations in CNOT gate error rates are accompanied by varying duration for a significant subset of qubits; see \cref{tab:montreal_cnot_data}.

\begin{table}[t]
    \centering
    \small
    \begin{tabular}{c C{1.75cm} C{1.75cm} C{1.8cm} C{1.8cm}}
    \hline
    \hline
    Qubit & ${T_1}$ $[{\mu}s]$& ${T_2}$ $[{\mu}s]$& 1Q Gate Error $[\times10^{-3}]$ & Readout Error $[\times10^{-2}]$\\
    \hline
    0 & $120.8 \pm 25.0$ & $71.3 \pm 18.4$ & $0.31 \pm 0.11$ & $2.93 \pm 0.65$\\ 
    1 & $173.9 \pm 42.9$ & $68.9 \pm 6.6$ & $0.22\pm 0.01$ & $2.57 \pm 0.26$\\ 
    2 & $136.6 \pm 20.4$ & $26.0 \pm 2.0$ & $0.31\pm 0.04$ & $2.50 \pm 0.30$\\ 
    3 & $200.0 \pm 20.6$ & $60.4 \pm 6.3$ & $0.20\pm 0.02$ & $2.54 \pm 0.25$\\ 
    4 & $148.7 \pm 17.0$ & $43.0 \pm 1.3$ & $0.49\pm 0.07$ & $3.51 \pm 1.42$\\ \hline
    \hline
    \end{tabular}
    \caption{Physical parameters for the Manila device averaged over five realizations of demonstrations collected during September 20-25, 2022. Uncertainties denote one standard deviation.}
    \label{tab:manila_calibration_data}
\end{table}

\begin{table}[t]
\centering
    \begin{tabular}{C{1cm} C{3cm} C{2.5cm}}
    \hline
    \hline
    Qubits (C,T)& CNOT Error Rate $[\times10^{-3}]$ & CNOT Duration [ns]\\
    \hline
    (0,1) & $6.91\pm 0.23$ &  277.33\\
    (1,2) & $13.78\pm 4.06$ &  469.33\\
    (2,3) & $7.61\pm 1.53$ &  355.55\\
    (4,3) & $5.61\pm 0.82$ & 298.67 \\
    \hline\hline
    \end{tabular}
    \caption{CNOT gate error rates and durations for specific control (C) and target (T) qubits for Manila. Values denote averages with one standard deviation determined from calibration data collected during September 20-25, 2022. Reversing control and target qubits requires an additional single qubit gate that increases the gate time by 35 ns and effectively increases the error rate based on \cref{tab:manila_calibration_data}.}
    \label{tab:manila_cnot_data}
\end{table}

\begin{table}[t]
    \centering
    \begin{tabular}{c C{1.75cm} C{1.75cm} C{1.8cm} C{1.8cm}}
    \hline
    \hline
    Qubit & ${T_1}$ $[{\mu}s]$& ${T_2}$ $[{\mu}s]$& 1Q Gate Error $[\times10^{-3}]$ & Readout Error $[\times10^{-2}]$\\
    \hline
    0 & $96.3\pm 14.2$ & $38.2\pm 10.4$ & $0.19\pm 0.02$ & $1.10\pm 0.12$\\ 
    1 & $99.5\pm 13.8$ & $22.7\pm 0.8$ & $0.21\pm 0.06$ & $1.63\pm 0.41$\\ 
    2 & $87.2\pm 7.1$ & $106.5\pm 6.4$ & $0.32\pm 0.1$ & $1.39\pm 0.38$\\ 
    3 & $80.3\pm 4.1$ & $71.5\pm 3.5$ & $0.2\pm 0.02$ & $0.93\pm 0.09$\\ 
    4 & $76.8\pm 8.6$ & $104.7\pm 13.0$ & $0.24\pm 0.07$ & $1.57\pm 0.36$\\ 
    5 & $85.5\pm 7.1$ & $95.5\pm 11.6$ & $0.27\pm 0.18$ & $1.21\pm 0.64$\\
    7 & $75.3\pm 14.8$ & $67.2\pm 11.2$ & $0.55\pm 0.11$ & $5.36\pm 0.82$\\ 
    8 & $95.1\pm 13.7$ & $120.4\pm 9.0$ & $0.27\pm 0.08$ & $1.14\pm 0.13$\\
    9 & $86.9\pm 7.6$ & $107.6\pm 9.9$ & $0.26\pm 0.03$ & $1.02\pm 0.29$\\
    10 & $86.5\pm 17.6$ & $86.3\pm 9.8$ & $0.32\pm 0.07$ & $0.82\pm 0.15$\\
    11 & $93.5\pm 5.3$ & $60.9\pm 9.8$ & $0.22\pm 0.04$ & $1.53\pm 0.31$\\
    12 & $101.9\pm 10.9$ & $140.0\pm 26.8$ & $0.29\pm 0.08$ & $2.02\pm 0.51$\\
    13 & $58.9\pm 18.9$ & $55.6\pm 11.0$ & $0.35\pm 0.09$ & $1.34\pm 0.31$\\
    14 & $79.4\pm 17.4$ & $100.2\pm 24.0$ & $0.41\pm 0.27$ & $1.08\pm 0.3$\\
    15 & $97.8\pm 4.3$ & $121.3\pm 9.6$ & $0.39\pm 0.17$ & $1.81\pm 0.75$\\
    16 & $82.2\pm 5.9$ & $87.3\pm 4.7$ & $0.3\pm 0.13$ & $1.24\pm 0.82$\\
    18 & $73.7\pm 12.6$ & $28.7\pm 2.2$ & $0.46\pm 0.17$ & $3.14\pm 0.73$\\
    19 & $89.9\pm 5.6$ & $137.8\pm 17.9$ & $0.24\pm 0.15$ & $1.11\pm 0.29$\\
    21 & $105.2\pm 11.3$ & $50.6\pm 5.4$ & $0.51\pm 0.1$ & $3.53\pm 0.56$\\
    23 & $55.6\pm 24.4$ & $55.9\pm 18.0$ & $0.41\pm 0.35$ & $2.03\pm 1.44$\\
    24 & $79.5\pm 7.5$ & $56.9\pm 3.6$ & $0.32\pm 0.09$ & $2.41\pm 1.34$\\
    \hline
    \hline
    \end{tabular}
    \caption{Qubit characteristic timescales and error rates for Montreal. Data shown for qubits used in demonstrations only. Averages and error bars (one standard deviation) determined from calibration data collected October 11-20, 2022.}
    \label{tab:montreal_calibration_data}
\end{table}

\begin{table}[t]
\centering
    \small
    \begin{tabular}{C{1cm} C{3cm} C{2.5cm}}
    \hline
    \hline
    Qubits (C,T)& CNOT Error Rate $[\times10^{-3}]$ & CNOT Duration [ns]\\
    \hline
    (0,1) & $7.6 \pm 3.95$ & $412.08 \pm 44.8$\\
    (3,2) & $7.53 \pm 1.12$ & $375.79 \pm 9.6$\\
    (1,4) & $10.19 \pm 2.7$ & $492.13 \pm 54.4$\\
    (5,3) & $7.4 \pm 1.8$& $350.63 \pm 19.2$\\
    (4,7) & $13.37 \pm 4.49$ & $294.47 \pm 16.0$\\
    (9,8) & $5.97 \pm 0.85$ & $373.79 \pm 6.4$\\
    (11,8) & $7.8 \pm 1.87$ & $470.06 \pm 35.2$\\
    (12,10) & $6.49 \pm 0.66$ & $374.88 \pm 3.2$\\
    (13,14) & $12.37 \pm 5.59$ & $502.7 \pm 19.2$\\
    (15,12) & $9.58 \pm 2.1$ & $369.78\pm 0.0$\\
    (16,14) & $9.89 \pm 3.43$ & $309.97 \pm 16.0$\\
    (15,18) & $18.62 \pm 7.05$ & 597.33\\
    (16,19) & $13.70 \pm 2.35$ & 270.22\\
    (23,21) & $12.50 \pm 3.61$ & 391.11\\
    (23,24) & $10.28 \pm 2.87$ & $397.31 \pm 12.8$\\
    \hline\hline
    \end{tabular}
    \caption{CNOT error rates and durations for IBMQP Montreal for qubits used in demonstrations. Average calibration values shown for data collected October 11-20, 2022. Error bars denote one standard deviation.}
    \label{tab:montreal_cnot_data}
\end{table}

\section{Data Collection and Analysis}
\label{app:data}

\subsection{Data Collection Practices}
The IBMQP devices are subject to recalibration every few hours. During calibration, the characterization of qubit transition frequencies, error rates, and decoherence times are performed alongside updates to single-qubit and two-qubit pulse waveforms. We observe that qubit performance can vary significantly between, and even within, calibration cycles. Fluctuations in qubit characteristic parameters typically manifest as large shifts in fidelity when data is collected across calibrations. Furthermore, variations in the fidelity are observed depending upon when one performs the experiment. For example, experiments performed soon after a calibration can be distinct from those performed just before a calibration. While this variability is likely due to drift in qubit frequencies and/or the control master clock, knowledge of the potential origin of the errors does not necessarily imply that it is straightforward to mitigate.

Our demonstrations require a large suite of quantum circuits to be executed and thus, we are subject to data collection across multiple calibration cycles. In order to address the effects of hardware variability, we incorporate three practices in our circuit execution. Let us describe each practice by first defining a demonstration $D=\{C_j\}^{N}_{j=1}$ consisting of $N$ sets of circuits $C_j$. Each set $C_j=\{c_{j,k}\}^{K}_{k=1}$ is composed of $K$ circuits each of approximately equivalent total time $T_j$. For example, $D$ could describe a DD experiment where $N$ different DD repetitions are applied and $C_j$ consists of different DD sequences of equivalent total time. 

The first practice is intrinsic to the definition of $D$. During circuit creation, we organize the circuits such that those with the same total time are performed immediately after each other; hence, $C_j$. In this manner, we aim to mitigate potential variability in the hardware noise environment as data is collected for different error protection protocols at the same $T_j$. In addition, the order in which circuits are implemented is randomized with respect to $j$. For example, a demonstration consisting of $N=4$ total times may be executed on hardware in the following order: $C_3, C_1, C_2, C_4$. We find that this approach effectively averages variability due to calibrations across all $T_j$ rather than isolating it to specific total times. Lastly, multiple realizations (or replications) of each demonstration are executed over many hours or even days to collect data under a variety of hardware conditions. The data is then compiled and used to estimate various statistical quantities via bootstrapping. We find that this approach enables a more reliable estimate of an error protection protocol's performance.

\subsection{Bootstrapping}
The results reported in the main text display the mean and confidence intervals estimated via the bootstrapping method described in Ref.~\cite{stine1989bootstrap}. This technique is implemented by randomly sampling $N$ data points (with replacement) from a data set of size $N$ and then computing the mean of this bootstrapped sample. By repeating this procedure $K$ times, a new, bootstrapped data set of size $K$ is generated. The mean and confidence interval (CI) can be calculated based on this bootstrapped data set. This approach is used to estimate mean fidelities and CIs for \cref{eq:fid,eq:avg-fid,eq:t-avg-fid}, which are used in the comparisons shown in Figs.~\ref{fig:theta_scan}-\ref{fig:scalability}.

\subsection{Fitting Protocol}
\label{app:subsec:fitting}
Data fits are performed for fidelity vs. time comparisons shown in \cref{fig:dfs_compare}. Bootstrapped estimates of fidelity are fit to the generic fit equation given in \cref{eq:fit}. Parameter reductions of the fit equation are also considered, most notably in cases where generic fits suggest parameters are inconsequential. Various fits derived from \cref{eq:fit} are compared using the Akaike information criterion (AIC)~\cite{stoica2004model}, an estimator of prediction error. Fits shown in the main text correspond to the cases where AIC is minimal among the fit variations considered.

\section{Post-Selection Analysis}
\label{app:sec:post-select-analysis}

A crucial component for the success of the DFS codes is the use of post-selection (PS) to detect errors in the logical states. In the DFS$_2$ protocol, post-selected states are aggregated based on the state of the ancilla qubit. Namely, combined two-qubit states in which the ancilla returns to the ground state after decoding are deemed viable. In the DFS$_3$ code, the gauge qubit ($q_2$ in \cref{fig:ns-enc-circ}) is used to identify valid states. In this section, we examine PS from a variety of different viewpoints to highlight its impact on fidelity and protocol resources.

\subsection{State-Dependence}
\label{app:subsec:ps-state-dependence}
\begin{figure}[t]
    \centering
    \includegraphics[width=\columnwidth]{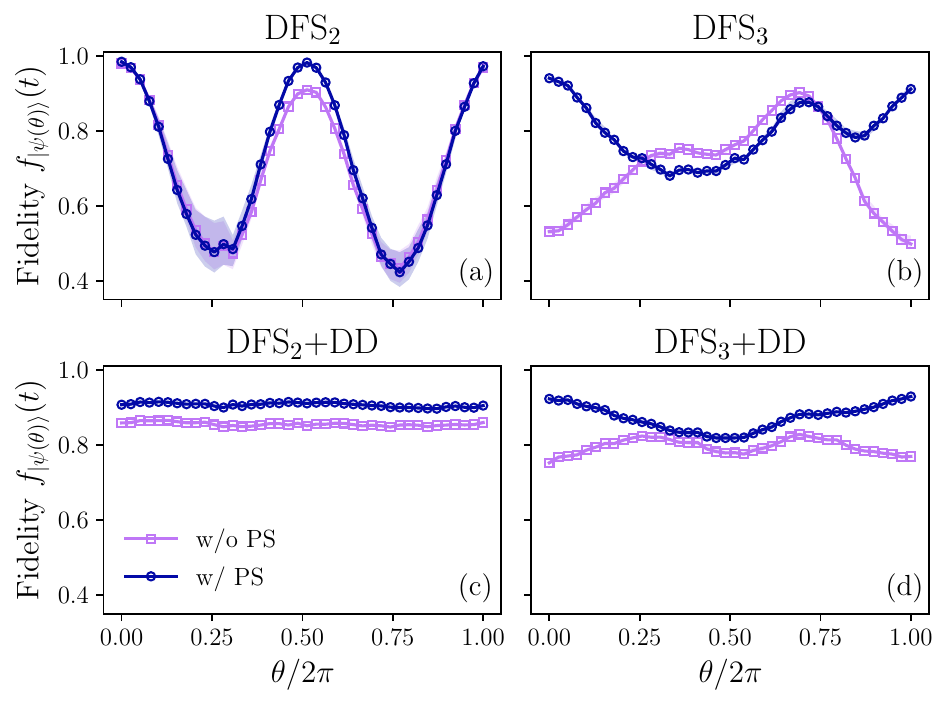}
    \caption{Fidelity as a function of initial state for DFS encodings with and without PS. Panels (a) and (b) denote DFS$_2$ and DFS$_3$ encodings alone, respectively. Encodings with DD are shown in panels (c) and (d). The most prominent impact on fidelity is observed for the DFS$_3$ case, specifically states that are near the poles of the logical Bloch sphere. Data points denote means and shaded regions are CIs. Statistical quantities are estimated from bootstrapping over five realizations of the demonstration using $8000$ shots.}
    \label{fig:angle-vary-ps-compare}
\end{figure}

The impact of PS on fidelity is strongly dependent upon the logical state and encoding. In the case of the DFS$_2$ code, only minor improvements in fidelity are found when using PS without error suppression. States near the $\ket{+}$ state are particularly enhanced by PS, as can be seen in \cref{fig:angle-vary-ps-compare}(a). Near equivalent fidelity between the DFS with and without PS suggests that the logical states are predominately plagued by logical errors or bit-flip errors rather than detectable single qubit phase-flip errors. Through DD, the logical fidelity greatly improves on average, and similarly, so does the effectiveness of error detection. The reduction of the logical error rate enables detectable errors to become more pronounced so that PS yields an average increase in the fidelity of 6.1\%; see \cref{fig:angle-vary-ps-compare}(c).

The DFS$_3$ code offers a striking contrast to the DFS$_2$ code when DD is not employed. States near the poles of the logical Bloch sphere are highly susceptible to noise that error detection can reduce considerably. This suggests that the primary error is of weight one. Conversely, states approaching the logical $\ket{+}$ state are negatively impacted by PS, implying that logical errors are dominant. Despite these distinctions between the DFS$_2$ and DFS$_3$ codes, similarities re-emerge upon the introduction of DD. Symmetrization leads to an overall improvement in logical state fidelity and detectable errors. On average, PS yields a 9.7\% improvement in the DFS$_3$+DD fidelity. Illustrations of the DFS$_3$ and DFS$_3$+DD state dependence are shown in panels (b) and (d) of \cref{fig:angle-vary-ps-compare}.

\label{app:subsec:ps-time-dependence}
\begin{figure}[t]
    \centering
    \includegraphics[width=\columnwidth]{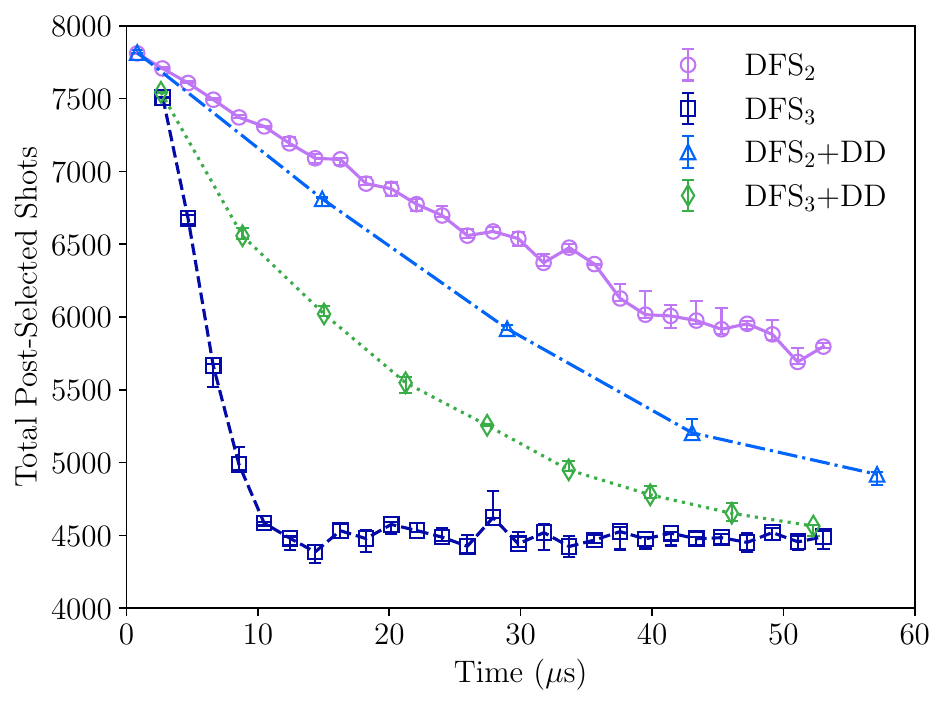}
    \caption{Total number of post-selected shots used to determine the state-averaged fidelity in \cref{fig:dfs_compare} as a function of time. Markers denote means, and error bars denote CIs, both of which are determined by bootstrapping. Overall, we observe that the inclusion of DD negatively impacts the PS shot count for the DFS$_2$ code, while improving the number of viable shots for the DFS$_3$ code. Despite this behavior, DD has an overall positive effect on code performance as it reduces logical errors, i.e., errors not detected by PS.}
    \label{fig:ps-compare}
\end{figure}

\subsection{Time-Dependence}
\begin{figure}[t]
    \centering
    \includegraphics[width=\columnwidth]{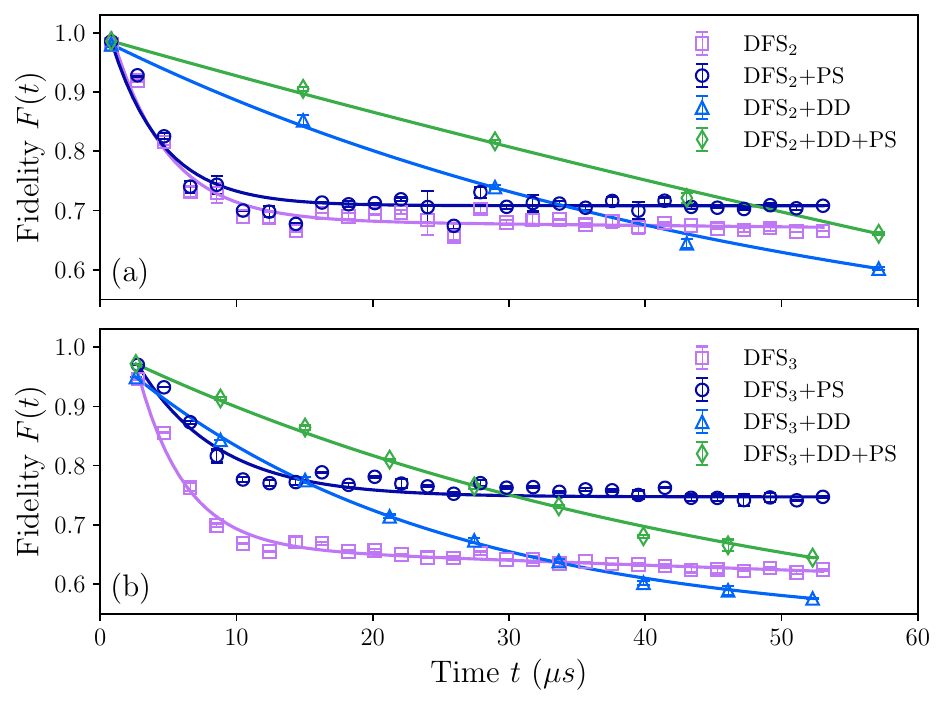}
    \caption{State-averaged fidelity as a function of time for one logical qubit. Panels (a) and (b) show results for the DFS$_2$ and DFS$_3$ codes, respectively. Data and fits are shown for each code with PS, DD, or both. The results shown here supplement those shown in \cref{fig:dfs_compare}, where the data was used to calculate the time-averaged fidelity in the short- and long-time limits. Data points denote means while error bars denote CIs, both obtained from bootstrapping over five experimental data sets, each using $8000$ measurement shots.}
    \label{fig:ps-gens-compare}
\end{figure}

Investigating the effectiveness of PS as a function of time provides an alternative perspective on each protection protocol. In this section, we study the number of post-selected shots and state-averaged fidelity as a function of time. The former is shown in \cref{fig:ps-compare}, while the latter is displayed in \cref{fig:ps-gens-compare}. Both are produced from the same data set used to create \cref{fig:dfs_compare} in the main text; hence, they focus on the time-dependent state preservation of a single logical qubit. 

The cost of PS is a reduction in the total number of experimental measurements (or shots) that can be used to estimate state fidelity. \cref{fig:ps-compare} illustrates the cost for each code with and without DD. Interestingly, DD does not always increase the total number of viable shots. In the case of the two-qubit protocol, the total number of post-selected shots reduces more slowly with the DFS alone. After one cycle of DD, the quantity of PS shots reduces by approximately 4\%. Of course, this does not imply an increase in state fidelity due to the presence of logical errors, as is indicated by \cref{fig:dfs_compare}. DD with PS is still more advantageous than PS alone but ultimately requires more shots to achieve a particular sampling threshold.

The three-qubit DFS code contrasts with the two-qubit case. Specifically, the DFS$_3$ code benefits from DD in regards to the total number of PS shots. The DFS$_3$ code alone is subject to a dramatic reduction after approximately 10 $\mu$s, where only about half of the total shots satisfy the PS criteria. DD affords a substantial improvement, increasing the total PS shot counts by over 31\%. As such, PS and DD improve the total viable shot count and fidelity.

\begin{figure}[t]
    \centering
    \includegraphics[width=\columnwidth]{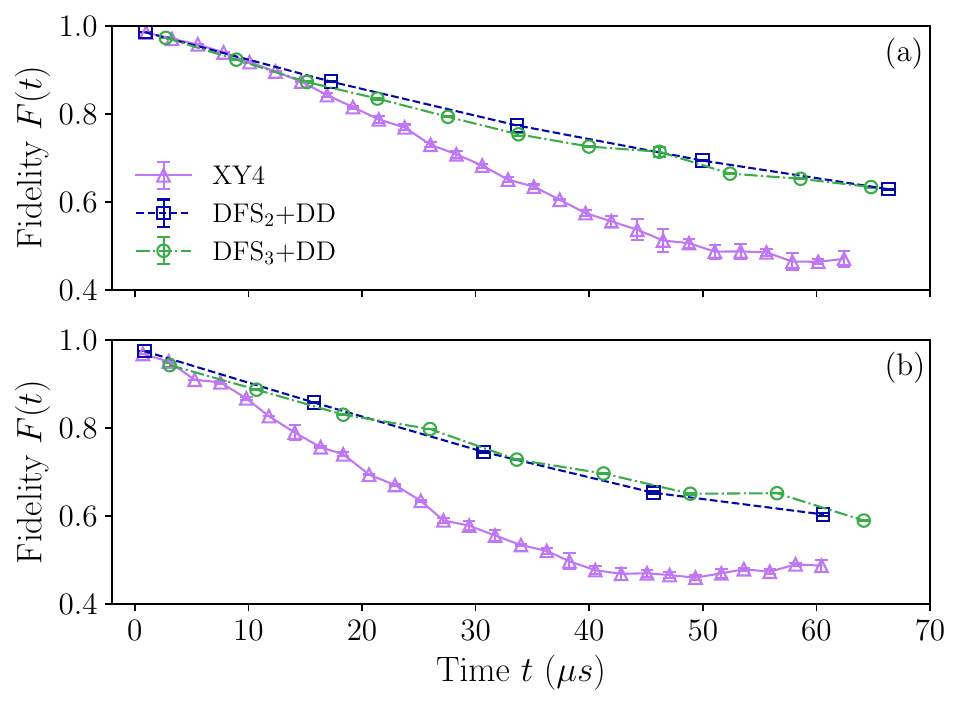}
    \caption{State-averaged fidelity as a function of time for two different qubit mappings on Manila. Panel (a) shows results for XY4, DFS$_2$+DD, and DFS$_3$+DD for $(q_0, q_1, q_2)=(3,4,2)$. Similar results are shown in panel (b) for $(q_0, q_1, q_2)=(2,3,1)$. Means (data points) and CIs are estimated from bootstrapping over one demonstration performed on October 20, 2022, using $8000$ shots. In both cases, logical encodings used in conjunction with DD and PS yield higher fidelity and slower fidelity decay than DD alone. Thus, the success of the protocol is independent of the qubit configuration.}
    \label{fig:subset-compare}
\end{figure}

\begin{figure}[t]
    \centering
    \includegraphics[width=\columnwidth]{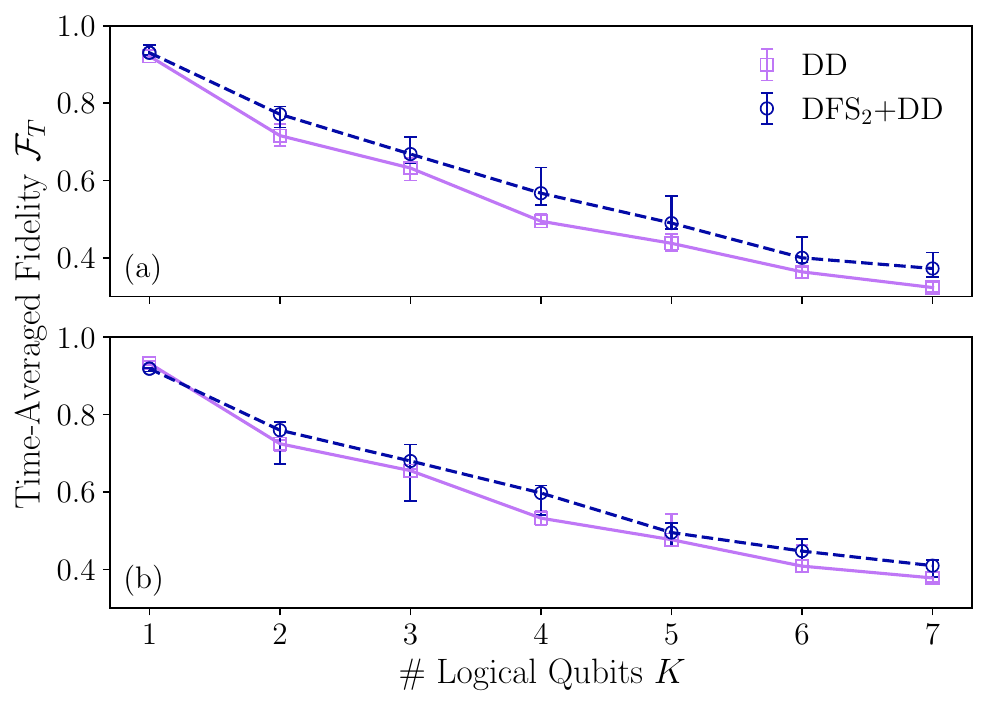}
    \caption{Time-averaged fidelity versus the number of logical qubits for distinct physical qubits configurations on Montreal. Panels (a) and (b) show comparisons for configurations 2 and 3 (\cref{tbl:qubit-configs}), respectively. The DD and DFS$_2$+DD protocols used for the comparison are equivalent to those described in \cref{subsec:multi-preserve}. In both cases, DFS$_2$+DD performs similarly or better than physical DD alone. Data is processed using bootstrapping to estimate means and CIs (error bars). Bootstrapping is performed over five realizations of the demonstration that were collected from October 20-22, 2022, each using $8000$ shots.}
    \label{fig:other-configs}
\end{figure}

Despite the dissimilarity in total PS shot count between the codes, the trend in fidelity is universal: DD and PS used together typically supply the greatest positive impact on code performance. In \cref{fig:ps-gens-compare}, the state-averaged fidelity as a function of time is shown for each code under a variety of conditions. The data shown in each panel is used to produce the time-averaged fidelity shown in \cref{fig:dfs_compare}. It serves an additional purpose here, giving an additional viewpoint on PS and DD over a range of times not limited to the short- and long-time limits. This is particularly useful for observing crossovers in fidelity between protocols. For example, while DD+PS yields the highest fidelity for both protocols at short times (one repetition), PS is generally better suited for periods of long state preservation. The transition in preferred protocol arises due to apparent steady-state behavior in the fidelity that is inconsistent with the completely mixed state. In fact, it is more consistent with a convergence towards a partial symmetric subspace, most notably in the case of the DFS$_3$ code. Further analysis is required to clarify this behavior.

\begin{table}[t]
\begin{tabularx}{\columnwidth}{c c}
\hline\hline
Configuration & Logical Qubits \\ \hline
1 & (15,18),(10,12),(4,7),(13,14),(11,8),(16,19),(5,3)\\
2 & (22,25),(20,19),(24,23),(21,18),(15,12),(13,14),(11,8)\\
3 & (4,1),(2,3),(10,7),(5,8),(15,12),(11,14),(17,18)\\
\hline\hline
\end{tabularx}
\caption{Logical qubit configurations for DFS protocol performed on IBMQP Montreal. Each pair $(q_0,q_1)$ denotes a logical qubit, with $q_i$ designating the qubit number based on the hardware topology graph shown in \cref{fig:hw-tops}. Results for configuration 1 are shown in \cref{fig:scalability}, while results for 2 and 3 are given in \cref{fig:other-configs}.}
\label{tbl:qubit-configs}
\end{table}

\begin{figure*}[t]
    \centering
    \includegraphics[width=\textwidth]{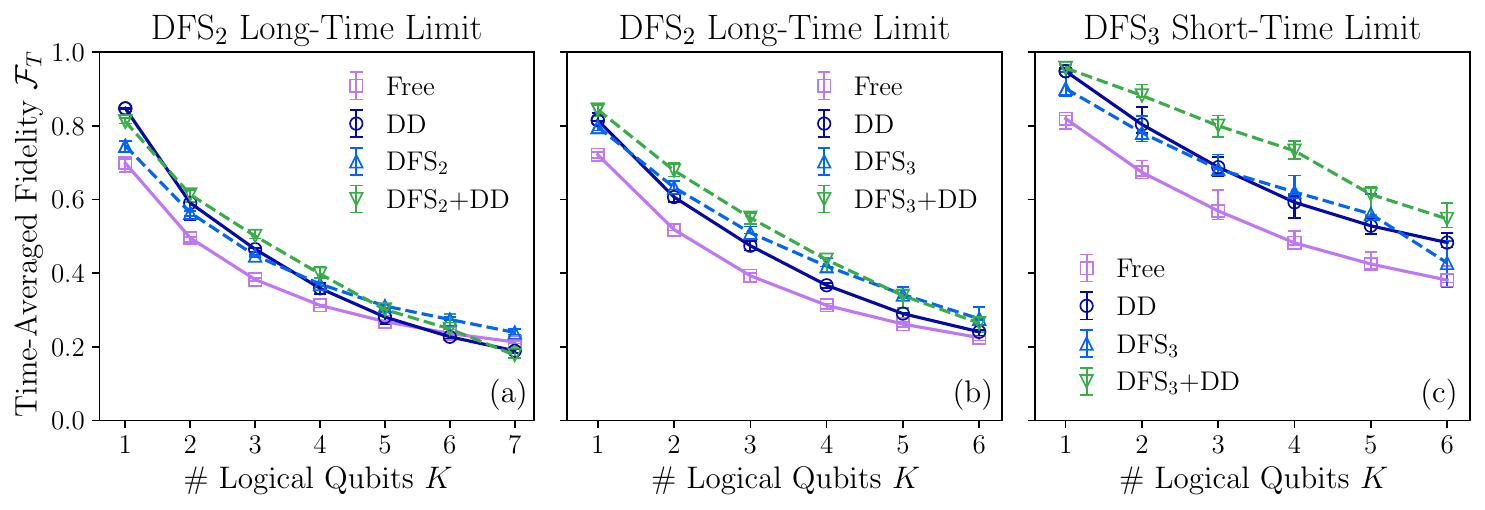}
    \caption{Time-averaged fidelity as a function of the number of logical qubits for different integration times. Comparisons between physical and logical protocols in the long-time limit (i.e., 3 repetitions of the DFS$_2$+DD sequence) are shown in panels (a) and (b) for the DFS$_2$ and DFS$_3$ code, respectively. Physical (solid lines) and logical (dashed lines) protocols are described in \cref{subsec:multi-preserve}. In both cases, the advantage of the logical protocols is observed, particularly when incorporating DD and PS; this is most notable for the DFS$_3$ code. In panel (c), the DFS$_3$ code again exhibits an advantage over physical encoding, but for a total integration time of one DFS$_3$+DD repetition. Data shown here was collected concurrently with the data used to produce \cref{fig:scalability} and therefore, follows the same data collection and processing practices.}
    \label{fig:scaling-ns-one-rep}
\end{figure*}

\section{Logical Qubit Fidelity and Qubit Variability}
\label{app:subsec:alt-q-subsets}

In the main text, we showcase demonstrations that provide evidence of logical encodings outperforming physical qubits via DFS codes combined with error detection and suppression. In \cref{subsec:one-qubit-pres,subsec:multi-preserve}, data is shown for specific subsets of qubits on Manila and Montreal. In this section, we show that the behavior observed from these devices is not limited to those subsets; it can also be found in other qubit configurations.

\subsection{One Logical Qubit}
First, we focus on the single logical qubit case discussed in \cref{subsec:one-qubit-pres}. The results shown in \cref{fig:dfs_compare} are for the qubit mapping $(q_0, q_1, q_2)=(3,4,2)$. While this subset of qubits yields the highest fidelity for the logical protection protocols, it is not the only subset that conveys an advantage from logical encoding. In \cref{fig:subset-compare}, results are shown for two additional qubit configurations (a) $(q_0, q_1, q_2)=(3,2,4)$ and (b) $(q_0, q_1, q_2)=(2,3,1)$; see \cref{fig:hw-tops} for device topology. Average qubit fidelity is determined by bootstrapping over the set of 20 states discussed in \cref{subsec:one-qubit-pres}. Mean fidelities and CIs are determined from one realization of the demonstration performed on October 20, 2022. Note that in both cases, the DFS encoding, DD, and PS together yield higher fidelities and slower fidelity decay rates than XY4 alone. Thus, empirical findings suggest that the relative improvements from the DFS protocol are robust to variability in qubit characteristics.

\subsection{Multiple Logical Qubits}
We further illustrate the robustness of the DFS code through additional studies of the time-averaged fidelity as a function of the number of logical qubits. In \cref{fig:other-configs}, the time-averaged fidelity in the short-time limit is shown for two additional configurations of physical qubits outlined in \cref{tbl:qubit-configs}. Panel (a) and (b) display results for configurations 2 and 3, respectively, with configuration 1 shown in the main text (\cref{fig:scalability}). Each panel contains a comparison between the physical DD and logical DFS$_2$+DD (with PS) protocols outlined in \cref{subsec:multi-preserve}. In both cases, the logical encoding performs similarly or somewhat better than physical-qubit DD, consistent with the results shown in the main text.

\section{Time-Averaged Fidelity at Different Integration Times}
In the main text, results were shown for the time-averaged fidelity with an integration time equivalent to one DFS$_2$+DD repetition, including encoding and decoding time. In this section, we consider additional integration times and examine the efficacy of each logical protocol. In \cref{fig:scaling-ns-one-rep}, the DFS$_2$ and DFS$_3$ codes are compared against physical encoding schemes in the long-time limit in panels (a) and (b). Panel (c) compares the DFS$_3$ code to physical encodings for an integration time of one DFS$_3$+DD repetition with encoding and decoding. As in the main text, all logical protocols utilize PS.

Longer integration time lowers the fidelity. This is particularly true for both the physical DD and DFS$_2$+DD protocols. Reduction in fidelity is accompanied by near equivalent performance among both DD-based protocols. Alternatively, the DFS$_2$ without DD exhibits improvements in this regime such that it begins to outperform all protocols for $K>5$. The steady-state behavior observed in \cref{fig:ps-gens-compare} in the long-time limit ultimately explains this crossover in performance.

PS continues to be an essential part of the DFS$_3$ protocol outperforming physical qubit error suppression. Both DFS$_3$ and DFS$_3$+DD achieve a higher fidelity than DD alone, with the advantage of the DFS$_3$ increasing with $K$. The performance improvement is so significant that both the DFS$_3$ and DFS$_3$+DD protocols perform nearly identically for $K>5$.

Lastly, we comment on an additional short-time limit scenario relative to the DFS$_3$ code. The main text defines the short-time limit as one cycle of the DFS$_2$+DD sequence or approximately two cycles of the DFS$_3$+DD sequence. Here, we consider an integration time of one DFS$_3$+DD sequence, with results of this comparison shown in Fig.~\ref{fig:scaling-ns-one-rep}(c). The result is a more pronounced disparity between the physical and logical protocols. The DFS$_3$ code achieves fidelities nearly equivalent to DD, while the DFS$_3$+DD protocol yields a maximum improvement in $\mathcal{F}_T$ of 23.6\% over DD. The success of both protocols further justifies employing passive QEC codes as a viable logical encoding option for near-term devices.

%


\begin{thebibliography}{90}%
\makeatletter
\providecommand \@ifxundefined [1]{%
 \@ifx{#1\undefined}
}%
\providecommand \@ifnum [1]{%
 \ifnum #1\expandafter \@firstoftwo
 \else \expandafter \@secondoftwo
 \fi
}%
\providecommand \@ifx [1]{%
 \ifx #1\expandafter \@firstoftwo
 \else \expandafter \@secondoftwo
 \fi
}%
\providecommand \natexlab [1]{#1}%
\providecommand \enquote  [1]{``#1''}%
\providecommand \bibnamefont  [1]{#1}%
\providecommand \bibfnamefont [1]{#1}%
\providecommand \citenamefont [1]{#1}%
\providecommand \href@noop [0]{\@secondoftwo}%
\providecommand \href [0]{\begingroup \@sanitize@url \@href}%
\providecommand \@href[1]{\@@startlink{#1}\@@href}%
\providecommand \@@href[1]{\endgroup#1\@@endlink}%
\providecommand \@sanitize@url [0]{\catcode `\\12\catcode `\$12\catcode
  `\&12\catcode `\#12\catcode `\^12\catcode `\_12\catcode `\%12\relax}%
\providecommand \@@startlink[1]{}%
\providecommand \@@endlink[0]{}%
\providecommand \url  [0]{\begingroup\@sanitize@url \@url }%
\providecommand \@url [1]{\endgroup\@href {#1}{\urlprefix }}%
\providecommand \urlprefix  [0]{URL }%
\providecommand \Eprint [0]{\href }%
\providecommand \doibase [0]{https://doi.org/}%
\providecommand \selectlanguage [0]{\@gobble}%
\providecommand \bibinfo  [0]{\@secondoftwo}%
\providecommand \bibfield  [0]{\@secondoftwo}%
\providecommand \translation [1]{[#1]}%
\providecommand \BibitemOpen [0]{}%
\providecommand \bibitemStop [0]{}%
\providecommand \bibitemNoStop [0]{.\EOS\space}%
\providecommand \EOS [0]{\spacefactor3000\relax}%
\providecommand \BibitemShut  [1]{\csname bibitem#1\endcsname}%
\let\auto@bib@innerbib\@empty
\bibitem [{\citenamefont {Lidar}\ and\ \citenamefont
  {Brun}(2013)}]{lidar_brun_2013}%
  \BibitemOpen
  \bibinfo {editor} {\bibfnamefont {D.}~\bibnamefont {Lidar}}\ and\ \bibinfo
  {editor} {\bibfnamefont {T.}~\bibnamefont {Brun}},\ eds.,\ \href
  {http://www.cambridge.org/9780521897877} {\emph {\bibinfo {title} {Quantum
  Error Correction}}}\ (\bibinfo  {publisher} {Cambridge University Press},\
  \bibinfo {address} {{Cambridge, UK}},\ \bibinfo {year} {2013})\BibitemShut
  {NoStop}%
\bibitem [{\citenamefont {Viola}\ and\ \citenamefont {Lloyd}(1998)}]{Viola:98}%
  \BibitemOpen
  \bibfield  {author} {\bibinfo {author} {\bibfnamefont {L.}~\bibnamefont
  {Viola}}\ and\ \bibinfo {author} {\bibfnamefont {S.}~\bibnamefont {Lloyd}},\
  }\bibfield  {title} {\bibinfo {title} {Dynamical suppression of decoherence
  in two-state quantum systems},\ }\href
  {https://link.aps.org/doi/10.1103/PhysRevA.58.2733} {\bibfield  {journal}
  {\bibinfo  {journal} {Phys. Rev. A}\ }\textbf {\bibinfo {volume} {58}},\
  \bibinfo {pages} {2733} (\bibinfo {year} {1998})}\BibitemShut {NoStop}%
\bibitem [{\citenamefont {Viola}\ \emph {et~al.}(1999)\citenamefont {Viola},
  \citenamefont {Knill},\ and\ \citenamefont {Lloyd}}]{viola1999dynamical}%
  \BibitemOpen
  \bibfield  {author} {\bibinfo {author} {\bibfnamefont {L.}~\bibnamefont
  {Viola}}, \bibinfo {author} {\bibfnamefont {E.}~\bibnamefont {Knill}},\ and\
  \bibinfo {author} {\bibfnamefont {S.}~\bibnamefont {Lloyd}},\ }\bibfield
  {title} {\bibinfo {title} {Dynamical decoupling of open quantum systems},\
  }\href@noop {} {\bibfield  {journal} {\bibinfo  {journal} {Physical Review
  Letters}\ }\textbf {\bibinfo {volume} {82}},\ \bibinfo {pages} {2417}
  (\bibinfo {year} {1999})}\BibitemShut {NoStop}%
\bibitem [{\citenamefont {Zanardi}(1999)}]{zanardiSymmetrizingEvolutions1999}%
  \BibitemOpen
  \bibfield  {author} {\bibinfo {author} {\bibfnamefont {P.}~\bibnamefont
  {Zanardi}},\ }\bibfield  {title} {\bibinfo {title} {Symmetrizing
  evolutions},\ }\href
  {http://www.sciencedirect.com/science/article/pii/S0375960199003655}
  {\bibfield  {journal} {\bibinfo  {journal} {Physics Letters A}\ }\textbf
  {\bibinfo {volume} {258}},\ \bibinfo {pages} {77} (\bibinfo {year}
  {1999})}\BibitemShut {NoStop}%
\bibitem [{\citenamefont {Gordon}\ \emph {et~al.}(2008)\citenamefont {Gordon},
  \citenamefont {Kurizki},\ and\ \citenamefont {Lidar}}]{Gordon:2008:010403}%
  \BibitemOpen
  \bibfield  {author} {\bibinfo {author} {\bibfnamefont {G.}~\bibnamefont
  {Gordon}}, \bibinfo {author} {\bibfnamefont {G.}~\bibnamefont {Kurizki}},\
  and\ \bibinfo {author} {\bibfnamefont {D.~A.}\ \bibnamefont {Lidar}},\
  }\bibfield  {title} {\bibinfo {title} {Optimal dynamical decoherence control
  of a qubit},\ }\href {https://doi.org/10.1103/PhysRevLett.101.010403}
  {\bibfield  {journal} {\bibinfo  {journal} {Phys. Rev. Lett.}\ }\textbf
  {\bibinfo {volume} {101}},\ \bibinfo {pages} {010403} (\bibinfo {year}
  {2008})}\BibitemShut {NoStop}%
\bibitem [{\citenamefont {Suter}\ and\ \citenamefont
  {{\'A}lvarez}(2016)}]{Suter:2016aa}%
  \BibitemOpen
  \bibfield  {author} {\bibinfo {author} {\bibfnamefont {D.}~\bibnamefont
  {Suter}}\ and\ \bibinfo {author} {\bibfnamefont {G.~A.}\ \bibnamefont
  {{\'A}lvarez}},\ }\bibfield  {title} {\bibinfo {title} {Colloquium:
  Protecting quantum information against environmental noise},\ }\href
  {https://link.aps.org/doi/10.1103/RevModPhys.88.041001} {\bibfield  {journal}
  {\bibinfo  {journal} {Rev. Mod. Phys.}\ }\textbf {\bibinfo {volume} {88}},\
  \bibinfo {pages} {041001} (\bibinfo {year} {2016})}\BibitemShut {NoStop}%
\bibitem [{\citenamefont {Shor}(1995)}]{shor1995scheme}%
  \BibitemOpen
  \bibfield  {author} {\bibinfo {author} {\bibfnamefont {P.~W.}\ \bibnamefont
  {Shor}},\ }\bibfield  {title} {\bibinfo {title} {Scheme for reducing
  decoherence in quantum computer memory},\ }\href@noop {} {\bibfield
  {journal} {\bibinfo  {journal} {Physical review A}\ }\textbf {\bibinfo
  {volume} {52}},\ \bibinfo {pages} {R2493} (\bibinfo {year}
  {1995})}\BibitemShut {NoStop}%
\bibitem [{\citenamefont {Steane}(1996)}]{Steane:96a}%
  \BibitemOpen
  \bibfield  {author} {\bibinfo {author} {\bibfnamefont {A.~M.}\ \bibnamefont
  {Steane}},\ }\bibfield  {title} {\bibinfo {title} {Error correcting codes in
  quantum theory},\ }\href {http://link.aps.org/doi/10.1103/PhysRevLett.77.793}
  {\bibfield  {journal} {\bibinfo  {journal} {Phys. Rev. Lett.}\ }\textbf
  {\bibinfo {volume} {77}},\ \bibinfo {pages} {793} (\bibinfo {year}
  {1996})}\BibitemShut {NoStop}%
\bibitem [{\citenamefont {Gottesman}(1996)}]{Gottesman:1996fk}%
  \BibitemOpen
  \bibfield  {author} {\bibinfo {author} {\bibfnamefont {D.}~\bibnamefont
  {Gottesman}},\ }\bibfield  {title} {\bibinfo {title} {Class of quantum
  error-correcting codes saturating the quantum hamming bound},\ }\href
  {https://doi.org/10.1103/PhysRevA.54.1862} {\bibfield  {journal} {\bibinfo
  {journal} {{Phys. Rev. A}}\ }\textbf {\bibinfo {volume} {54}},\ \bibinfo
  {pages} {1862} (\bibinfo {year} {1996})}\BibitemShut {NoStop}%
\bibitem [{\citenamefont {Gaitan}(2008)}]{Gaitan:book}%
  \BibitemOpen
  \bibfield  {author} {\bibinfo {author} {\bibfnamefont {F.}~\bibnamefont
  {Gaitan}},\ }\href {http://books.google.com/books?id=zwvlqspyOK8C} {\emph
  {\bibinfo {title} {Quantum Error Correction and Fault Tolerant Quantum
  Computing}}}\ (\bibinfo  {publisher} {Taylor \& Francis Group},\ \bibinfo
  {address} {Boca Raton},\ \bibinfo {year} {2008})\BibitemShut {NoStop}%
\bibitem [{\citenamefont {Alicki}(1988)}]{Alicki:88}%
  \BibitemOpen
  \bibfield  {author} {\bibinfo {author} {\bibfnamefont {R.}~\bibnamefont
  {Alicki}},\ }\bibfield  {title} {\bibinfo {title} {Limited thermalization for
  the {M}arkov mean-field model of ${N}$ atoms in thermal field},\ }\href
  {http://www.sciencedirect.com/science/article/pii/037843718890163X}
  {\bibfield  {journal} {\bibinfo  {journal} {Physica A: Statistical Mechanics
  and its Applications}\ }\textbf {\bibinfo {volume} {150}},\ \bibinfo {pages}
  {455} (\bibinfo {year} {1988})}\BibitemShut {NoStop}%
\bibitem [{\citenamefont {{G.M. Palma, K.-A. Suominen and A.K.
  Ekert}}(1996)}]{Palma:96}%
  \BibitemOpen
  \bibfield  {author} {\bibinfo {author} {\bibnamefont {{G.M. Palma, K.-A.
  Suominen and A.K. Ekert}}},\ }\bibfield  {title} {\bibinfo {title} {{Quantum
  Computers and Dissipation}},\ }\href
  {https://royalsocietypublishing.org/doi/10.1098/rspa.1996.0029} {\bibfield
  {journal} {\bibinfo  {journal} {Proc. R.. Soc. London Ser. A}\ }\textbf
  {\bibinfo {volume} {452}},\ \bibinfo {pages} {567} (\bibinfo {year}
  {1996})}\BibitemShut {NoStop}%
\bibitem [{\citenamefont {Zanardi}\ and\ \citenamefont
  {Rasetti}(1997)}]{zanardiNoiselessQuantumCodes1997}%
  \BibitemOpen
  \bibfield  {author} {\bibinfo {author} {\bibfnamefont {P.}~\bibnamefont
  {Zanardi}}\ and\ \bibinfo {author} {\bibfnamefont {M.}~\bibnamefont
  {Rasetti}},\ }\bibfield  {title} {\bibinfo {title} {Noiseless {{Quantum
  Codes}}},\ }\href {https://doi.org/10.1103/PhysRevLett.79.3306} {\bibfield
  {journal} {\bibinfo  {journal} {Physical Review Letters}\ }\textbf {\bibinfo
  {volume} {79}},\ \bibinfo {pages} {3306} (\bibinfo {year}
  {1997})}\BibitemShut {NoStop}%
\bibitem [{\citenamefont {Duan}\ and\ \citenamefont {Guo}(1997)}]{Duan:1997aa}%
  \BibitemOpen
  \bibfield  {author} {\bibinfo {author} {\bibfnamefont {L.-M.}\ \bibnamefont
  {Duan}}\ and\ \bibinfo {author} {\bibfnamefont {G.-C.}\ \bibnamefont {Guo}},\
  }\bibfield  {title} {\bibinfo {title} {Preserving coherence in quantum
  computation by pairing quantum bits},\ }\href
  {https://doi.org/10.1103/PhysRevLett.79.1953} {\bibfield  {journal} {\bibinfo
   {journal} {Physical Review Letters}\ }\textbf {\bibinfo {volume} {79}},\
  \bibinfo {pages} {1953} (\bibinfo {year} {1997})}\BibitemShut {NoStop}%
\bibitem [{\citenamefont {Lidar}\ \emph {et~al.}(1998)\citenamefont {Lidar},
  \citenamefont {Chuang},\ and\ \citenamefont {Whaley}}]{lidar1998decoherence}%
  \BibitemOpen
  \bibfield  {author} {\bibinfo {author} {\bibfnamefont {D.~A.}\ \bibnamefont
  {Lidar}}, \bibinfo {author} {\bibfnamefont {I.~L.}\ \bibnamefont {Chuang}},\
  and\ \bibinfo {author} {\bibfnamefont {K.~B.}\ \bibnamefont {Whaley}},\
  }\bibfield  {title} {\bibinfo {title} {Decoherence-free subspaces for quantum
  computation},\ }\href@noop {} {\bibfield  {journal} {\bibinfo  {journal}
  {Physical Review Letters}\ }\textbf {\bibinfo {volume} {81}},\ \bibinfo
  {pages} {2594} (\bibinfo {year} {1998})}\BibitemShut {NoStop}%
\bibitem [{\citenamefont {Knill}\ \emph {et~al.}(2000)\citenamefont {Knill},
  \citenamefont {Laflamme},\ and\ \citenamefont {Viola}}]{Knill:2000dq}%
  \BibitemOpen
  \bibfield  {author} {\bibinfo {author} {\bibfnamefont {E.}~\bibnamefont
  {Knill}}, \bibinfo {author} {\bibfnamefont {R.}~\bibnamefont {Laflamme}},\
  and\ \bibinfo {author} {\bibfnamefont {L.}~\bibnamefont {Viola}},\ }\bibfield
   {title} {\bibinfo {title} {Theory of quantum error correction for general
  noise},\ }\href {http://link.aps.org/doi/10.1103/PhysRevLett.84.2525}
  {\bibfield  {journal} {\bibinfo  {journal} {{Phys.~Rev.~Lett.}}\ }\textbf
  {\bibinfo {volume} {84}},\ \bibinfo {pages} {2525} (\bibinfo {year}
  {2000})}\BibitemShut {NoStop}%
\bibitem [{\citenamefont {Zanardi}(2000)}]{Zanardi:99d}%
  \BibitemOpen
  \bibfield  {author} {\bibinfo {author} {\bibfnamefont {P.}~\bibnamefont
  {Zanardi}},\ }\bibfield  {title} {\bibinfo {title} {Stabilizing quantum
  information},\ }\href {http://link.aps.org/doi/10.1103/PhysRevA.63.012301}
  {\bibfield  {journal} {\bibinfo  {journal} {Physical Review A}\ }\textbf
  {\bibinfo {volume} {63}},\ \bibinfo {pages} {012301} (\bibinfo {year}
  {2000})}\BibitemShut {NoStop}%
\bibitem [{\citenamefont {Cai}\ \emph {et~al.}(2022)\citenamefont {Cai},
  \citenamefont {Babbush}, \citenamefont {Benjamin}, \citenamefont {Endo},
  \citenamefont {Huggins}, \citenamefont {Li}, \citenamefont {McClean},\ and\
  \citenamefont {O'Brien}}]{cai2022quantum}%
  \BibitemOpen
  \bibfield  {author} {\bibinfo {author} {\bibfnamefont {Z.}~\bibnamefont
  {Cai}}, \bibinfo {author} {\bibfnamefont {R.}~\bibnamefont {Babbush}},
  \bibinfo {author} {\bibfnamefont {S.~C.}\ \bibnamefont {Benjamin}}, \bibinfo
  {author} {\bibfnamefont {S.}~\bibnamefont {Endo}}, \bibinfo {author}
  {\bibfnamefont {W.~J.}\ \bibnamefont {Huggins}}, \bibinfo {author}
  {\bibfnamefont {Y.}~\bibnamefont {Li}}, \bibinfo {author} {\bibfnamefont
  {J.~R.}\ \bibnamefont {McClean}},\ and\ \bibinfo {author} {\bibfnamefont
  {T.~E.}\ \bibnamefont {O'Brien}},\ }\bibfield  {title} {\bibinfo {title}
  {Quantum error mitigation},\ }\href@noop {} {\bibfield  {journal} {\bibinfo
  {journal} {arXiv preprint arXiv:2210.00921}\ } (\bibinfo {year}
  {2022})}\BibitemShut {NoStop}%
\bibitem [{\citenamefont {Lidar}\ \emph {et~al.}(1999)\citenamefont {Lidar},
  \citenamefont {Bacon},\ and\ \citenamefont {Whaley}}]{lidar1999qec-dfs}%
  \BibitemOpen
  \bibfield  {author} {\bibinfo {author} {\bibfnamefont {D.~A.}\ \bibnamefont
  {Lidar}}, \bibinfo {author} {\bibfnamefont {D.}~\bibnamefont {Bacon}},\ and\
  \bibinfo {author} {\bibfnamefont {K.~B.}\ \bibnamefont {Whaley}},\ }\bibfield
   {title} {\bibinfo {title} {Concatenating decoherence-free subspaces with
  quantum error correcting codes},\ }\href
  {https://doi.org/10.1103/PhysRevLett.82.4556} {\bibfield  {journal} {\bibinfo
   {journal} {Phys. Rev. Lett.}\ }\textbf {\bibinfo {volume} {82}},\ \bibinfo
  {pages} {4556} (\bibinfo {year} {1999})}\BibitemShut {NoStop}%
\bibitem [{\citenamefont {Lidar}\ \emph {et~al.}(2000)\citenamefont {Lidar},
  \citenamefont {Bacon}, \citenamefont {Kempe},\ and\ \citenamefont
  {Birgitta~Whaley}}]{Lidar:PRA00Exchange}%
  \BibitemOpen
  \bibfield  {author} {\bibinfo {author} {\bibfnamefont {D.~A.}\ \bibnamefont
  {Lidar}}, \bibinfo {author} {\bibfnamefont {D.}~\bibnamefont {Bacon}},
  \bibinfo {author} {\bibfnamefont {J.}~\bibnamefont {Kempe}},\ and\ \bibinfo
  {author} {\bibfnamefont {K.}~\bibnamefont {Birgitta~Whaley}},\ }\bibfield
  {title} {\bibinfo {title} {Protecting quantum information encoded in
  decoherence-free states against exchange errors},\ }\href
  {https://doi.org/10.1103/PhysRevA.61.052307} {\bibfield  {journal} {\bibinfo
  {journal} {Physical Review A}\ }\textbf {\bibinfo {volume} {61}},\ \bibinfo
  {pages} {052307} (\bibinfo {year} {2000})}\BibitemShut {NoStop}%
\bibitem [{\citenamefont {Viola}\ \emph {et~al.}(2000)\citenamefont {Viola},
  \citenamefont {Knill},\ and\ \citenamefont
  {Lloyd}}]{violaDynamicalGenerationNoiseless2000}%
  \BibitemOpen
  \bibfield  {author} {\bibinfo {author} {\bibfnamefont {L.}~\bibnamefont
  {Viola}}, \bibinfo {author} {\bibfnamefont {E.}~\bibnamefont {Knill}},\ and\
  \bibinfo {author} {\bibfnamefont {S.}~\bibnamefont {Lloyd}},\ }\bibfield
  {title} {\bibinfo {title} {Dynamical generation of noiseless quantum
  subsystems},\ }\href {http://link.aps.org/doi/10.1103/PhysRevLett.85.3520}
  {\bibfield  {journal} {\bibinfo  {journal} {Phys. Rev. Lett.}\ }\textbf
  {\bibinfo {volume} {85}},\ \bibinfo {pages} {3520} (\bibinfo {year}
  {2000})}\BibitemShut {NoStop}%
\bibitem [{\citenamefont {Alber}\ \emph {et~al.}(2001)\citenamefont {Alber},
  \citenamefont {Beth}, \citenamefont {Charnes}, \citenamefont {Delgado},
  \citenamefont {Grassl},\ and\ \citenamefont {Mussinger}}]{Alber:01}%
  \BibitemOpen
  \bibfield  {author} {\bibinfo {author} {\bibfnamefont {G.}~\bibnamefont
  {Alber}}, \bibinfo {author} {\bibfnamefont {T.}~\bibnamefont {Beth}},
  \bibinfo {author} {\bibfnamefont {C.}~\bibnamefont {Charnes}}, \bibinfo
  {author} {\bibfnamefont {A.}~\bibnamefont {Delgado}}, \bibinfo {author}
  {\bibfnamefont {M.}~\bibnamefont {Grassl}},\ and\ \bibinfo {author}
  {\bibfnamefont {M.}~\bibnamefont {Mussinger}},\ }\bibfield  {title} {\bibinfo
  {title} {Stabilizing distinguishable qubits against spontaneous decay by
  detected-jump correcting quantum codes},\ }\href
  {https://doi.org/10.1103/PhysRevLett.86.4402} {\bibfield  {journal} {\bibinfo
   {journal} {Physical Review Letters}\ }\textbf {\bibinfo {volume} {86}},\
  \bibinfo {pages} {4402} (\bibinfo {year} {2001})}\BibitemShut {NoStop}%
\bibitem [{\citenamefont {Alber}\ \emph {et~al.}(2003)\citenamefont {Alber},
  \citenamefont {Beth}, \citenamefont {Charnes}, \citenamefont {Delgado},
  \citenamefont {Grassl},\ and\ \citenamefont {Mussinger}}]{Alber:02a}%
  \BibitemOpen
  \bibfield  {author} {\bibinfo {author} {\bibfnamefont {G.}~\bibnamefont
  {Alber}}, \bibinfo {author} {\bibfnamefont {T.}~\bibnamefont {Beth}},
  \bibinfo {author} {\bibfnamefont {C.}~\bibnamefont {Charnes}}, \bibinfo
  {author} {\bibfnamefont {A.}~\bibnamefont {Delgado}}, \bibinfo {author}
  {\bibfnamefont {M.}~\bibnamefont {Grassl}},\ and\ \bibinfo {author}
  {\bibfnamefont {M.}~\bibnamefont {Mussinger}},\ }\bibfield  {title} {\bibinfo
  {title} {Detected-jump-error-correcting quantum codes, quantum error designs,
  and quantum computation},\ }\href
  {https://doi.org/10.1103/PhysRevA.68.012316} {\bibfield  {journal} {\bibinfo
  {journal} {Physical Review A}\ }\textbf {\bibinfo {volume} {68}},\ \bibinfo
  {pages} {012316} (\bibinfo {year} {2003})}\BibitemShut {NoStop}%
\bibitem [{\citenamefont {Khodjasteh}\ and\ \citenamefont
  {Lidar}(2002)}]{KhodjastehLidar:02}%
  \BibitemOpen
  \bibfield  {author} {\bibinfo {author} {\bibfnamefont {K.}~\bibnamefont
  {Khodjasteh}}\ and\ \bibinfo {author} {\bibfnamefont {D.~A.}\ \bibnamefont
  {Lidar}},\ }\bibfield  {title} {\bibinfo {title} {Universal fault-tolerant
  quantum computation in the presence of spontaneous emission and collective
  dephasing},\ }\href {https://link.aps.org/doi/10.1103/PhysRevLett.89.197904}
  {\bibfield  {journal} {\bibinfo  {journal} {Physical Review Letters}\
  }\textbf {\bibinfo {volume} {89}},\ \bibinfo {pages} {197904} (\bibinfo
  {year} {2002})}\BibitemShut {NoStop}%
\bibitem [{\citenamefont {Khodjasteh}\ and\ \citenamefont
  {Lidar}(2003)}]{KhodjastehLidar:03}%
  \BibitemOpen
  \bibfield  {author} {\bibinfo {author} {\bibfnamefont {K.}~\bibnamefont
  {Khodjasteh}}\ and\ \bibinfo {author} {\bibfnamefont {D.~A.}\ \bibnamefont
  {Lidar}},\ }\bibfield  {title} {\bibinfo {title} {Quantum computing in the
  presence of spontaneous emission by a combined dynamical decoupling and
  quantum-error-correction strategy},\ }\href
  {https://link.aps.org/doi/10.1103/PhysRevA.68.022322} {\bibfield  {journal}
  {\bibinfo  {journal} {Physical Review A}\ }\textbf {\bibinfo {volume} {68}},\
  \bibinfo {pages} {022322} (\bibinfo {year} {2003})},\ \bibinfo {note}
  {erratum: {\it ibid}, Phys. Rev. A {\bf 72}, 029905 (2005).}\BibitemShut
  {Stop}%
\bibitem [{\citenamefont {Ng}\ \emph {et~al.}(2011)\citenamefont {Ng},
  \citenamefont {Lidar},\ and\ \citenamefont
  {Preskill}}]{ngCombiningDynamicalDecoupling2011}%
  \BibitemOpen
  \bibfield  {author} {\bibinfo {author} {\bibfnamefont {H.~K.}\ \bibnamefont
  {Ng}}, \bibinfo {author} {\bibfnamefont {D.~A.}\ \bibnamefont {Lidar}},\ and\
  \bibinfo {author} {\bibfnamefont {J.}~\bibnamefont {Preskill}},\ }\bibfield
  {title} {\bibinfo {title} {Combining dynamical decoupling with fault-tolerant
  quantum computation},\ }\href
  {http://link.aps.org/doi/10.1103/PhysRevA.84.012305} {\bibfield  {journal}
  {\bibinfo  {journal} {Phys. Rev. A}\ }\textbf {\bibinfo {volume} {84}},\
  \bibinfo {pages} {012305} (\bibinfo {year} {2011})}\BibitemShut {NoStop}%
\bibitem [{\citenamefont {Paz-Silva}\ and\ \citenamefont
  {Lidar}(2013)}]{Paz-Silva:2013tt}%
  \BibitemOpen
  \bibfield  {author} {\bibinfo {author} {\bibfnamefont {G.~A.}\ \bibnamefont
  {Paz-Silva}}\ and\ \bibinfo {author} {\bibfnamefont {D.~A.}\ \bibnamefont
  {Lidar}},\ }\bibfield  {title} {\bibinfo {title} {Optimally combining
  dynamical decoupling and quantum error correction},\ }\href
  {https://www.nature.com/articles/srep01530} {\bibfield  {journal} {\bibinfo
  {journal} {Sci. Rep.}\ }\textbf {\bibinfo {volume} {3}},\ \bibinfo {pages}
  {1530} (\bibinfo {year} {2013})}\BibitemShut {NoStop}%
\bibitem [{\citenamefont {Campbell}\ \emph {et~al.}(2017)\citenamefont
  {Campbell}, \citenamefont {Terhal},\ and\ \citenamefont
  {Vuillot}}]{Campbell:2017aa}%
  \BibitemOpen
  \bibfield  {author} {\bibinfo {author} {\bibfnamefont {E.~T.}\ \bibnamefont
  {Campbell}}, \bibinfo {author} {\bibfnamefont {B.~M.}\ \bibnamefont
  {Terhal}},\ and\ \bibinfo {author} {\bibfnamefont {C.}~\bibnamefont
  {Vuillot}},\ }\bibfield  {title} {\bibinfo {title} {Roads towards
  fault-tolerant universal quantum computation},\ }\href
  {http://dx.doi.org/10.1038/nature23460} {\bibfield  {journal} {\bibinfo
  {journal} {Nature}\ }\textbf {\bibinfo {volume} {549}},\ \bibinfo {pages}
  {172 EP } (\bibinfo {year} {2017})}\BibitemShut {NoStop}%
\bibitem [{\citenamefont {Dumitrescu}\ \emph {et~al.}(2018)\citenamefont
  {Dumitrescu}, \citenamefont {McCaskey}, \citenamefont {Hagen}, \citenamefont
  {Jansen}, \citenamefont {Morris}, \citenamefont {Papenbrock}, \citenamefont
  {Pooser}, \citenamefont {Dean},\ and\ \citenamefont
  {Lougovski}}]{Dumitrescu2018vqe}%
  \BibitemOpen
  \bibfield  {author} {\bibinfo {author} {\bibfnamefont {E.~F.}\ \bibnamefont
  {Dumitrescu}}, \bibinfo {author} {\bibfnamefont {A.~J.}\ \bibnamefont
  {McCaskey}}, \bibinfo {author} {\bibfnamefont {G.}~\bibnamefont {Hagen}},
  \bibinfo {author} {\bibfnamefont {G.~R.}\ \bibnamefont {Jansen}}, \bibinfo
  {author} {\bibfnamefont {T.~D.}\ \bibnamefont {Morris}}, \bibinfo {author}
  {\bibfnamefont {T.}~\bibnamefont {Papenbrock}}, \bibinfo {author}
  {\bibfnamefont {R.~C.}\ \bibnamefont {Pooser}}, \bibinfo {author}
  {\bibfnamefont {D.~J.}\ \bibnamefont {Dean}},\ and\ \bibinfo {author}
  {\bibfnamefont {P.}~\bibnamefont {Lougovski}},\ }\bibfield  {title} {\bibinfo
  {title} {Cloud quantum computing of an atomic nucleus},\ }\href
  {https://doi.org/10.1103/PhysRevLett.120.210501} {\bibfield  {journal}
  {\bibinfo  {journal} {Phys. Rev. Lett.}\ }\textbf {\bibinfo {volume} {120}},\
  \bibinfo {pages} {210501} (\bibinfo {year} {2018})}\BibitemShut {NoStop}%
\bibitem [{\citenamefont {Kandala}\ \emph {et~al.}(2019)\citenamefont
  {Kandala}, \citenamefont {Temme}, \citenamefont {C{\'o}rcoles}, \citenamefont
  {Mezzacapo}, \citenamefont {Chow},\ and\ \citenamefont
  {Gambetta}}]{kandala2019error}%
  \BibitemOpen
  \bibfield  {author} {\bibinfo {author} {\bibfnamefont {A.}~\bibnamefont
  {Kandala}}, \bibinfo {author} {\bibfnamefont {K.}~\bibnamefont {Temme}},
  \bibinfo {author} {\bibfnamefont {A.~D.}\ \bibnamefont {C{\'o}rcoles}},
  \bibinfo {author} {\bibfnamefont {A.}~\bibnamefont {Mezzacapo}}, \bibinfo
  {author} {\bibfnamefont {J.~M.}\ \bibnamefont {Chow}},\ and\ \bibinfo
  {author} {\bibfnamefont {J.~M.}\ \bibnamefont {Gambetta}},\ }\bibfield
  {title} {\bibinfo {title} {Error mitigation extends the computational reach
  of a noisy quantum processor},\ }\href@noop {} {\bibfield  {journal}
  {\bibinfo  {journal} {Nature}\ }\textbf {\bibinfo {volume} {567}},\ \bibinfo
  {pages} {491} (\bibinfo {year} {2019})}\BibitemShut {NoStop}%
\bibitem [{\citenamefont {Kim}\ \emph {et~al.}(2023)\citenamefont {Kim},
  \citenamefont {Wood}, \citenamefont {Yoder}, \citenamefont {Merkel},
  \citenamefont {Gambetta}, \citenamefont {Temme},\ and\ \citenamefont
  {Kandala}}]{kim2023scalable}%
  \BibitemOpen
  \bibfield  {author} {\bibinfo {author} {\bibfnamefont {Y.}~\bibnamefont
  {Kim}}, \bibinfo {author} {\bibfnamefont {C.~J.}\ \bibnamefont {Wood}},
  \bibinfo {author} {\bibfnamefont {T.~J.}\ \bibnamefont {Yoder}}, \bibinfo
  {author} {\bibfnamefont {S.~T.}\ \bibnamefont {Merkel}}, \bibinfo {author}
  {\bibfnamefont {J.~M.}\ \bibnamefont {Gambetta}}, \bibinfo {author}
  {\bibfnamefont {K.}~\bibnamefont {Temme}},\ and\ \bibinfo {author}
  {\bibfnamefont {A.}~\bibnamefont {Kandala}},\ }\bibfield  {title} {\bibinfo
  {title} {Scalable error mitigation for noisy quantum circuits produces
  competitive expectation values},\ }\href@noop {} {\bibfield  {journal}
  {\bibinfo  {journal} {Nature Physics}\ ,\ \bibinfo {pages} {1}} (\bibinfo
  {year} {2023})}\BibitemShut {NoStop}%
\bibitem [{\citenamefont {Pokharel}\ and\ \citenamefont
  {Lidar}(2023)}]{pokharelDemonstrationAlgorithmicQuantum2022}%
  \BibitemOpen
  \bibfield  {author} {\bibinfo {author} {\bibfnamefont {B.}~\bibnamefont
  {Pokharel}}\ and\ \bibinfo {author} {\bibfnamefont {D.~A.}\ \bibnamefont
  {Lidar}},\ }\bibfield  {title} {\bibinfo {title} {Demonstration of
  algorithmic quantum speedup},\ }\href
  {https://doi.org/10.1103/PhysRevLett.130.210602} {\bibfield  {journal}
  {\bibinfo  {journal} {Physical Review Letters}\ }\textbf {\bibinfo {volume}
  {130}},\ \bibinfo {pages} {210602} (\bibinfo {year} {2023})}\BibitemShut
  {NoStop}%
\bibitem [{\citenamefont {Carr}\ and\ \citenamefont
  {Purcell}(1954)}]{carr1954effects}%
  \BibitemOpen
  \bibfield  {author} {\bibinfo {author} {\bibfnamefont {H.~Y.}\ \bibnamefont
  {Carr}}\ and\ \bibinfo {author} {\bibfnamefont {E.~M.}\ \bibnamefont
  {Purcell}},\ }\bibfield  {title} {\bibinfo {title} {Effects of diffusion on
  free precession in nuclear magnetic resonance experiments},\ }\href
  {https://doi.org/10.1103/PhysRev.94.630} {\bibfield  {journal} {\bibinfo
  {journal} {Phys. Rev.}\ }\textbf {\bibinfo {volume} {94}},\ \bibinfo {pages}
  {630} (\bibinfo {year} {1954})}\BibitemShut {NoStop}%
\bibitem [{\citenamefont {Meiboom}\ and\ \citenamefont
  {Gill}(1958)}]{meiboom1958modified}%
  \BibitemOpen
  \bibfield  {author} {\bibinfo {author} {\bibfnamefont {S.}~\bibnamefont
  {Meiboom}}\ and\ \bibinfo {author} {\bibfnamefont {D.}~\bibnamefont {Gill}},\
  }\bibfield  {title} {\bibinfo {title} {Modified spin-echo method for
  measuring nuclear relaxation times},\ }\href@noop {} {\bibfield  {journal}
  {\bibinfo  {journal} {Review of scientific instruments}\ }\textbf {\bibinfo
  {volume} {29}},\ \bibinfo {pages} {688} (\bibinfo {year} {1958})}\BibitemShut
  {NoStop}%
\bibitem [{\citenamefont {Haeberlen}\ and\ \citenamefont
  {Waugh}(1968)}]{haeberlen1968coherent}%
  \BibitemOpen
  \bibfield  {author} {\bibinfo {author} {\bibfnamefont {U.}~\bibnamefont
  {Haeberlen}}\ and\ \bibinfo {author} {\bibfnamefont {J.~S.}\ \bibnamefont
  {Waugh}},\ }\bibfield  {title} {\bibinfo {title} {Coherent averaging effects
  in magnetic resonance},\ }\href@noop {} {\bibfield  {journal} {\bibinfo
  {journal} {Physical Review}\ }\textbf {\bibinfo {volume} {175}},\ \bibinfo
  {pages} {453} (\bibinfo {year} {1968})}\BibitemShut {NoStop}%
\bibitem [{\citenamefont {Biercuk}\ \emph
  {et~al.}(2009{\natexlab{a}})\citenamefont {Biercuk}, \citenamefont {Uys},
  \citenamefont {VanDevender}, \citenamefont {Shiga}, \citenamefont {Itano},\
  and\ \citenamefont {Bollinger}}]{biercuk2009experimental}%
  \BibitemOpen
  \bibfield  {author} {\bibinfo {author} {\bibfnamefont {M.~J.}\ \bibnamefont
  {Biercuk}}, \bibinfo {author} {\bibfnamefont {H.}~\bibnamefont {Uys}},
  \bibinfo {author} {\bibfnamefont {A.~P.}\ \bibnamefont {VanDevender}},
  \bibinfo {author} {\bibfnamefont {N.}~\bibnamefont {Shiga}}, \bibinfo
  {author} {\bibfnamefont {W.~M.}\ \bibnamefont {Itano}},\ and\ \bibinfo
  {author} {\bibfnamefont {J.~J.}\ \bibnamefont {Bollinger}},\ }\bibfield
  {title} {\bibinfo {title} {Experimental uhrig dynamical decoupling using
  trapped ions},\ }\href@noop {} {\bibfield  {journal} {\bibinfo  {journal}
  {Physical Review A}\ }\textbf {\bibinfo {volume} {79}},\ \bibinfo {pages}
  {062324} (\bibinfo {year} {2009}{\natexlab{a}})}\BibitemShut {NoStop}%
\bibitem [{\citenamefont {Biercuk}\ \emph
  {et~al.}(2009{\natexlab{b}})\citenamefont {Biercuk}, \citenamefont {Uys},
  \citenamefont {VanDevender}, \citenamefont {Shiga}, \citenamefont {Itano},\
  and\ \citenamefont {Bollinger}}]{Biercuk:09}%
  \BibitemOpen
  \bibfield  {author} {\bibinfo {author} {\bibfnamefont {M.~J.}\ \bibnamefont
  {Biercuk}}, \bibinfo {author} {\bibfnamefont {H.}~\bibnamefont {Uys}},
  \bibinfo {author} {\bibfnamefont {A.~P.}\ \bibnamefont {VanDevender}},
  \bibinfo {author} {\bibfnamefont {N.}~\bibnamefont {Shiga}}, \bibinfo
  {author} {\bibfnamefont {W.~M.}\ \bibnamefont {Itano}},\ and\ \bibinfo
  {author} {\bibfnamefont {J.~J.}\ \bibnamefont {Bollinger}},\ }\bibfield
  {title} {\bibinfo {title} {Optimized dynamical decoupling in a model quantum
  memory},\ }\href {https://doi.org/10.1038/nature07951} {\bibfield  {journal}
  {\bibinfo  {journal} {Nature}\ }\textbf {\bibinfo {volume} {458}},\ \bibinfo
  {pages} {996} (\bibinfo {year} {2009}{\natexlab{b}})}\BibitemShut {NoStop}%
\bibitem [{\citenamefont {Sagi}\ \emph {et~al.}(2010)\citenamefont {Sagi},
  \citenamefont {Almog},\ and\ \citenamefont {Davidson}}]{Sagi:2010aa}%
  \BibitemOpen
  \bibfield  {author} {\bibinfo {author} {\bibfnamefont {Y.}~\bibnamefont
  {Sagi}}, \bibinfo {author} {\bibfnamefont {I.}~\bibnamefont {Almog}},\ and\
  \bibinfo {author} {\bibfnamefont {N.}~\bibnamefont {Davidson}},\ }\bibfield
  {title} {\bibinfo {title} {Process tomography of dynamical decoupling in a
  dense cold atomic ensemble},\ }\href
  {https://doi.org/10.1103/PhysRevLett.105.053201} {\bibfield  {journal}
  {\bibinfo  {journal} {Physical Review Letters}\ }\textbf {\bibinfo {volume}
  {105}},\ \bibinfo {pages} {053201} (\bibinfo {year} {2010})}\BibitemShut
  {NoStop}%
\bibitem [{\citenamefont {Naydenov}\ \emph {et~al.}(2011)\citenamefont
  {Naydenov}, \citenamefont {Dolde}, \citenamefont {Hall}, \citenamefont
  {Shin}, \citenamefont {Fedder}, \citenamefont {Hollenberg}, \citenamefont
  {Jelezko},\ and\ \citenamefont {Wrachtrup}}]{naydenov2011dynamical}%
  \BibitemOpen
  \bibfield  {author} {\bibinfo {author} {\bibfnamefont {B.}~\bibnamefont
  {Naydenov}}, \bibinfo {author} {\bibfnamefont {F.}~\bibnamefont {Dolde}},
  \bibinfo {author} {\bibfnamefont {L.~T.}\ \bibnamefont {Hall}}, \bibinfo
  {author} {\bibfnamefont {C.}~\bibnamefont {Shin}}, \bibinfo {author}
  {\bibfnamefont {H.}~\bibnamefont {Fedder}}, \bibinfo {author} {\bibfnamefont
  {L.~C.}\ \bibnamefont {Hollenberg}}, \bibinfo {author} {\bibfnamefont
  {F.}~\bibnamefont {Jelezko}},\ and\ \bibinfo {author} {\bibfnamefont
  {J.}~\bibnamefont {Wrachtrup}},\ }\bibfield  {title} {\bibinfo {title}
  {Dynamical decoupling of a single-electron spin at room temperature},\
  }\href@noop {} {\bibfield  {journal} {\bibinfo  {journal} {Physical Review
  B}\ }\textbf {\bibinfo {volume} {83}},\ \bibinfo {pages} {081201} (\bibinfo
  {year} {2011})}\BibitemShut {NoStop}%
\bibitem [{\citenamefont {van~der Sar}\ \emph {et~al.}(2012)\citenamefont
  {van~der Sar}, \citenamefont {Wang}, \citenamefont {Blok}, \citenamefont
  {Bernien}, \citenamefont {Taminiau}, \citenamefont {Toyli}, \citenamefont
  {Lidar}, \citenamefont {Awschalom}, \citenamefont {Hanson},\ and\
  \citenamefont {Dobrovitski}}]{Sar:2012km}%
  \BibitemOpen
  \bibfield  {author} {\bibinfo {author} {\bibfnamefont {T.}~\bibnamefont
  {van~der Sar}}, \bibinfo {author} {\bibfnamefont {Z.~H.}\ \bibnamefont
  {Wang}}, \bibinfo {author} {\bibfnamefont {M.~S.}\ \bibnamefont {Blok}},
  \bibinfo {author} {\bibfnamefont {H.}~\bibnamefont {Bernien}}, \bibinfo
  {author} {\bibfnamefont {T.~H.}\ \bibnamefont {Taminiau}}, \bibinfo {author}
  {\bibfnamefont {D.~M.}\ \bibnamefont {Toyli}}, \bibinfo {author}
  {\bibfnamefont {D.~A.}\ \bibnamefont {Lidar}}, \bibinfo {author}
  {\bibfnamefont {D.~D.}\ \bibnamefont {Awschalom}}, \bibinfo {author}
  {\bibfnamefont {R.}~\bibnamefont {Hanson}},\ and\ \bibinfo {author}
  {\bibfnamefont {V.~V.}\ \bibnamefont {Dobrovitski}},\ }\bibfield  {title}
  {\bibinfo {title} {Decoherence-protected quantum gates for a hybrid
  solid-state spin register},\ }\href {https://doi.org/10.1038/nature10900}
  {\bibfield  {journal} {\bibinfo  {journal} {Nature}\ }\textbf {\bibinfo
  {volume} {484}},\ \bibinfo {pages} {82} (\bibinfo {year} {2012})}\BibitemShut
  {NoStop}%
\bibitem [{\citenamefont {Souza}\ \emph {et~al.}(2012)\citenamefont {Souza},
  \citenamefont {{\'A}lvarez},\ and\ \citenamefont {Suter}}]{souza2012robust}%
  \BibitemOpen
  \bibfield  {author} {\bibinfo {author} {\bibfnamefont {A.~M.}\ \bibnamefont
  {Souza}}, \bibinfo {author} {\bibfnamefont {G.~A.}\ \bibnamefont
  {{\'A}lvarez}},\ and\ \bibinfo {author} {\bibfnamefont {D.}~\bibnamefont
  {Suter}},\ }\bibfield  {title} {\bibinfo {title} {Robust dynamical
  decoupling},\ }\href@noop {} {\bibfield  {journal} {\bibinfo  {journal}
  {Philosophical Transactions of the Royal Society A: Mathematical, Physical
  and Engineering Sciences}\ }\textbf {\bibinfo {volume} {370}},\ \bibinfo
  {pages} {4748} (\bibinfo {year} {2012})}\BibitemShut {NoStop}%
\bibitem [{\citenamefont {Pokharel}\ \emph {et~al.}(2018)\citenamefont
  {Pokharel}, \citenamefont {Anand}, \citenamefont {Fortman},\ and\
  \citenamefont {Lidar}}]{pokharelDemonstrationFidelityImprovement2018}%
  \BibitemOpen
  \bibfield  {author} {\bibinfo {author} {\bibfnamefont {B.}~\bibnamefont
  {Pokharel}}, \bibinfo {author} {\bibfnamefont {N.}~\bibnamefont {Anand}},
  \bibinfo {author} {\bibfnamefont {B.}~\bibnamefont {Fortman}},\ and\ \bibinfo
  {author} {\bibfnamefont {D.~A.}\ \bibnamefont {Lidar}},\ }\bibfield  {title}
  {\bibinfo {title} {Demonstration of {{Fidelity Improvement Using Dynamical
  Decoupling}} with {{Superconducting Qubits}}},\ }\bibfield  {journal}
  {\bibinfo  {journal} {Physical Review Letters}\ }\textbf {\bibinfo {volume}
  {121}},\ \href {https://doi.org/10.1103/PhysRevLett.121.220502}
  {10.1103/PhysRevLett.121.220502} (\bibinfo {year} {2018})\BibitemShut
  {NoStop}%
\bibitem [{\citenamefont {Jurcevic}\ \emph {et~al.}(2021)\citenamefont
  {Jurcevic}, \citenamefont {{Javadi-Abhari}}, \citenamefont {Bishop},
  \citenamefont {Lauer}, \citenamefont {Bogorin}, \citenamefont {Brink},
  \citenamefont {Capelluto}, \citenamefont {G{\"u}nl{\"u}k}, \citenamefont
  {Itoko}, \citenamefont {Kanazawa}, \citenamefont {Kandala}, \citenamefont
  {Keefe}, \citenamefont {Krsulich}, \citenamefont {Landers}, \citenamefont
  {Lewandowski}, \citenamefont {McClure}, \citenamefont {Nannicini},
  \citenamefont {Narasgond}, \citenamefont {Nayfeh}, \citenamefont {Pritchett},
  \citenamefont {Rothwell}, \citenamefont {Srinivasan}, \citenamefont
  {Sundaresan}, \citenamefont {Wang}, \citenamefont {Wei}, \citenamefont
  {Wood}, \citenamefont {Yau}, \citenamefont {Zhang}, \citenamefont {Dial},
  \citenamefont {Chow},\ and\ \citenamefont
  {Gambetta}}]{jurcevicDemonstrationQuantumVolume2020}%
  \BibitemOpen
  \bibfield  {author} {\bibinfo {author} {\bibfnamefont {P.}~\bibnamefont
  {Jurcevic}}, \bibinfo {author} {\bibfnamefont {A.}~\bibnamefont
  {{Javadi-Abhari}}}, \bibinfo {author} {\bibfnamefont {L.~S.}\ \bibnamefont
  {Bishop}}, \bibinfo {author} {\bibfnamefont {I.}~\bibnamefont {Lauer}},
  \bibinfo {author} {\bibfnamefont {D.~F.}\ \bibnamefont {Bogorin}}, \bibinfo
  {author} {\bibfnamefont {M.}~\bibnamefont {Brink}}, \bibinfo {author}
  {\bibfnamefont {L.}~\bibnamefont {Capelluto}}, \bibinfo {author}
  {\bibfnamefont {O.}~\bibnamefont {G{\"u}nl{\"u}k}}, \bibinfo {author}
  {\bibfnamefont {T.}~\bibnamefont {Itoko}}, \bibinfo {author} {\bibfnamefont
  {N.}~\bibnamefont {Kanazawa}}, \bibinfo {author} {\bibfnamefont
  {A.}~\bibnamefont {Kandala}}, \bibinfo {author} {\bibfnamefont {G.~A.}\
  \bibnamefont {Keefe}}, \bibinfo {author} {\bibfnamefont {K.}~\bibnamefont
  {Krsulich}}, \bibinfo {author} {\bibfnamefont {W.}~\bibnamefont {Landers}},
  \bibinfo {author} {\bibfnamefont {E.~P.}\ \bibnamefont {Lewandowski}},
  \bibinfo {author} {\bibfnamefont {D.~T.}\ \bibnamefont {McClure}}, \bibinfo
  {author} {\bibfnamefont {G.}~\bibnamefont {Nannicini}}, \bibinfo {author}
  {\bibfnamefont {A.}~\bibnamefont {Narasgond}}, \bibinfo {author}
  {\bibfnamefont {H.~M.}\ \bibnamefont {Nayfeh}}, \bibinfo {author}
  {\bibfnamefont {E.}~\bibnamefont {Pritchett}}, \bibinfo {author}
  {\bibfnamefont {M.~B.}\ \bibnamefont {Rothwell}}, \bibinfo {author}
  {\bibfnamefont {S.}~\bibnamefont {Srinivasan}}, \bibinfo {author}
  {\bibfnamefont {N.}~\bibnamefont {Sundaresan}}, \bibinfo {author}
  {\bibfnamefont {C.}~\bibnamefont {Wang}}, \bibinfo {author} {\bibfnamefont
  {K.~X.}\ \bibnamefont {Wei}}, \bibinfo {author} {\bibfnamefont {C.~J.}\
  \bibnamefont {Wood}}, \bibinfo {author} {\bibfnamefont {J.-B.}\ \bibnamefont
  {Yau}}, \bibinfo {author} {\bibfnamefont {E.~J.}\ \bibnamefont {Zhang}},
  \bibinfo {author} {\bibfnamefont {O.~E.}\ \bibnamefont {Dial}}, \bibinfo
  {author} {\bibfnamefont {J.~M.}\ \bibnamefont {Chow}},\ and\ \bibinfo
  {author} {\bibfnamefont {J.~M.}\ \bibnamefont {Gambetta}},\ }\bibfield
  {title} {\bibinfo {title} {Demonstration of quantum volume 64 on a
  superconducting quantum computing system},\ }\href
  {https://iopscience.iop.org/article/10.1088/2058-9565/abe519} {\bibfield
  {journal} {\bibinfo  {journal} {Quantum Sci. Technol.}\ }\textbf {\bibinfo
  {volume} {6}},\ \bibinfo {pages} {025020} (\bibinfo {year}
  {2021})}\BibitemShut {NoStop}%
\bibitem [{\citenamefont {Aharony}\ \emph {et~al.}(2021)\citenamefont
  {Aharony}, \citenamefont {Akerman}, \citenamefont {Ozeri}, \citenamefont
  {Perez}, \citenamefont {Savoray},\ and\ \citenamefont
  {Shaniv}}]{Aharony:2021aa}%
  \BibitemOpen
  \bibfield  {author} {\bibinfo {author} {\bibfnamefont {S.}~\bibnamefont
  {Aharony}}, \bibinfo {author} {\bibfnamefont {N.}~\bibnamefont {Akerman}},
  \bibinfo {author} {\bibfnamefont {R.}~\bibnamefont {Ozeri}}, \bibinfo
  {author} {\bibfnamefont {G.}~\bibnamefont {Perez}}, \bibinfo {author}
  {\bibfnamefont {I.}~\bibnamefont {Savoray}},\ and\ \bibinfo {author}
  {\bibfnamefont {R.}~\bibnamefont {Shaniv}},\ }\bibfield  {title} {\bibinfo
  {title} {Constraining rapidly oscillating scalar dark matter using dynamic
  decoupling},\ }\href {https://doi.org/10.1103/PhysRevD.103.075017} {\bibfield
   {journal} {\bibinfo  {journal} {Physical Review D}\ }\textbf {\bibinfo
  {volume} {103}},\ \bibinfo {pages} {075017} (\bibinfo {year}
  {2021})}\BibitemShut {NoStop}%
\bibitem [{\citenamefont {Tripathi}\ \emph {et~al.}(2022)\citenamefont
  {Tripathi}, \citenamefont {Chen}, \citenamefont {Khezri}, \citenamefont
  {Yip}, \citenamefont {{Levenson-Falk}},\ and\ \citenamefont
  {Lidar}}]{tripathiSuppressionCrosstalkSuperconducting2022}%
  \BibitemOpen
  \bibfield  {author} {\bibinfo {author} {\bibfnamefont {V.}~\bibnamefont
  {Tripathi}}, \bibinfo {author} {\bibfnamefont {H.}~\bibnamefont {Chen}},
  \bibinfo {author} {\bibfnamefont {M.}~\bibnamefont {Khezri}}, \bibinfo
  {author} {\bibfnamefont {K.-W.}\ \bibnamefont {Yip}}, \bibinfo {author}
  {\bibfnamefont {E.}~\bibnamefont {{Levenson-Falk}}},\ and\ \bibinfo {author}
  {\bibfnamefont {D.~A.}\ \bibnamefont {Lidar}},\ }\bibfield  {title} {\bibinfo
  {title} {Suppression of {{Crosstalk}} in {{Superconducting Qubits Using
  Dynamical Decoupling}}},\ }\href
  {https://doi.org/10.1103/PhysRevApplied.18.024068} {\bibfield  {journal}
  {\bibinfo  {journal} {Physical Review Applied}\ }\textbf {\bibinfo {volume}
  {18}},\ \bibinfo {pages} {024068} (\bibinfo {year} {2022})}\BibitemShut
  {NoStop}%
\bibitem [{\citenamefont {Ezzell}\ \emph {et~al.}(2023)\citenamefont {Ezzell},
  \citenamefont {Pokharel}, \citenamefont {Tewala}, \citenamefont {Quiroz},\
  and\ \citenamefont {Lidar}}]{ezzellDynamicalDecouplingSuperconducting2022}%
  \BibitemOpen
  \bibfield  {author} {\bibinfo {author} {\bibfnamefont {N.}~\bibnamefont
  {Ezzell}}, \bibinfo {author} {\bibfnamefont {B.}~\bibnamefont {Pokharel}},
  \bibinfo {author} {\bibfnamefont {L.}~\bibnamefont {Tewala}}, \bibinfo
  {author} {\bibfnamefont {G.}~\bibnamefont {Quiroz}},\ and\ \bibinfo {author}
  {\bibfnamefont {D.~A.}\ \bibnamefont {Lidar}},\ }\bibfield  {title} {\bibinfo
  {title} {Dynamical decoupling for superconducting qubits: A performance
  survey},\ }\href {https://doi.org/10.1103/PhysRevApplied.20.064027}
  {\bibfield  {journal} {\bibinfo  {journal} {Physical Review Applied}\
  }\textbf {\bibinfo {volume} {20}},\ \bibinfo {pages} {064027} (\bibinfo
  {year} {2023})}\BibitemShut {NoStop}%
\bibitem [{\citenamefont {Zhou}\ \emph {et~al.}(2023)\citenamefont {Zhou},
  \citenamefont {Sitler}, \citenamefont {Oda}, \citenamefont {Schultz},\ and\
  \citenamefont {Quiroz}}]{zhouQuantumCrosstalkRobust2022}%
  \BibitemOpen
  \bibfield  {author} {\bibinfo {author} {\bibfnamefont {Z.}~\bibnamefont
  {Zhou}}, \bibinfo {author} {\bibfnamefont {R.}~\bibnamefont {Sitler}},
  \bibinfo {author} {\bibfnamefont {Y.}~\bibnamefont {Oda}}, \bibinfo {author}
  {\bibfnamefont {K.}~\bibnamefont {Schultz}},\ and\ \bibinfo {author}
  {\bibfnamefont {G.}~\bibnamefont {Quiroz}},\ }\bibfield  {title} {\bibinfo
  {title} {Quantum crosstalk robust quantum control},\ }\href
  {https://doi.org/10.1103/PhysRevLett.131.210802} {\bibfield  {journal}
  {\bibinfo  {journal} {Physical Review Letters}\ }\textbf {\bibinfo {volume}
  {131}},\ \bibinfo {pages} {210802} (\bibinfo {year} {2023})}\BibitemShut
  {NoStop}%
\bibitem [{\citenamefont {Boulant}\ \emph {et~al.}(2005)\citenamefont
  {Boulant}, \citenamefont {Viola}, \citenamefont {Fortunato},\ and\
  \citenamefont {Cory}}]{PhysRevLett.94.130501}%
  \BibitemOpen
  \bibfield  {author} {\bibinfo {author} {\bibfnamefont {N.}~\bibnamefont
  {Boulant}}, \bibinfo {author} {\bibfnamefont {L.}~\bibnamefont {Viola}},
  \bibinfo {author} {\bibfnamefont {E.~M.}\ \bibnamefont {Fortunato}},\ and\
  \bibinfo {author} {\bibfnamefont {D.~G.}\ \bibnamefont {Cory}},\ }\bibfield
  {title} {\bibinfo {title} {Experimental implementation of a concatenated
  quantum error-correcting code},\ }\href
  {https://doi.org/10.1103/PhysRevLett.94.130501} {\bibfield  {journal}
  {\bibinfo  {journal} {Phys. Rev. Lett.}\ }\textbf {\bibinfo {volume} {94}},\
  \bibinfo {pages} {130501} (\bibinfo {year} {2005})}\BibitemShut {NoStop}%
\bibitem [{\citenamefont {Harper}\ and\ \citenamefont
  {Flammia}(2019)}]{harper2019ftgates}%
  \BibitemOpen
  \bibfield  {author} {\bibinfo {author} {\bibfnamefont {R.}~\bibnamefont
  {Harper}}\ and\ \bibinfo {author} {\bibfnamefont {S.~T.}\ \bibnamefont
  {Flammia}},\ }\bibfield  {title} {\bibinfo {title} {Fault-tolerant logical
  gates in the ibm quantum experience},\ }\href
  {https://doi.org/10.1103/PhysRevLett.122.080504} {\bibfield  {journal}
  {\bibinfo  {journal} {Phys. Rev. Lett.}\ }\textbf {\bibinfo {volume} {122}},\
  \bibinfo {pages} {080504} (\bibinfo {year} {2019})}\BibitemShut {NoStop}%
\bibitem [{\citenamefont {Andersen}\ \emph {et~al.}(2020)\citenamefont
  {Andersen}, \citenamefont {Remm}, \citenamefont {Lazar}, \citenamefont
  {Krinner}, \citenamefont {Lacroix}, \citenamefont {Norris}, \citenamefont
  {Gabureac}, \citenamefont {Eichler},\ and\ \citenamefont
  {Wallraff}}]{andersen2020repeated}%
  \BibitemOpen
  \bibfield  {author} {\bibinfo {author} {\bibfnamefont {C.~K.}\ \bibnamefont
  {Andersen}}, \bibinfo {author} {\bibfnamefont {A.}~\bibnamefont {Remm}},
  \bibinfo {author} {\bibfnamefont {S.}~\bibnamefont {Lazar}}, \bibinfo
  {author} {\bibfnamefont {S.}~\bibnamefont {Krinner}}, \bibinfo {author}
  {\bibfnamefont {N.}~\bibnamefont {Lacroix}}, \bibinfo {author} {\bibfnamefont
  {G.~J.}\ \bibnamefont {Norris}}, \bibinfo {author} {\bibfnamefont
  {M.}~\bibnamefont {Gabureac}}, \bibinfo {author} {\bibfnamefont
  {C.}~\bibnamefont {Eichler}},\ and\ \bibinfo {author} {\bibfnamefont
  {A.}~\bibnamefont {Wallraff}},\ }\bibfield  {title} {\bibinfo {title}
  {Repeated quantum error detection in a surface code},\ }\href@noop {}
  {\bibfield  {journal} {\bibinfo  {journal} {Nature Physics}\ }\textbf
  {\bibinfo {volume} {16}},\ \bibinfo {pages} {875} (\bibinfo {year}
  {2020})}\BibitemShut {NoStop}%
\bibitem [{\citenamefont {Chen}\ \emph {et~al.}(2021)\citenamefont {Chen},
  \citenamefont {Satzinger}, \citenamefont {Atalaya}, \citenamefont {Korotkov},
  \citenamefont {Dunsworth}, \citenamefont {Sank}, \citenamefont {Quintana},
  \citenamefont {McEwen}, \citenamefont {Barends}, \citenamefont {Klimov} \emph
  {et~al.}}]{chen2021qec}%
  \BibitemOpen
  \bibfield  {author} {\bibinfo {author} {\bibfnamefont {Z.}~\bibnamefont
  {Chen}}, \bibinfo {author} {\bibfnamefont {K.~J.}\ \bibnamefont {Satzinger}},
  \bibinfo {author} {\bibfnamefont {J.}~\bibnamefont {Atalaya}}, \bibinfo
  {author} {\bibfnamefont {A.~N.}\ \bibnamefont {Korotkov}}, \bibinfo {author}
  {\bibfnamefont {A.}~\bibnamefont {Dunsworth}}, \bibinfo {author}
  {\bibfnamefont {D.}~\bibnamefont {Sank}}, \bibinfo {author} {\bibfnamefont
  {C.}~\bibnamefont {Quintana}}, \bibinfo {author} {\bibfnamefont
  {M.}~\bibnamefont {McEwen}}, \bibinfo {author} {\bibfnamefont
  {R.}~\bibnamefont {Barends}}, \bibinfo {author} {\bibfnamefont {P.~V.}\
  \bibnamefont {Klimov}}, \emph {et~al.},\ }\bibfield  {title} {\bibinfo
  {title} {Exponential suppression of bit or phase errors with cyclic error
  correction},\ }\href@noop {} {\bibfield  {journal} {\bibinfo  {journal}
  {Nature}\ }\textbf {\bibinfo {volume} {595}},\ \bibinfo {pages} {383}
  (\bibinfo {year} {2021})}\BibitemShut {NoStop}%
\bibitem [{\citenamefont {Krinner}\ \emph {et~al.}(2022)\citenamefont
  {Krinner}, \citenamefont {Lacroix}, \citenamefont {Remm}, \citenamefont
  {Di~Paolo}, \citenamefont {Genois}, \citenamefont {Leroux}, \citenamefont
  {Hellings}, \citenamefont {Lazar}, \citenamefont {Swiadek}, \citenamefont
  {Herrmann} \emph {et~al.}}]{krinner2022realizing}%
  \BibitemOpen
  \bibfield  {author} {\bibinfo {author} {\bibfnamefont {S.}~\bibnamefont
  {Krinner}}, \bibinfo {author} {\bibfnamefont {N.}~\bibnamefont {Lacroix}},
  \bibinfo {author} {\bibfnamefont {A.}~\bibnamefont {Remm}}, \bibinfo {author}
  {\bibfnamefont {A.}~\bibnamefont {Di~Paolo}}, \bibinfo {author}
  {\bibfnamefont {E.}~\bibnamefont {Genois}}, \bibinfo {author} {\bibfnamefont
  {C.}~\bibnamefont {Leroux}}, \bibinfo {author} {\bibfnamefont
  {C.}~\bibnamefont {Hellings}}, \bibinfo {author} {\bibfnamefont
  {S.}~\bibnamefont {Lazar}}, \bibinfo {author} {\bibfnamefont
  {F.}~\bibnamefont {Swiadek}}, \bibinfo {author} {\bibfnamefont
  {J.}~\bibnamefont {Herrmann}}, \emph {et~al.},\ }\bibfield  {title} {\bibinfo
  {title} {Realizing repeated quantum error correction in a distance-three
  surface code},\ }\href@noop {} {\bibfield  {journal} {\bibinfo  {journal}
  {Nature}\ }\textbf {\bibinfo {volume} {605}},\ \bibinfo {pages} {669}
  (\bibinfo {year} {2022})}\BibitemShut {NoStop}%
\bibitem [{\citenamefont {Sivak}\ \emph {et~al.}(2023)\citenamefont {Sivak},
  \citenamefont {Eickbusch}, \citenamefont {Royer}, \citenamefont {Singh},
  \citenamefont {Tsioutsios}, \citenamefont {Ganjam}, \citenamefont {Miano},
  \citenamefont {Brock}, \citenamefont {Ding}, \citenamefont {Frunzio} \emph
  {et~al.}}]{sivak2023real}%
  \BibitemOpen
  \bibfield  {author} {\bibinfo {author} {\bibfnamefont {V.}~\bibnamefont
  {Sivak}}, \bibinfo {author} {\bibfnamefont {A.}~\bibnamefont {Eickbusch}},
  \bibinfo {author} {\bibfnamefont {B.}~\bibnamefont {Royer}}, \bibinfo
  {author} {\bibfnamefont {S.}~\bibnamefont {Singh}}, \bibinfo {author}
  {\bibfnamefont {I.}~\bibnamefont {Tsioutsios}}, \bibinfo {author}
  {\bibfnamefont {S.}~\bibnamefont {Ganjam}}, \bibinfo {author} {\bibfnamefont
  {A.}~\bibnamefont {Miano}}, \bibinfo {author} {\bibfnamefont
  {B.}~\bibnamefont {Brock}}, \bibinfo {author} {\bibfnamefont
  {A.}~\bibnamefont {Ding}}, \bibinfo {author} {\bibfnamefont {L.}~\bibnamefont
  {Frunzio}}, \emph {et~al.},\ }\bibfield  {title} {\bibinfo {title} {Real-time
  quantum error correction beyond break-even},\ }\href@noop {} {\bibfield
  {journal} {\bibinfo  {journal} {Nature}\ }\textbf {\bibinfo {volume} {616}},\
  \bibinfo {pages} {50} (\bibinfo {year} {2023})}\BibitemShut {NoStop}%
\bibitem [{\citenamefont {Miao}\ \emph {et~al.}(2022)\citenamefont {Miao},
  \citenamefont {McEwen}, \citenamefont {Atalaya}, \citenamefont {Kafri},
  \citenamefont {Pryadko}, \citenamefont {Bengtsson}, \citenamefont {Opremcak},
  \citenamefont {Satzinger}, \citenamefont {Chen}, \citenamefont {Klimov} \emph
  {et~al.}}]{miao2022overcoming}%
  \BibitemOpen
  \bibfield  {author} {\bibinfo {author} {\bibfnamefont {K.~C.}\ \bibnamefont
  {Miao}}, \bibinfo {author} {\bibfnamefont {M.}~\bibnamefont {McEwen}},
  \bibinfo {author} {\bibfnamefont {J.}~\bibnamefont {Atalaya}}, \bibinfo
  {author} {\bibfnamefont {D.}~\bibnamefont {Kafri}}, \bibinfo {author}
  {\bibfnamefont {L.~P.}\ \bibnamefont {Pryadko}}, \bibinfo {author}
  {\bibfnamefont {A.}~\bibnamefont {Bengtsson}}, \bibinfo {author}
  {\bibfnamefont {A.}~\bibnamefont {Opremcak}}, \bibinfo {author}
  {\bibfnamefont {K.~J.}\ \bibnamefont {Satzinger}}, \bibinfo {author}
  {\bibfnamefont {Z.}~\bibnamefont {Chen}}, \bibinfo {author} {\bibfnamefont
  {P.~V.}\ \bibnamefont {Klimov}}, \emph {et~al.},\ }\bibfield  {title}
  {\bibinfo {title} {Overcoming leakage in scalable quantum error correction},\
  }\href@noop {} {\bibfield  {journal} {\bibinfo  {journal} {arXiv preprint
  arXiv:2211.04728}\ } (\bibinfo {year} {2022})}\BibitemShut {NoStop}%
\bibitem [{\citenamefont {{Google Quantum AI}}(2023)}]{ai2023suppressing}%
  \BibitemOpen
  \bibfield  {author} {\bibinfo {author} {\bibnamefont {{Google Quantum AI}}},\
  }\bibfield  {title} {\bibinfo {title} {Suppressing quantum errors by scaling
  a surface code logical qubit},\ }\href
  {https://doi.org/10.1038/s41586-022-05434-1} {\bibfield  {journal} {\bibinfo
  {journal} {Nature}\ }\textbf {\bibinfo {volume} {614}},\ \bibinfo {pages}
  {676} (\bibinfo {year} {2023})}\BibitemShut {NoStop}%
\bibitem [{\citenamefont {Urbanek}\ \emph {et~al.}(2020)\citenamefont
  {Urbanek}, \citenamefont {Nachman},\ and\ \citenamefont
  {de~Jong}}]{urbanek2020vqe}%
  \BibitemOpen
  \bibfield  {author} {\bibinfo {author} {\bibfnamefont {M.}~\bibnamefont
  {Urbanek}}, \bibinfo {author} {\bibfnamefont {B.}~\bibnamefont {Nachman}},\
  and\ \bibinfo {author} {\bibfnamefont {W.~A.}\ \bibnamefont {de~Jong}},\
  }\bibfield  {title} {\bibinfo {title} {Error detection on quantum computers
  improving the accuracy of chemical calculations},\ }\href
  {https://doi.org/10.1103/PhysRevA.102.022427} {\bibfield  {journal} {\bibinfo
   {journal} {Phys. Rev. A}\ }\textbf {\bibinfo {volume} {102}},\ \bibinfo
  {pages} {022427} (\bibinfo {year} {2020})}\BibitemShut {NoStop}%
\bibitem [{\citenamefont {Postler}\ \emph {et~al.}(2023)\citenamefont
  {Postler}, \citenamefont {Butt}, \citenamefont {Pogorelov}, \citenamefont
  {Marciniak}, \citenamefont {Heu{\ss}en}, \citenamefont {Blatt}, \citenamefont
  {Schindler}, \citenamefont {Rispler}, \citenamefont {M{\"u}ller},\ and\
  \citenamefont {Monz}}]{postler2023demonstration}%
  \BibitemOpen
  \bibfield  {author} {\bibinfo {author} {\bibfnamefont {L.}~\bibnamefont
  {Postler}}, \bibinfo {author} {\bibfnamefont {F.}~\bibnamefont {Butt}},
  \bibinfo {author} {\bibfnamefont {I.}~\bibnamefont {Pogorelov}}, \bibinfo
  {author} {\bibfnamefont {C.~D.}\ \bibnamefont {Marciniak}}, \bibinfo {author}
  {\bibfnamefont {S.}~\bibnamefont {Heu{\ss}en}}, \bibinfo {author}
  {\bibfnamefont {R.}~\bibnamefont {Blatt}}, \bibinfo {author} {\bibfnamefont
  {P.}~\bibnamefont {Schindler}}, \bibinfo {author} {\bibfnamefont
  {M.}~\bibnamefont {Rispler}}, \bibinfo {author} {\bibfnamefont
  {M.}~\bibnamefont {M{\"u}ller}},\ and\ \bibinfo {author} {\bibfnamefont
  {T.}~\bibnamefont {Monz}},\ }\bibfield  {title} {\bibinfo {title}
  {Demonstration of fault-tolerant steane quantum error correction},\
  }\href@noop {} {\bibfield  {journal} {\bibinfo  {journal} {arXiv preprint
  arXiv:2312.09745}\ } (\bibinfo {year} {2023})}\BibitemShut {NoStop}%
\bibitem [{\citenamefont {Bluvstein}\ \emph {et~al.}(2023)\citenamefont
  {Bluvstein}, \citenamefont {Evered}, \citenamefont {Geim}, \citenamefont
  {Li}, \citenamefont {Zhou}, \citenamefont {Manovitz}, \citenamefont {Ebadi},
  \citenamefont {Cain}, \citenamefont {Kalinowski}, \citenamefont {Hangleiter}
  \emph {et~al.}}]{bluvstein2023logical}%
  \BibitemOpen
  \bibfield  {author} {\bibinfo {author} {\bibfnamefont {D.}~\bibnamefont
  {Bluvstein}}, \bibinfo {author} {\bibfnamefont {S.~J.}\ \bibnamefont
  {Evered}}, \bibinfo {author} {\bibfnamefont {A.~A.}\ \bibnamefont {Geim}},
  \bibinfo {author} {\bibfnamefont {S.~H.}\ \bibnamefont {Li}}, \bibinfo
  {author} {\bibfnamefont {H.}~\bibnamefont {Zhou}}, \bibinfo {author}
  {\bibfnamefont {T.}~\bibnamefont {Manovitz}}, \bibinfo {author}
  {\bibfnamefont {S.}~\bibnamefont {Ebadi}}, \bibinfo {author} {\bibfnamefont
  {M.}~\bibnamefont {Cain}}, \bibinfo {author} {\bibfnamefont {M.}~\bibnamefont
  {Kalinowski}}, \bibinfo {author} {\bibfnamefont {D.}~\bibnamefont
  {Hangleiter}}, \emph {et~al.},\ }\bibfield  {title} {\bibinfo {title}
  {Logical quantum processor based on reconfigurable atom arrays},\ }\href@noop
  {} {\bibfield  {journal} {\bibinfo  {journal} {Nature}\ ,\ \bibinfo {pages}
  {1}} (\bibinfo {year} {2023})}\BibitemShut {NoStop}%
\bibitem [{\citenamefont {Pokharel}\ and\ \citenamefont
  {Lidar}(2022)}]{Pokharel:better-than-classical-Grover}%
  \BibitemOpen
  \bibfield  {author} {\bibinfo {author} {\bibfnamefont {B.}~\bibnamefont
  {Pokharel}}\ and\ \bibinfo {author} {\bibfnamefont {D.}~\bibnamefont
  {Lidar}},\ }\href {https://arxiv.org/abs/2211.04543} {\bibinfo {title}
  {{Better-than-classical Grover search via quantum error detection and
  suppression}}} (\bibinfo {year} {2022}),\ \Eprint
  {https://arxiv.org/abs/2211.04543} {arXiv:2211.04543 [quant-ph]} \BibitemShut
  {NoStop}%
\bibitem [{\citenamefont {Kwiat}\ \emph {et~al.}(2000)\citenamefont {Kwiat},
  \citenamefont {Berglund}, \citenamefont {Altepeter},\ and\ \citenamefont
  {White}}]{kwiat2000experimental}%
  \BibitemOpen
  \bibfield  {author} {\bibinfo {author} {\bibfnamefont {P.~G.}\ \bibnamefont
  {Kwiat}}, \bibinfo {author} {\bibfnamefont {A.~J.}\ \bibnamefont {Berglund}},
  \bibinfo {author} {\bibfnamefont {J.~B.}\ \bibnamefont {Altepeter}},\ and\
  \bibinfo {author} {\bibfnamefont {A.~G.}\ \bibnamefont {White}},\ }\bibfield
  {title} {\bibinfo {title} {Experimental verification of decoherence-free
  subspaces},\ }\href@noop {} {\bibfield  {journal} {\bibinfo  {journal}
  {Science}\ }\textbf {\bibinfo {volume} {290}},\ \bibinfo {pages} {498}
  (\bibinfo {year} {2000})}\BibitemShut {NoStop}%
\bibitem [{\citenamefont {Viola}\ \emph {et~al.}(2001)\citenamefont {Viola},
  \citenamefont {Fortunato}, \citenamefont {Pravia}, \citenamefont {Knill},
  \citenamefont {Laflamme},\ and\ \citenamefont
  {Cory}}]{viola2001experimental}%
  \BibitemOpen
  \bibfield  {author} {\bibinfo {author} {\bibfnamefont {L.}~\bibnamefont
  {Viola}}, \bibinfo {author} {\bibfnamefont {E.~M.}\ \bibnamefont
  {Fortunato}}, \bibinfo {author} {\bibfnamefont {M.~A.}\ \bibnamefont
  {Pravia}}, \bibinfo {author} {\bibfnamefont {E.}~\bibnamefont {Knill}},
  \bibinfo {author} {\bibfnamefont {R.}~\bibnamefont {Laflamme}},\ and\
  \bibinfo {author} {\bibfnamefont {D.~G.}\ \bibnamefont {Cory}},\ }\bibfield
  {title} {\bibinfo {title} {Experimental realization of noiseless subsystems
  for quantum information processing},\ }\href
  {http://science.sciencemag.org/content/293/5537/2059} {\bibfield  {journal}
  {\bibinfo  {journal} {Science}\ }\textbf {\bibinfo {volume} {293}},\ \bibinfo
  {pages} {2059} (\bibinfo {year} {2001})}\BibitemShut {NoStop}%
\bibitem [{\citenamefont {Ollerenshaw}\ \emph {et~al.}(2003)\citenamefont
  {Ollerenshaw}, \citenamefont {Lidar},\ and\ \citenamefont
  {Kay}}]{PhysRevLett.91.217904}%
  \BibitemOpen
  \bibfield  {author} {\bibinfo {author} {\bibfnamefont {J.~E.}\ \bibnamefont
  {Ollerenshaw}}, \bibinfo {author} {\bibfnamefont {D.~A.}\ \bibnamefont
  {Lidar}},\ and\ \bibinfo {author} {\bibfnamefont {L.~E.}\ \bibnamefont
  {Kay}},\ }\bibfield  {title} {\bibinfo {title} {Magnetic resonance
  realization of decoherence-free quantum computation},\ }\href
  {https://doi.org/10.1103/PhysRevLett.91.217904} {\bibfield  {journal}
  {\bibinfo  {journal} {Phys. Rev. Lett.}\ }\textbf {\bibinfo {volume} {91}},\
  \bibinfo {pages} {217904} (\bibinfo {year} {2003})}\BibitemShut {NoStop}%
\bibitem [{\citenamefont {Mohseni}\ \emph {et~al.}(2003)\citenamefont
  {Mohseni}, \citenamefont {Lundeen}, \citenamefont {Resch},\ and\
  \citenamefont {Steinberg}}]{mohseni2003dfs-exp}%
  \BibitemOpen
  \bibfield  {author} {\bibinfo {author} {\bibfnamefont {M.}~\bibnamefont
  {Mohseni}}, \bibinfo {author} {\bibfnamefont {J.~S.}\ \bibnamefont
  {Lundeen}}, \bibinfo {author} {\bibfnamefont {K.~J.}\ \bibnamefont {Resch}},\
  and\ \bibinfo {author} {\bibfnamefont {A.~M.}\ \bibnamefont {Steinberg}},\
  }\bibfield  {title} {\bibinfo {title} {Experimental application of
  decoherence-free subspaces in an optical quantum-computing algorithm},\
  }\href {https://doi.org/10.1103/PhysRevLett.91.187903} {\bibfield  {journal}
  {\bibinfo  {journal} {Phys. Rev. Lett.}\ }\textbf {\bibinfo {volume} {91}},\
  \bibinfo {pages} {187903} (\bibinfo {year} {2003})}\BibitemShut {NoStop}%
\bibitem [{\citenamefont {Fortunato}\ \emph {et~al.}(2003)\citenamefont
  {Fortunato}, \citenamefont {Viola}, \citenamefont {Pravia}, \citenamefont
  {Knill}, \citenamefont {Laflamme}, \citenamefont {Havel},\ and\ \citenamefont
  {Cory}}]{fortunato2003ns-exp}%
  \BibitemOpen
  \bibfield  {author} {\bibinfo {author} {\bibfnamefont {E.~M.}\ \bibnamefont
  {Fortunato}}, \bibinfo {author} {\bibfnamefont {L.}~\bibnamefont {Viola}},
  \bibinfo {author} {\bibfnamefont {M.~A.}\ \bibnamefont {Pravia}}, \bibinfo
  {author} {\bibfnamefont {E.}~\bibnamefont {Knill}}, \bibinfo {author}
  {\bibfnamefont {R.}~\bibnamefont {Laflamme}}, \bibinfo {author}
  {\bibfnamefont {T.~F.}\ \bibnamefont {Havel}},\ and\ \bibinfo {author}
  {\bibfnamefont {D.~G.}\ \bibnamefont {Cory}},\ }\bibfield  {title} {\bibinfo
  {title} {Exploring noiseless subsystems via nuclear magnetic resonance},\
  }\href {https://doi.org/10.1103/PhysRevA.67.062303} {\bibfield  {journal}
  {\bibinfo  {journal} {Phys. Rev. A}\ }\textbf {\bibinfo {volume} {67}},\
  \bibinfo {pages} {062303} (\bibinfo {year} {2003})}\BibitemShut {NoStop}%
\bibitem [{\citenamefont {Altepeter}\ \emph {et~al.}(2004)\citenamefont
  {Altepeter}, \citenamefont {Hadley}, \citenamefont {Wendelken}, \citenamefont
  {Berglund},\ and\ \citenamefont {Kwiat}}]{altepeter2004dfs-exp}%
  \BibitemOpen
  \bibfield  {author} {\bibinfo {author} {\bibfnamefont {J.~B.}\ \bibnamefont
  {Altepeter}}, \bibinfo {author} {\bibfnamefont {P.~G.}\ \bibnamefont
  {Hadley}}, \bibinfo {author} {\bibfnamefont {S.~M.}\ \bibnamefont
  {Wendelken}}, \bibinfo {author} {\bibfnamefont {A.~J.}\ \bibnamefont
  {Berglund}},\ and\ \bibinfo {author} {\bibfnamefont {P.~G.}\ \bibnamefont
  {Kwiat}},\ }\bibfield  {title} {\bibinfo {title} {Experimental investigation
  of a two-qubit decoherence-free subspace},\ }\href
  {https://doi.org/10.1103/PhysRevLett.92.147901} {\bibfield  {journal}
  {\bibinfo  {journal} {Phys. Rev. Lett.}\ }\textbf {\bibinfo {volume} {92}},\
  \bibinfo {pages} {147901} (\bibinfo {year} {2004})}\BibitemShut {NoStop}%
\bibitem [{\citenamefont {Pushin}\ \emph {et~al.}(2011)\citenamefont {Pushin},
  \citenamefont {Huber}, \citenamefont {Arif},\ and\ \citenamefont
  {Cory}}]{pushin2011dfs-exp}%
  \BibitemOpen
  \bibfield  {author} {\bibinfo {author} {\bibfnamefont {D.~A.}\ \bibnamefont
  {Pushin}}, \bibinfo {author} {\bibfnamefont {M.~G.}\ \bibnamefont {Huber}},
  \bibinfo {author} {\bibfnamefont {M.}~\bibnamefont {Arif}},\ and\ \bibinfo
  {author} {\bibfnamefont {D.~G.}\ \bibnamefont {Cory}},\ }\bibfield  {title}
  {\bibinfo {title} {Experimental realization of decoherence-free subspace in
  neutron interferometry},\ }\href
  {https://doi.org/10.1103/PhysRevLett.107.150401} {\bibfield  {journal}
  {\bibinfo  {journal} {Phys. Rev. Lett.}\ }\textbf {\bibinfo {volume} {107}},\
  \bibinfo {pages} {150401} (\bibinfo {year} {2011})}\BibitemShut {NoStop}%
\bibitem [{\citenamefont {Preskill}(2018)}]{Preskill2018}%
  \BibitemOpen
  \bibfield  {author} {\bibinfo {author} {\bibfnamefont {J.}~\bibnamefont
  {Preskill}},\ }\bibfield  {title} {\bibinfo {title} {Quantum {C}omputing in
  the {NISQ} era and beyond},\ }\href
  {https://doi.org/10.22331/q-2018-08-06-79} {\bibfield  {journal} {\bibinfo
  {journal} {{Quantum}}\ }\textbf {\bibinfo {volume} {2}},\ \bibinfo {pages}
  {79} (\bibinfo {year} {2018})}\BibitemShut {NoStop}%
\bibitem [{\citenamefont {Wu}\ and\ \citenamefont
  {Lidar}(2002)}]{wuCreatingDecoherenceFreeSubspaces2002}%
  \BibitemOpen
  \bibfield  {author} {\bibinfo {author} {\bibfnamefont {L.~A.}\ \bibnamefont
  {Wu}}\ and\ \bibinfo {author} {\bibfnamefont {D.~A.}\ \bibnamefont {Lidar}},\
  }\bibfield  {title} {\bibinfo {title} {Creating decoherence-free subspaces
  using strong and fast pulses},\ }\href
  {http://link.aps.org/doi/10.1103/PhysRevLett.88.207902} {\bibfield  {journal}
  {\bibinfo  {journal} {Phys. Rev. Lett.}\ }\textbf {\bibinfo {volume} {88}},\
  \bibinfo {pages} {207902} (\bibinfo {year} {2002})}\BibitemShut {NoStop}%
\bibitem [{\citenamefont {Wu}\ \emph {et~al.}(2002)\citenamefont {Wu},
  \citenamefont {Byrd},\ and\ \citenamefont
  {Lidar}}]{wuEfficientUniversalLeakage2002}%
  \BibitemOpen
  \bibfield  {author} {\bibinfo {author} {\bibfnamefont {L.~A.}\ \bibnamefont
  {Wu}}, \bibinfo {author} {\bibfnamefont {M.~S.}\ \bibnamefont {Byrd}},\ and\
  \bibinfo {author} {\bibfnamefont {D.~A.}\ \bibnamefont {Lidar}},\ }\bibfield
  {title} {\bibinfo {title} {Efficient universal leakage elimination for
  physical and encoded qubits},\ }\href
  {http://link.aps.org/doi/10.1103/PhysRevLett.89.127901} {\bibfield  {journal}
  {\bibinfo  {journal} {{Phys.~Rev.~Lett.}}\ }\textbf {\bibinfo {volume}
  {89}},\ \bibinfo {pages} {127901} (\bibinfo {year} {2002})}\BibitemShut
  {NoStop}%
\bibitem [{\citenamefont {Byrd}\ and\ \citenamefont
  {Lidar}(2002)}]{Byrd:2002:047901}%
  \BibitemOpen
  \bibfield  {author} {\bibinfo {author} {\bibfnamefont {M.~S.}\ \bibnamefont
  {Byrd}}\ and\ \bibinfo {author} {\bibfnamefont {D.~A.}\ \bibnamefont
  {Lidar}},\ }\bibfield  {title} {\bibinfo {title} {Comprehensive encoding and
  decoupling solution to problems of decoherence and design in solid-state
  quantum computing},\ }\href {https://doi.org/10.1103/PhysRevLett.89.047901}
  {\bibfield  {journal} {\bibinfo  {journal} {Phys. Rev. Lett.}\ }\textbf
  {\bibinfo {volume} {89}},\ \bibinfo {pages} {047901} (\bibinfo {year}
  {2002})}\BibitemShut {NoStop}%
\bibitem [{\citenamefont {Lidar}\ and\ \citenamefont
  {Wu}(2003{\natexlab{a}})}]{LidarWu:02}%
  \BibitemOpen
  \bibfield  {author} {\bibinfo {author} {\bibfnamefont {D.~A.}\ \bibnamefont
  {Lidar}}\ and\ \bibinfo {author} {\bibfnamefont {L.~A.}\ \bibnamefont {Wu}},\
  }\bibfield  {title} {\bibinfo {title} {Encoded recoupling and decoupling: An
  alternative to quantum error-correcting codes applied to trapped-ion quantum
  computation},\ }\href {https://doi.org/10.1103/PhysRevA.67.032313} {\bibfield
   {journal} {\bibinfo  {journal} {Physical Review A}\ }\textbf {\bibinfo
  {volume} {67}},\ \bibinfo {pages} {032313} (\bibinfo {year}
  {2003}{\natexlab{a}})}\BibitemShut {NoStop}%
\bibitem [{\citenamefont {Lidar}\ and\ \citenamefont
  {Wu}(2003{\natexlab{b}})}]{lidarQuantumComputersDecoherence2003}%
  \BibitemOpen
  \bibfield  {author} {\bibinfo {author} {\bibfnamefont {D.~A.}\ \bibnamefont
  {Lidar}}\ and\ \bibinfo {author} {\bibfnamefont {L.-A.}\ \bibnamefont {Wu}},\
  }\bibfield  {title} {\bibinfo {title} {Quantum computers and decoherence:
  exorcising the demon from the machine},\ }in\ \href
  {https://doi.org/10.1117/12.488754} {\emph {\bibinfo {booktitle} {Proc. SPIE,
  Noise and Information in Nanoelectronics, Sensors, and Standards}}},\ Vol.\
  \bibinfo {volume} {5115}\ (\bibinfo {year} {2003})\ pp.\ \bibinfo {pages}
  {256--270}\BibitemShut {NoStop}%
\bibitem [{\citenamefont {Byrd}\ \emph {et~al.}(2005)\citenamefont {Byrd},
  \citenamefont {Lidar}, \citenamefont {Wu},\ and\ \citenamefont
  {Zanardi}}]{ByrdLidarWuZanardi:05}%
  \BibitemOpen
  \bibfield  {author} {\bibinfo {author} {\bibfnamefont {M.~S.}\ \bibnamefont
  {Byrd}}, \bibinfo {author} {\bibfnamefont {D.~A.}\ \bibnamefont {Lidar}},
  \bibinfo {author} {\bibfnamefont {L.-A.}\ \bibnamefont {Wu}},\ and\ \bibinfo
  {author} {\bibfnamefont {P.}~\bibnamefont {Zanardi}},\ }\bibfield  {title}
  {\bibinfo {title} {Universal leakage elimination},\ }\href
  {http://link.aps.org/doi/10.1103/PhysRevA.71.052301} {\bibfield  {journal}
  {\bibinfo  {journal} {Phys. Rev. A}\ }\textbf {\bibinfo {volume} {71}},\
  \bibinfo {pages} {052301} (\bibinfo {year} {2005})}\BibitemShut {NoStop}%
\bibitem [{\citenamefont {Fong}\ and\ \citenamefont
  {Wandzura}(2011)}]{fongUniversalQuantumComputation2011}%
  \BibitemOpen
  \bibfield  {author} {\bibinfo {author} {\bibfnamefont {B.~H.}\ \bibnamefont
  {Fong}}\ and\ \bibinfo {author} {\bibfnamefont {S.~M.}\ \bibnamefont
  {Wandzura}},\ }\bibfield  {title} {\bibinfo {title} {Universal quantum
  computation and leakage reduction in the 3-qubit decoherence free
  subsystem},\ }\href {http://arXiv.org/abs/1102.2909} {\bibfield  {journal}
  {\bibinfo  {journal} {{Quantum. Inf. Comput.}}\ }\textbf {\bibinfo {volume}
  {11}},\ \bibinfo {pages} {1003} (\bibinfo {year} {2011})}\BibitemShut
  {NoStop}%
\bibitem [{\citenamefont {West}\ and\ \citenamefont
  {Fong}(2012)}]{west2012exchange}%
  \BibitemOpen
  \bibfield  {author} {\bibinfo {author} {\bibfnamefont {J.~R.}\ \bibnamefont
  {West}}\ and\ \bibinfo {author} {\bibfnamefont {B.~H.}\ \bibnamefont
  {Fong}},\ }\bibfield  {title} {\bibinfo {title} {Exchange-only dynamical
  decoupling in the three-qubit decoherence free subsystem},\ }\href
  {https://doi.org/10.1088/1367-2630/14/8/083002} {\bibfield  {journal}
  {\bibinfo  {journal} {New Journal of Physics}\ }\textbf {\bibinfo {volume}
  {14}},\ \bibinfo {pages} {083002} (\bibinfo {year} {2012})}\BibitemShut
  {NoStop}%
\bibitem [{\citenamefont {Kempe}\ \emph {et~al.}(2001)\citenamefont {Kempe},
  \citenamefont {Bacon}, \citenamefont {Lidar},\ and\ \citenamefont
  {Whaley}}]{kempe2001theory}%
  \BibitemOpen
  \bibfield  {author} {\bibinfo {author} {\bibfnamefont {J.}~\bibnamefont
  {Kempe}}, \bibinfo {author} {\bibfnamefont {D.}~\bibnamefont {Bacon}},
  \bibinfo {author} {\bibfnamefont {D.~A.}\ \bibnamefont {Lidar}},\ and\
  \bibinfo {author} {\bibfnamefont {K.~B.}\ \bibnamefont {Whaley}},\ }\bibfield
   {title} {\bibinfo {title} {Theory of decoherence-free fault-tolerant
  universal quantum computation},\ }\href
  {https://doi.org/10.1103/PhysRevA.63.042307} {\bibfield  {journal} {\bibinfo
  {journal} {Phys. Rev. A}\ }\textbf {\bibinfo {volume} {63}},\ \bibinfo
  {pages} {042307} (\bibinfo {year} {2001})}\BibitemShut {NoStop}%
\bibitem [{\citenamefont {Fortunato}\ \emph {et~al.}(2002)\citenamefont
  {Fortunato}, \citenamefont {Viola}, \citenamefont {Hodges}, \citenamefont
  {Teklemariam},\ and\ \citenamefont {Cory}}]{fortunato2002implementation}%
  \BibitemOpen
  \bibfield  {author} {\bibinfo {author} {\bibfnamefont {E.~M.}\ \bibnamefont
  {Fortunato}}, \bibinfo {author} {\bibfnamefont {L.}~\bibnamefont {Viola}},
  \bibinfo {author} {\bibfnamefont {J.}~\bibnamefont {Hodges}}, \bibinfo
  {author} {\bibfnamefont {G.}~\bibnamefont {Teklemariam}},\ and\ \bibinfo
  {author} {\bibfnamefont {D.~G.}\ \bibnamefont {Cory}},\ }\bibfield  {title}
  {\bibinfo {title} {Implementation of universal control on a decoherence-free
  qubit},\ }\href@noop {} {\bibfield  {journal} {\bibinfo  {journal} {New
  Journal of Physics}\ }\textbf {\bibinfo {volume} {4}},\ \bibinfo {pages} {5}
  (\bibinfo {year} {2002})}\BibitemShut {NoStop}%
\bibitem [{\citenamefont {Yang}\ and\ \citenamefont
  {Gea-Banacloche}(2001)}]{Yang:01}%
  \BibitemOpen
  \bibfield  {author} {\bibinfo {author} {\bibfnamefont {C.-P.}\ \bibnamefont
  {Yang}}\ and\ \bibinfo {author} {\bibfnamefont {J.}~\bibnamefont
  {Gea-Banacloche}},\ }\bibfield  {title} {\bibinfo {title} {Three-qubit
  quantum error-correction scheme for collective decoherence},\ }\href
  {https://doi.org/10.1103/PhysRevA.63.022311} {\bibfield  {journal} {\bibinfo
  {journal} {Physical Review A}\ }\textbf {\bibinfo {volume} {63}},\ \bibinfo
  {pages} {022311} (\bibinfo {year} {2001})}\BibitemShut {NoStop}%
\bibitem [{\citenamefont {{D. Bacon, J. Kempe, D. P. DiVincenzo, D. A. Lidar,
  and K. B. Whaley}}(2001)}]{Bacon:Sydney}%
  \BibitemOpen
  \bibfield  {author} {\bibinfo {author} {\bibnamefont {{D. Bacon, J. Kempe, D.
  P. DiVincenzo, D. A. Lidar, and K. B. Whaley}}},\ }\bibfield  {title}
  {\bibinfo {title} {Encoded universality in physical implementations of a
  quantum computer},\ }in\ \href@noop {} {\emph {\bibinfo {booktitle}
  {Proceedings of the 1st International Conference on Experimental
  Implementations of Quantum Computation, Sydney, Australia}}},\ \bibinfo
  {editor} {edited by\ \bibinfo {editor} {\bibfnamefont {R.}~\bibnamefont
  {Clark}}}\ (\bibinfo  {publisher} {Rinton},\ \bibinfo {address} {Princeton,
  NJ},\ \bibinfo {year} {2001})\ p.\ \bibinfo {pages} {257},\ \Eprint
  {https://arxiv.org/abs/quant-ph/0102140} {quant-ph/0102140} \BibitemShut
  {NoStop}%
\bibitem [{\citenamefont {{J. Kempe, D. Bacon, D.P. DiVincenzo and K.B.
  Whaley}}(2001)}]{Kempe:01}%
  \BibitemOpen
  \bibfield  {author} {\bibinfo {author} {\bibnamefont {{J. Kempe, D. Bacon,
  D.P. DiVincenzo and K.B. Whaley}}},\ }\bibfield  {title} {\bibinfo {title}
  {{Encoded Universality from a Single Physical Interaction}},\ }\href
  {https://arxiv.org/abs/quant-ph/0112013} {\bibfield  {journal} {\bibinfo
  {journal} {Quant. Inf. Comput.}\ }\textbf {\bibinfo {volume} {1}},\ \bibinfo
  {pages} {33} (\bibinfo {year} {2001})}\BibitemShut {NoStop}%
\bibitem [{\citenamefont {DiVincenzo}\ \emph {et~al.}(2000)\citenamefont
  {DiVincenzo}, \citenamefont {Bacon}, \citenamefont {Kempe}, \citenamefont
  {Burkard},\ and\ \citenamefont {Whaley}}]{DiVincenzo:2000kx}%
  \BibitemOpen
  \bibfield  {author} {\bibinfo {author} {\bibfnamefont {D.~P.}\ \bibnamefont
  {DiVincenzo}}, \bibinfo {author} {\bibfnamefont {D.}~\bibnamefont {Bacon}},
  \bibinfo {author} {\bibfnamefont {J.}~\bibnamefont {Kempe}}, \bibinfo
  {author} {\bibfnamefont {G.}~\bibnamefont {Burkard}},\ and\ \bibinfo {author}
  {\bibfnamefont {K.~B.}\ \bibnamefont {Whaley}},\ }\bibfield  {title}
  {\bibinfo {title} {Universal quantum computation with the exchange
  interaction},\ }\href {https://doi.org/10.1038/35042541} {\bibfield
  {journal} {\bibinfo  {journal} {Nature}\ }\textbf {\bibinfo {volume} {408}},\
  \bibinfo {pages} {339} (\bibinfo {year} {2000})}\BibitemShut {NoStop}%
\bibitem [{\citenamefont {Weinstein}\ \emph {et~al.}(2023)\citenamefont
  {Weinstein}, \citenamefont {Reed}, \citenamefont {Jones}, \citenamefont
  {Andrews}, \citenamefont {Barnes}, \citenamefont {Blumoff}, \citenamefont
  {Euliss}, \citenamefont {Eng}, \citenamefont {Fong}, \citenamefont {Ha},
  \citenamefont {Hulbert}, \citenamefont {Jackson}, \citenamefont {Jura},
  \citenamefont {Keating}, \citenamefont {Kerckhoff}, \citenamefont {Kiselev},
  \citenamefont {Matten}, \citenamefont {Sabbir}, \citenamefont {Smith},
  \citenamefont {Wright}, \citenamefont {Rakher}, \citenamefont {Ladd},\ and\
  \citenamefont {Borselli}}]{weinstein2022universal}%
  \BibitemOpen
  \bibfield  {author} {\bibinfo {author} {\bibfnamefont {A.~J.}\ \bibnamefont
  {Weinstein}}, \bibinfo {author} {\bibfnamefont {M.~D.}\ \bibnamefont {Reed}},
  \bibinfo {author} {\bibfnamefont {A.~M.}\ \bibnamefont {Jones}}, \bibinfo
  {author} {\bibfnamefont {R.~W.}\ \bibnamefont {Andrews}}, \bibinfo {author}
  {\bibfnamefont {D.}~\bibnamefont {Barnes}}, \bibinfo {author} {\bibfnamefont
  {J.~Z.}\ \bibnamefont {Blumoff}}, \bibinfo {author} {\bibfnamefont {L.~E.}\
  \bibnamefont {Euliss}}, \bibinfo {author} {\bibfnamefont {K.}~\bibnamefont
  {Eng}}, \bibinfo {author} {\bibfnamefont {B.~H.}\ \bibnamefont {Fong}},
  \bibinfo {author} {\bibfnamefont {S.~D.}\ \bibnamefont {Ha}}, \bibinfo
  {author} {\bibfnamefont {D.~R.}\ \bibnamefont {Hulbert}}, \bibinfo {author}
  {\bibfnamefont {C.~A.~C.}\ \bibnamefont {Jackson}}, \bibinfo {author}
  {\bibfnamefont {M.}~\bibnamefont {Jura}}, \bibinfo {author} {\bibfnamefont
  {T.~E.}\ \bibnamefont {Keating}}, \bibinfo {author} {\bibfnamefont
  {J.}~\bibnamefont {Kerckhoff}}, \bibinfo {author} {\bibfnamefont {A.~A.}\
  \bibnamefont {Kiselev}}, \bibinfo {author} {\bibfnamefont {J.}~\bibnamefont
  {Matten}}, \bibinfo {author} {\bibfnamefont {G.}~\bibnamefont {Sabbir}},
  \bibinfo {author} {\bibfnamefont {A.}~\bibnamefont {Smith}}, \bibinfo
  {author} {\bibfnamefont {J.}~\bibnamefont {Wright}}, \bibinfo {author}
  {\bibfnamefont {M.~T.}\ \bibnamefont {Rakher}}, \bibinfo {author}
  {\bibfnamefont {T.~D.}\ \bibnamefont {Ladd}},\ and\ \bibinfo {author}
  {\bibfnamefont {M.~G.}\ \bibnamefont {Borselli}},\ }\bibfield  {title}
  {\bibinfo {title} {Universal logic with encoded spin qubits in silicon},\
  }\href {https://doi.org/10.1038/s41586-023-05777-3} {\bibfield  {journal}
  {\bibinfo  {journal} {Nature}\ }\textbf {\bibinfo {volume} {615}},\ \bibinfo
  {pages} {817} (\bibinfo {year} {2023})}\BibitemShut {NoStop}%
\bibitem [{\citenamefont {Xia}\ \emph {et~al.}(2011)\citenamefont {Xia},
  \citenamefont {Uhrig},\ and\ \citenamefont {Lidar}}]{Xia:2011uq}%
  \BibitemOpen
  \bibfield  {author} {\bibinfo {author} {\bibfnamefont {Y.}~\bibnamefont
  {Xia}}, \bibinfo {author} {\bibfnamefont {G.~S.}\ \bibnamefont {Uhrig}},\
  and\ \bibinfo {author} {\bibfnamefont {D.~A.}\ \bibnamefont {Lidar}},\
  }\bibfield  {title} {\bibinfo {title} {Rigorous performance bounds for
  quadratic and nested dynamical decoupling},\ }\href
  {http://link.aps.org/doi/10.1103/PhysRevA.84.062332} {\bibfield  {journal}
  {\bibinfo  {journal} {Phys. Rev. A}\ }\textbf {\bibinfo {volume} {84}},\
  \bibinfo {pages} {062332} (\bibinfo {year} {2011})}\BibitemShut {NoStop}%
\bibitem [{\citenamefont {Lidar}(2014)}]{lidarReviewDecoherenceFree2014}%
  \BibitemOpen
  \bibfield  {author} {\bibinfo {author} {\bibfnamefont {D.~A.}\ \bibnamefont
  {Lidar}},\ }\bibinfo {title} {Review of decoherence-free subspaces, noiseless
  subsystems, and dynamical decoupling},\ in\ \href
  {http://dx.doi.org/10.1002/9781118742631.ch11} {\emph {\bibinfo {booktitle}
  {Quantum Information and Computation for Chemistry}}}\ (\bibinfo  {publisher}
  {John Wiley \& Sons, Inc.},\ \bibinfo {year} {2014})\ pp.\ \bibinfo {pages}
  {295--354}\BibitemShut {NoStop}%
\bibitem [{\citenamefont {McKay}\ \emph {et~al.}(2018)\citenamefont {McKay},
  \citenamefont {Alexander}, \citenamefont {Bello}, \citenamefont {Biercuk},
  \citenamefont {Bishop}, \citenamefont {Chen}, \citenamefont {Chow},
  \citenamefont {C{\'o}rcoles}, \citenamefont {Egger}, \citenamefont {Filipp},
  \citenamefont {Gomez}, \citenamefont {Hush}, \citenamefont {{Javadi-Abhari}},
  \citenamefont {Moreda}, \citenamefont {Nation}, \citenamefont {Paulovicks},
  \citenamefont {Winston}, \citenamefont {Wood}, \citenamefont {Wootton},\ and\
  \citenamefont {Gambetta}}]{mckayQiskitBackendSpecifications2018}%
  \BibitemOpen
  \bibfield  {author} {\bibinfo {author} {\bibfnamefont {D.~C.}\ \bibnamefont
  {McKay}}, \bibinfo {author} {\bibfnamefont {T.}~\bibnamefont {Alexander}},
  \bibinfo {author} {\bibfnamefont {L.}~\bibnamefont {Bello}}, \bibinfo
  {author} {\bibfnamefont {M.~J.}\ \bibnamefont {Biercuk}}, \bibinfo {author}
  {\bibfnamefont {L.}~\bibnamefont {Bishop}}, \bibinfo {author} {\bibfnamefont
  {J.}~\bibnamefont {Chen}}, \bibinfo {author} {\bibfnamefont {J.~M.}\
  \bibnamefont {Chow}}, \bibinfo {author} {\bibfnamefont {A.~D.}\ \bibnamefont
  {C{\'o}rcoles}}, \bibinfo {author} {\bibfnamefont {D.}~\bibnamefont {Egger}},
  \bibinfo {author} {\bibfnamefont {S.}~\bibnamefont {Filipp}}, \bibinfo
  {author} {\bibfnamefont {J.}~\bibnamefont {Gomez}}, \bibinfo {author}
  {\bibfnamefont {M.}~\bibnamefont {Hush}}, \bibinfo {author} {\bibfnamefont
  {A.}~\bibnamefont {{Javadi-Abhari}}}, \bibinfo {author} {\bibfnamefont
  {D.}~\bibnamefont {Moreda}}, \bibinfo {author} {\bibfnamefont
  {P.}~\bibnamefont {Nation}}, \bibinfo {author} {\bibfnamefont
  {B.}~\bibnamefont {Paulovicks}}, \bibinfo {author} {\bibfnamefont
  {E.}~\bibnamefont {Winston}}, \bibinfo {author} {\bibfnamefont {C.~J.}\
  \bibnamefont {Wood}}, \bibinfo {author} {\bibfnamefont {J.}~\bibnamefont
  {Wootton}},\ and\ \bibinfo {author} {\bibfnamefont {J.~M.}\ \bibnamefont
  {Gambetta}},\ }\bibfield  {title} {\bibinfo {title} {Qiskit {{Backend
  Specifications}} for {{OpenQASM}} and {{OpenPulse Experiments}}},\ }\href
  {http://arxiv.org/abs/1809.03452} {\bibfield  {journal} {\bibinfo  {journal}
  {arXiv:1809.03452 [quant-ph]}\ } (\bibinfo {year} {2018})},\ \Eprint
  {https://arxiv.org/abs/1809.03452} {arXiv:1809.03452 [quant-ph]} \BibitemShut
  {NoStop}%
\bibitem [{\citenamefont {Tripathi}\ \emph {et~al.}(2023)\citenamefont
  {Tripathi}, \citenamefont {Chen}, \citenamefont {Levenson-Falk},\ and\
  \citenamefont {Lidar}}]{tripathi2023modeling}%
  \BibitemOpen
  \bibfield  {author} {\bibinfo {author} {\bibfnamefont {V.}~\bibnamefont
  {Tripathi}}, \bibinfo {author} {\bibfnamefont {H.}~\bibnamefont {Chen}},
  \bibinfo {author} {\bibfnamefont {E.}~\bibnamefont {Levenson-Falk}},\ and\
  \bibinfo {author} {\bibfnamefont {D.~A.}\ \bibnamefont {Lidar}},\ }\href@noop
  {} {\bibinfo {title} {Modeling low- and high-frequency noise in transmon
  qubits with resource-efficient measurement}} (\bibinfo {year} {2023}),\
  \Eprint {https://arxiv.org/abs/2303.00095} {arXiv:2303.00095 [quant-ph]}
  \BibitemShut {NoStop}%
\bibitem [{\citenamefont {Nielsen}\ and\ \citenamefont
  {Chuang}(2011)}]{Nielsen2011quantumcomputation}%
  \BibitemOpen
  \bibfield  {author} {\bibinfo {author} {\bibfnamefont {M.~A.}\ \bibnamefont
  {Nielsen}}\ and\ \bibinfo {author} {\bibfnamefont {I.~L.}\ \bibnamefont
  {Chuang}},\ }\href@noop {} {\emph {\bibinfo {title} {Quantum Computation and
  Quantum Information: 10th Anniversary Edition}}},\ \bibinfo {edition} {10th}\
  ed.\ (\bibinfo  {publisher} {Cambridge University Press},\ \bibinfo {address}
  {USA},\ \bibinfo {year} {2011})\BibitemShut {NoStop}%
\bibitem [{\citenamefont {Li}\ \emph {et~al.}(2011)\citenamefont {Li},
  \citenamefont {Nakahara}, \citenamefont {Poon}, \citenamefont {Sze},\ and\
  \citenamefont {Tomita}}]{liRecursiveEncodingDecoding2011a}%
  \BibitemOpen
  \bibfield  {author} {\bibinfo {author} {\bibfnamefont {C.-K.}\ \bibnamefont
  {Li}}, \bibinfo {author} {\bibfnamefont {M.}~\bibnamefont {Nakahara}},
  \bibinfo {author} {\bibfnamefont {Y.-T.}\ \bibnamefont {Poon}}, \bibinfo
  {author} {\bibfnamefont {N.-S.}\ \bibnamefont {Sze}},\ and\ \bibinfo {author}
  {\bibfnamefont {H.}~\bibnamefont {Tomita}},\ }\bibfield  {title} {\bibinfo
  {title} {Recursive encoding and decoding of the noiseless subsystem and
  decoherence-free subspace},\ }\href
  {https://doi.org/10.1103/PhysRevA.84.044301} {\bibfield  {journal} {\bibinfo
  {journal} {Physical Review A}\ }\textbf {\bibinfo {volume} {84}},\ \bibinfo
  {pages} {044301} (\bibinfo {year} {2011})}\BibitemShut {NoStop}%
\bibitem [{\citenamefont {Stine}(1989)}]{stine1989bootstrap}%
  \BibitemOpen
  \bibfield  {author} {\bibinfo {author} {\bibfnamefont {R.}~\bibnamefont
  {Stine}},\ }\bibfield  {title} {\bibinfo {title} {An introduction to
  bootstrap methods: Examples and ideas},\ }\href
  {https://doi.org/10.1177/0049124189018002003} {\bibfield  {journal} {\bibinfo
   {journal} {Sociological Methods \& Research}\ }\textbf {\bibinfo {volume}
  {18}},\ \bibinfo {pages} {243} (\bibinfo {year} {1989})}\BibitemShut
  {NoStop}%
\bibitem [{\citenamefont {Stoica}\ and\ \citenamefont
  {Selen}(2004)}]{stoica2004model}%
  \BibitemOpen
  \bibfield  {author} {\bibinfo {author} {\bibfnamefont {P.}~\bibnamefont
  {Stoica}}\ and\ \bibinfo {author} {\bibfnamefont {Y.}~\bibnamefont {Selen}},\
  }\bibfield  {title} {\bibinfo {title} {Model-order selection: a review of
  information criterion rules},\ }\href@noop {} {\bibfield  {journal} {\bibinfo
   {journal} {IEEE Signal Processing Magazine}\ }\textbf {\bibinfo {volume}
  {21}},\ \bibinfo {pages} {36} (\bibinfo {year} {2004})}\BibitemShut {NoStop}%
\end{thebibliography}
\end{document}